\documentclass[twocolummn,traditabstract]{aa}
\usepackage{txfonts}
\usepackage{graphicx}
\usepackage{natbib}
\usepackage{placeins}
\renewcommand{\deg}{\mbox{$^{\circ}$}}
\newcommand{\kms}{\mbox{km~s$^{-1}$}}
\newcommand{\mols}{\mbox{molec.~s$^{-1}$}}
\newcommand{\psj}{\textit{Planetary Science Journal}}
\usepackage[colorlinks=true, linkcolor=blue, citecolor=blue, filecolor=blue, urlcolor=blue]{hyperref}

\begin{document}

\title{Chemical composition of comets C/2021~A1 (Leonard) and C/2022~E3 (ZTF) from radio spectroscopy and the abundance of HCOOH and HNCO in comets
\thanks{Based on observations carried out with the IRAM-30m telescope.
IRAM is supported by INSU/CNRS (France), MPG (Germany), and IGN (Spain).}
\thanks{Odin is a Swedish-led satellite project funded jointly 
        by the Swedish National Space Board (SNSB), the Canadian Space Agency 
        (CSA), the National Technology Agency of Finland (Tekes), the
        Centre National d'\'Etudes Spatiales (CNES, France), and the European
        Space Agency (ESA). 
        The Swedish Space Corporation, today OHB Sweden, is the
        prime contractor, also responsible for Odin operations.}
\thanks{The radio spectra are available at the CDS via anonymous 
ftp to cdsarc.cds.unistra.fr (130.79.128.5)
or via http://cdsarc.cds.unistra.fr/cgi-bin/qcat?J/A+A/}
}
 
\author{N. Biver\inst{1}
  \and D. Bockel\'ee-Morvan\inst{1}
   \and B. Handzlik\inst{2}
   \and Aa. Sandqvist\inst{3}
   \and J. Boissier\inst{4}
   \and M.~N. Drozdovskaya\inst{5}
   \and R. Moreno\inst{1}
   \and J. Crovisier\inst{1}
   \and D.C. Lis\inst{6}
   \and M. Cordiner\inst{7,8}
   \and S. Milam\inst{7}
   \and N.X. Roth\inst{7,9} 
   \and B.P. Bonev\inst{9}
   \and N. Dello Russo\inst{10}
   \and R. Vervack\inst{10}
   \and C. Opitom\inst{11}
   \and H. Kawakita\inst{12}
   }
\institute{LESIA, Observatoire de Paris, PSL Research University, CNRS, 
  Sorbonne Universit\'e, Universit\'e Paris-Cit\'e,
  5 place Jules Janssen, F-92195 Meudon, France
 \and Astronomical Observatory, Jagiellonian University, ul. Orla 171, 30-244, Krak\'ow, Poland
 \and Stockholm Observatory, Stockholm University, SCFAB-AlbaNova, SE-10691 Stockholm, Sweden  
 \and IRAM, 300, rue de la Piscine, F-38406 Saint Martin d'H\`eres, France
 \and Physikalish-Meteorologisches Observatorium Davos und Weltstrahlungszentrum
      (PMOD/WRC) Dorfstrasse 33, CH-7260, Davos Dorf, Switzerland
 \and Jet Propulsion Laboratory, California Institute of Technology,
      4800 Oak Grove Drive, Pasadena, CA, 91109, USA 
 \and Solar System Exploration Division, Astrochemistry Laboratory Code 691,
      NASA-GSFC, Greenbelt, MD 20771, USA
 \and Department of Physics, Catholic University of America,
      Washington, DC 20064, USA
 \and Department of Physics, American University, Washington, DC, USA 
 \and Johns Hopkins University Applied Physics Laboratory,
      11100 Johns Hopkins Rd., Laurel, MD 20723, USA
 \and Institute for Astronomy, University of Edinburgh, Royal Observatory,
      Edinburgh, EH9 3HJ, UK
 \and Koyama Astronomical Observatory, Kyoto Sangyo University, Motoyama,
      Kamitamo, Kita, Kyoto 603-8555, Japan
}

   \titlerunning{Composition of comets C/2021 A1 (Leonard) and C/2022~E3 (ZTF)}
   \authorrunning{Biver et al.}
   \date{Received 30 May 2024, Accepted 26 July 2024}

   \abstract{We present the results of a molecular survey of long period comets
     C/2021~A1 (Leonard) and C/2022~E3 (ZTF). Comet C/2021~A1 was observed
     with the Institut de radioastronomie millim\'etrique (IRAM) 30-m radio
     telescope in November-December 2021 before
     perihelion (heliocentric distance 1.22 to 0.76~au) when it was
     closest to the Earth ($\approx0.24$~au).
     We observed C/2022~E3 in January-February 2023 with the {\it Odin}~1-m space
     telescope and IRAM~30-m, shortly after its perihelion at 1.11~au from the
     Sun, and when it was closest to the Earth ($\approx0.30$~au).
     Snapshots were obtained during 12--16 November 2021 period for comet
     C/2021~A1.
     Spectral surveys were undertaken over the 8--13 December 2021 period
     for comet C/2021~A1 (8~GHz bandwidth at 3~mm, 16~GHz at 2~mm and 61~GHz
     in the 1~mm window) and over the 3--7 February 2023 period for comet
     C/2022~E3 (25~GHz at 2~mm and 61~GHz at 1~mm).
     We report detections of 14 molecular species (HCN, HNC, CH$_3$CN,
     HNCO, NH$_2$CHO, CH$_3$OH, H$_2$CO, HCOOH, CH$_3$CHO,
     H$_2$S, CS, OCS, C$_2$H$_5$OH and aGg'-(CH$_2$OH)$_2$) in both comets.
     In addition, HC$_3$N and CH$_2$OHCHO were marginally detected in C/2021~A1
     and CO and H$_2$O (with {\it Odin}) were detected in C/2022~E3.
     The spatial distribution of several species (HCN, HNC, CS, H$_2$CO, HNCO,
     HCOOH, NH$_2$CHO and CH$_3$CHO) is investigated.
     Significant upper limits on the abundances of other
     molecules and isotopic ratios are also presented.
     The activity of comet C/2021~A1 did not vary significantly between 13
     November and 13 December 2021, when observations stopped, just before it started
     to exhibit major outbursts seen in the visible and from observations of the
     OH radical. Short-term variability in the outgassing of comet C/2022~E3 on
     the order of $\pm$20\% is present and possibly linked to its ~8h rotation
     period. Both comets exhibit rather low abundances relative to water for
     volatiles species such as CO ($<2$\%) and H$_2$S (0.15\%).
     Methanol is also rather depleted in comet C/2021~A1 (0.9\%).
     Following their revised photo-destruction rates, HNCO and HCOOH
     abundances in comets observed at millimetre wavelengths have
     been reevaluated. Both molecules are relatively enriched in these
     two comets ($\sim$0.2\% relative to water).
     Since the combined abundance of these two acids (0.1 to 1\%) is close
     to that of ammonia in comets, we cannot exclude that these species
     could be produced by the dissociation of ammonium formate and
     ammonium cyanate if present in comets.}

\maketitle

\section{Introduction}
Comets are the most pristine remnants of the formation of the
Solar System 4.6 billion years ago. They sample some of the oldest and most
primitive material in the Solar System, including ices, and are
thus our best window on the volatile composition of the solar
proto-planetary disk. Comets may also have played a role in the
delivery of water and organic material to the early Earth
\citep[see][and references therein]{Har11}.
The latest simulations of the early Solar System's evolution \citep{Bra13,Obr14}
suggest a more complex scenario. On the one hand, ice-rich bodies formed beyond
Jupiter may have been implanted early in the outer asteroid belt and
participated in the supply of water to the Earth. On the other hand, current
comets are coming from either the Oort Cloud or the scattered disk of the Kuiper
belt and may have formed in the same trans-Neptunian region, sampling the same
diversity of formation conditions.
Understanding the diversity in composition and isotopic
ratios of cometary materials is thus essential for assessing such
scenarios \citep{Alt03,Boc15}.

Comet C/2021~A1 (Leonard) is a long-period dynamically old Oort-Cloud Comet
(OCC, initial semi-major axis of 2000~au, inclination of 132.7\deg) that reached
perihelion at a heliocentric distance $r_h$=0.615~au on 3.3 January 2022 UT.
It came as close as 0.234~au from the Earth on 12.6 December 2021.
It was anticipated to become a bright
comet visible to the naked eye, but it under-performed until 14 December
when it was in solar conjunction. Then monitoring of the outgassing rate via
observations of the OH radical \citep{Cro21} and visual magnitudes showed
strong outbursts of activity, repeating on a $\sim$5~days period from
15 December to 7 January
2022\footnote{http://www.aerith.net/comet/catalog/2021A1/2021A1.html}.
Then, it developed spectacular ion and dust tails while being mostly
observable from the southern hemisphere. Later on, the comet faded rapidly
and images taken in April--May 2022 suggested that it was disintegrating
during this outbursting phase. The derived pre-disintegration nucleus radius
was $0.6\pm0.2$~km \citep{Jew23}.

We observed comet C/2021~A1 with the Institut de radioastronomie millim\'etrique
(IRAM) 30-m telescope briefly between 12.0 and 16.4 November, and extensively
between 8.4 and 13.4 December 2021 UT, before it reached too low declinations
for the northern observatories.

\begin{figure*}[]
\centering
\resizebox{\hsize}{!}{
  \includegraphics[angle=0,width=0.8\textwidth]{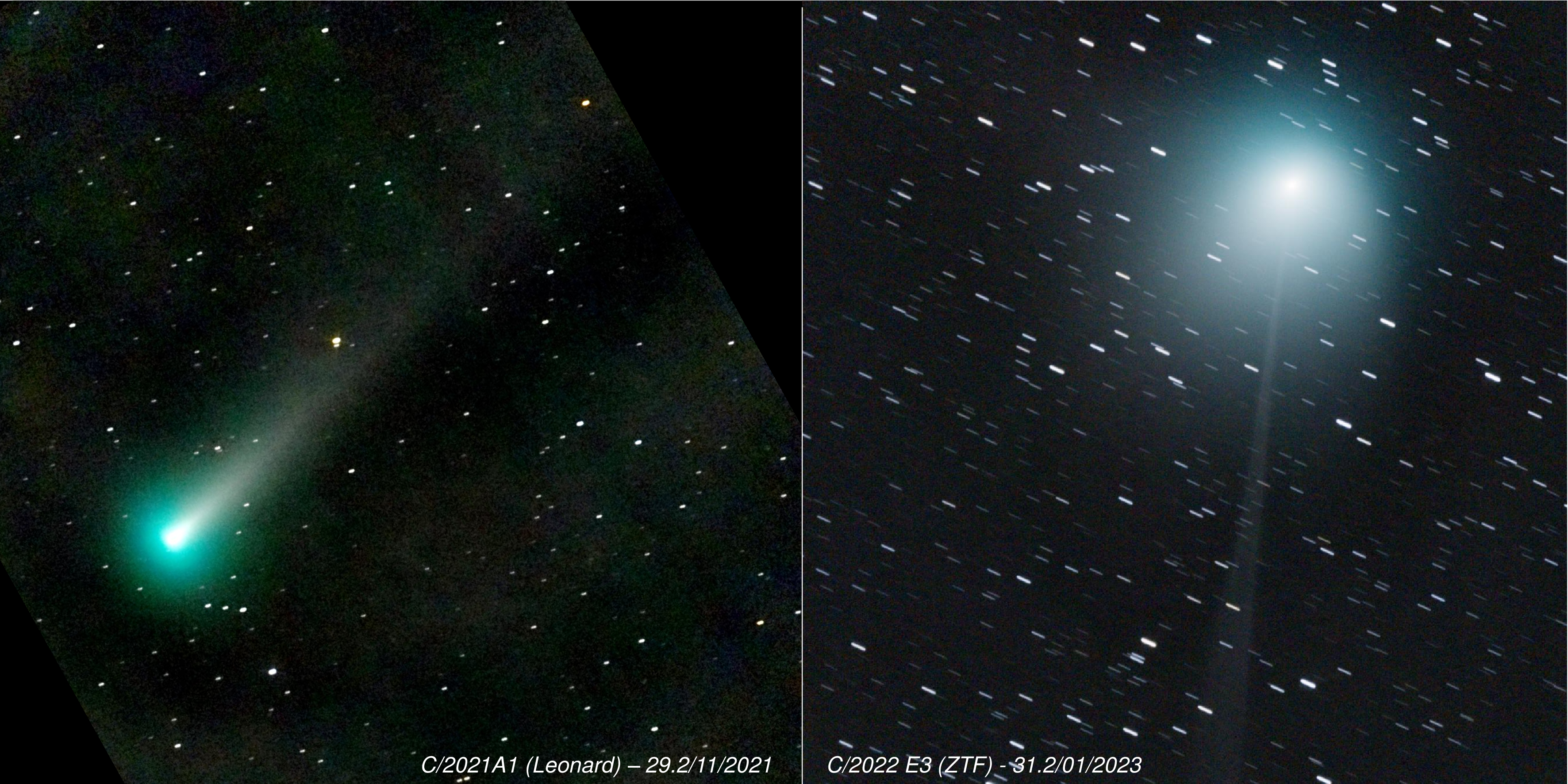}
}
\caption{Optical images of comets  C/2021~A1 (Leonard) (left) and C/2022~E3 (ZTF) (right), \copyright N. Biver.\protect\\
  Left panel: telescopic image of comet C/2021~A1 (Leonard) on
  29 November 2021 at 4:41 UTC (94~s exposure).\protect\\
  Right panel: telescopic image of comet C/2022~E3 (ZTF) on 31 January 2023
  at 4:53 UTC (126~s exposure).
  Images were taken at the focus of a 40.7-cm telescope at F/D=4.3 from Eure-et-Loir (France).
  Field of view is 45$\times$45\arcmin, North is up.}
\label{figimgleonardztf}
\end{figure*}

Comet C/2022~E3 (ZTF) is also a long-period dynamically old OCC
(initial semi-major axis of 1400~au, inclination of 109.2\deg). It reached
perihelion at $r_h$=1.112~au on 12.8 January 2023 UT. It was closest to the
Earth on 1.8 February 2023 at 0.284 au, and reached naked eye visibility
(total magnitude of 4.8). It attracted public attention and was the target
of a worldwide campaign because it was discovered 11 months ahead of its
peak brightness, expected to happen at the end of January 2023 when the comet
was circumpolar for northern hemisphere and visible to the naked eye all night
(Fig.~\ref{figimgleonardztf}).
The comet was further advertised in NASA news releases at the end of 2021\footnote{https://science.nasa.gov/resource/whats-up-december-2021/, https://www.nasa.gov/image-article/views-of-comet-leonard-from-two-sun-watching-spacecraft/}.

We observed comet C/2022~E3 with the {\it Odin} space telescope from 19.3 to 20.3
January 2023 \citep{Biv23b} and with the IRAM-30m radio telescope between 3.7 and 7.1 February.
Due to its high activity and a favourable
apparition, this comet was the focus of an international observing campaign,
from the radio (OH observed with the Nan\c{c}ay Radio Telescope (NRT) and the
Green Bank Telescope) to the infrared (ground based observations with IRTF and
Keck-NIRSPEC and from space with the {\it James Webb} Space Telescope on 28 February
and 4 March, \citet{Mil23}).

In this paper, we report clear detections of HCN, HNC, CH$_3$CN,
HNCO, NH$_2$CHO, CH$_3$OH, H$_2$CO, HCOOH, H$_2$S, and CS in both comets, the
more marginal detections of CO, HC$_3$N, CH$_3$CHO, OCS, (CH$_2$OH)$_2$,
CH$_2$OHCHO and C$_2$H$_5$OH, obtained by averaging several lines,
as well as significant upper limits on the abundances of SO, SO$_2$,
H$_2$CS, CH$_2$CO, PH$_3$ and other species.

In addition, following the revised photo-destruction rates for several
molecules by \citet{Hro23}, especially of HNCO and HCOOH for which the lifetime
has been reduced by nearly one order of magnitude in comparison to
previously published or assumed destruction rates \citep{Hea17,Hue15,Biv21a},
we have reevaluated their abundances in all comets in which they were
observed or searched for. 

In Sect.~\ref{sect-obs} we present the observations and spectra of the
detected molecules.
The information extracted from the observations to analyse the data and
compute production rates is provided in Sect.~\ref{sect-analysis}.
In Sect.~\ref{sect-results}, we present the retrieved
production rates and abundances or upper limits, which are discussed and
compared to other comets. The new analysis of HNCO and HCOOH observations
are detailed in Sect.~\ref{sect-hncohcooh} followed by the conclusions in
Sect.~\ref{sect-discussion}.

\section{Observations}
\label{sect-obs}

\subsection{Observations of comet C/2021~A1 with the IRAM-30m}\label{sect-iramleonard}

Comet C/2021~A1 (Leonard) (Fig.~\ref{figimgleonardztf}) was the
focus of a worldwide campaign as it was
expected to become very bright at its closest approach to Earth (total visual
magnitude $m_1\sim4$). It was less active than anticipated until 14 December
when it started to undergo a series of recurrent outbursts bringing it to a
maximum brightness of around $m_1=3$ between 15 and 25 December 2021
 
It was the target of the observing proposal 100-21
scheduled at the IRAM 30-m telescope between 8 and 13 December 2021.
Weather conditions were marginal during the first four days (half of the
day too icy, windy or foggy to observe, with precipitable
water vapour ($pwv$) in the 3 to 7~mm range otherwise).
On the first day (observations were taking place during daytime),
when observations resumed after an ice storm, the leftover ice or the impact of
de-icing on the antenna reduced the beam efficiency by a factor of 3
(at 1~mm wavelength) to 1.5 (at 2~mm).
The last two days offered good weather conditions with $pwv$ = 1 to 3~mm
(Table~\ref{tablogleonard}).
A few other snapshots (less than 0.5~h of observations) were obtained
between 12 and 16 November during observing run 001-21 when comet
67P/Churyumov-Gerasimenko was too high in elevation for tracking \citep{Biv23}.

Observations were obtained in wobbler switching mode with the secondary
(wobbling) mirror alternating pointing between the ON and 
OFF positions separated by 180\arcsec~ every 2 seconds.
The wobbler was not working on 14--16 November 2021 and we had to use to the
position switching mode (PSW). The reference OFF positions in
PSW mode were at 300\arcsec~ from the source, alternating ON and OFF every
15 seconds.
We used the EMIR \citep[Eight MIxer Receiver,][]{Car12} 3~mm, 2~mm, and
1~mm band receivers in 2SB mode connected to the FTS (Fast Fourier
Transform Spectrometer) and the VESPA (VErsatile SPectrometer Array)
high-resolution spectrometer (Table~\ref{tablogleonard}). The FTS offers
an instantaneous bandwidth of 16~GHz in two polarisations with 200~kHz
spectral resolution. The VESPA autocorrelator was optimised to provide
4--6 windows of 20 to 40~kHz spectral resolution on lines of interest in the
centre of the Intermediate Frequency windows ($6.25\pm0.25$~GHz).

\begin{figure}[]
\centering
\resizebox{\hsize}{!}{
  \includegraphics[angle=270,width=0.4\textwidth]{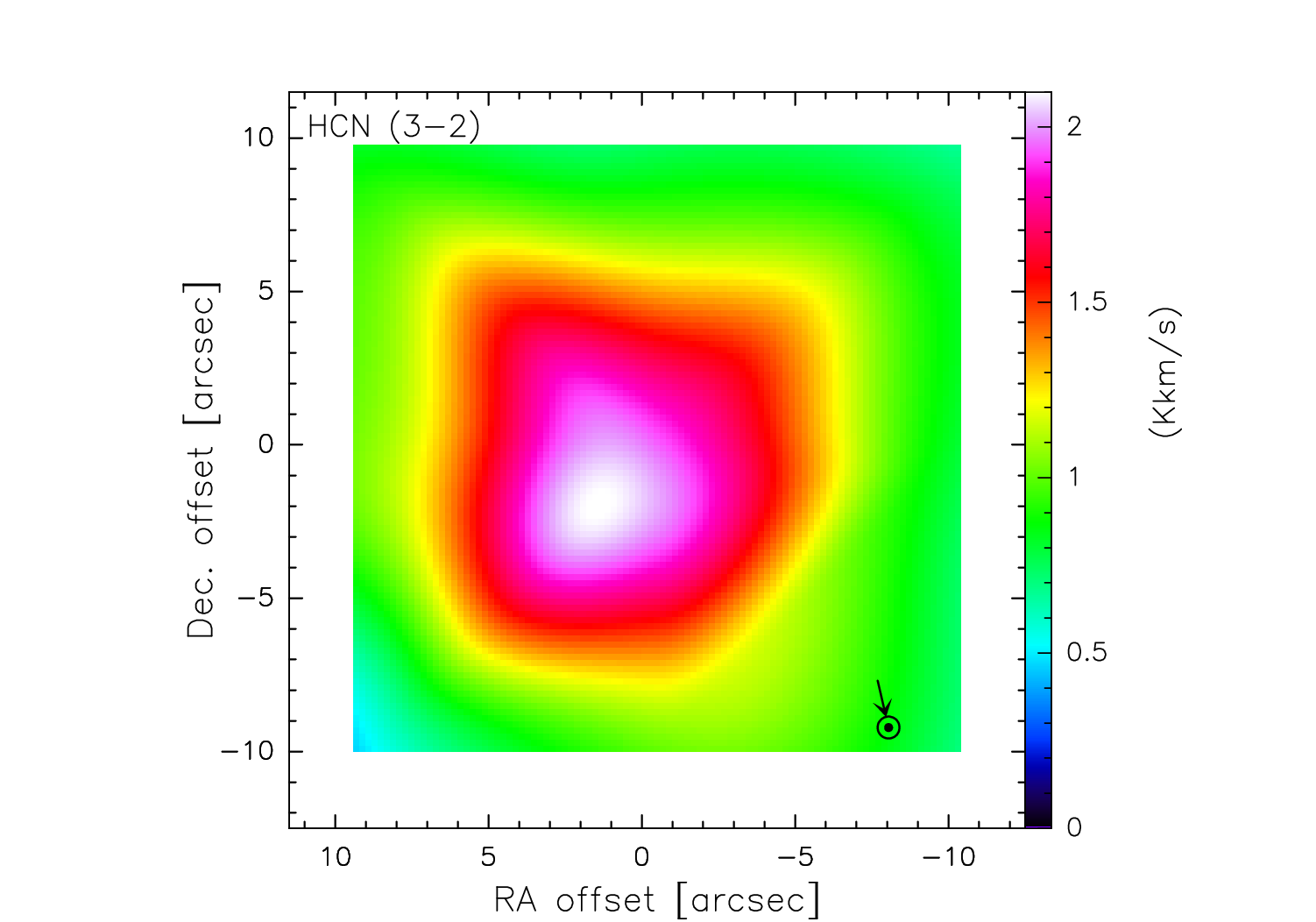}
}
\caption{Coarse map of the HCN(3-2) line integrated intensity in comet
  C/2021~A1 on 13.4 December 2021. The beam size is 9.3\arcsec.
  Solar phase angle was 158.6\deg.
  The direction of the Sun (PA=193.1\deg) is indicated at the lower right.}
\label{figmaphcn32}
\end{figure}

Comet C/2021~A1 was tracked with the IRAM-30m control software (NCS)
using the latest JPL Horizons\footnote{https://ssd.jpl.nasa.gov/horizons.cgi}
orbital elements available:\#15 in November, \#17 on 8 December,
and \#18 on 9--13 December 2021. Offsets (up to 7\arcsec) were added during
the observation to take into account the difference between the position
computed by the NCS and the very latest (1--3 days old) astrometric
measurements.
This was a critical point as for such a new comet relatively close to Earth,
positions errors can easily reach the beam size (10\arcsec~ at 240~GHz).
The comet position was also checked in real time using coarse (5--9 points)
mapping of HCN $J$=3--2 (Fig.\ref{figmaphcn32}).
Ephemeris offsets have been computed afterwards using the JPL\#19 orbit solution
(yielding values in agreement with the astrometry that was used).
Final pointing offsets for each observation were computed after taking into
account reconstruction of pointing errors and finally from intensity maps
of HCN, that generally yielded residual offsets of less than 1.2\arcsec.

\setcounter{table}{0}
\begin{table*}
\caption[]{Log of observations of comet C/2021~A1 (Leonard) in 2021.}\label{tablogleonard}\vspace{-0.2cm}
\begin{center}
\begin{tabular}{rllcccccc}
\hline\hline
UT date & $<r_{h}>$ & $<\Delta>$ & Phase\tablefootmark{a} & Tel. & Integ. time & pwv\tablefootmark{b} & Mode\tablefootmark{d} & Freq. range \\
$($mm/dd.d--dd.d) & (au)  & (au) &  &  & (min)\tablefootmark{c} & (mm) & & (GHz)  \\
\hline
 11/12.18--12.20 & 1.218 & 1.192 & 48.5\deg & IRAM &  28 & 4.8 & WSW & 248.7-256.5, 264.4-272.2 \\
 11/13.19--13.22 & 1.203 & 1.157 & 49.5\deg & IRAM &  33 & 6.6 & WSW & 146.9-154.7, 162.6-170.4 \\
 11/14.16--14.20 & 1.188 & 1.123 & 50.6\deg & IRAM &  34 & 4.4 & PSW & 248.7-256.5, 264.4-272.2 \\
    14.22--14.24 & 1.187 & 1.121 & 50.7\deg & IRAM &  16 & 5.0 & PSW & 209.7-217.5, 225.4-233.1 \\
 11/15.18--15.22 & 1.172 & 1.087 & 51.8\deg & IRAM &  36 & 3.1 & PSW & 146.9-154.7, 162.6-170.4 \\
 11/16.15--16.20 & 1.157 & 1.053 & 52.9\deg & IRAM &  44 & 1.3 & PSW & 248.7-256.5, 264.4-272.2 \\
\hline
 12/08.44--08.47 & 0.826 & 0.288 & 115.5\deg & IRAM &  42 &  2.0 &  WSW & 248.4-256.5, 264.4-272.5 \\
    08.52--08.58 & 0.824 & 0.286 & 116.4\deg & IRAM &  64 &  2.5 &  WSW & 146.6-154.8, 162.6-170.7 \\
 12/09.41--09.58 & 0.813 & 0.266 & 124.0\deg & IRAM & 150 &  4-6 &  WSW & 248.4-256.5, 264.4-272.5 \\
 12/10.29--10.44 & 0.800 & 0.250 & 131.8\deg & IRAM & 136 &  4-7 &  WSW & 248.4-256.5, 264.4-272.5 \\
    10.46--10.54 & 0.798 & 0.248 & 133.0\deg & IRAM & 124 & 8-19 &  WSW & 146.6-154.8, 162.6-170.7 \\
    10.54--10.59 & 0.798 & 0.248 & 133.5\deg & IRAM & 124 &  4-8 &  WSW & 146.6-154.8, 162.6-170.7 \\
 12/11.51--11.53 & 0.785 & 0.237 & 142.7\deg & IRAM &  28 &  6.0 &  WSW & 82.9-90.9,224.8-232.8 \\
    11.57--11.59 & 0.785 & 0.237 & 143.5\deg & IRAM &  28 &  5.7 &  WSW & 248.4-256.5, 264.4-272.5 \\
    11.59--11.66 & 0.784 & 0.237 & 143.7\deg & IRAM &  65 &  3-5 &  WSW & 248.4-256.5, 264.4-272.5 \\
 12/12.38--12.41 & 0.774 & 0.233 & 150.7\deg & IRAM &   9 &  2.8 &  WSW & 248.4-256.5, 264.4-272.5 \\
    12.42--12.54 & 0.773 & 0.233 & 151.4\deg & IRAM & 113 &  1.8 &  WSW & 240.0-248.1, 256.0-264.1 \\
    12.56--12.65 & 0.772 & 0.233 & 152.5\deg & IRAM & 109 &  2.0 &  WSW & 209.4-217.5, 225.4-233.4 \\
 12/13.38--13.41 & 0.762 & 0.236 & 158.0\deg & IRAM &  19 &  0.7 &  WSW & 248.4-256.5, 264.4-272.5 \\
    13.43--13.50 & 0.761 & 0.236 & 158.4\deg & IRAM &  79 &  0.4 &  WSW & 217.5-225.5, 234.5-241.5 \\
    13.52--13.57 & 0.760 & 0.237 & 158.7\deg & IRAM &  56 &  1.0 &  WSW & 240.0-248.1, 256.0-264.1 \\
    13.60--13.66 & 0.759 & 0.237 & 159.1\deg & IRAM &  62 &  2.1 &  WSW & 146.6-154.8, 162.6-170.7 \\
\hline
\end{tabular}
\end{center}
\tablefoot{
  \tablefoottext{a}{Phase angle.}
  \tablefoottext{b}{Mean precipitable water vapour in the atmosphere above the telescope.}\\
  \tablefoottext{c}{Total (offset positions included) integration time (ON+OFF) on the source.}
  \tablefoottext{d}{Observing mode: WSW = Wobbler Switching (reference at $\pm3$\arcmin);
    PSW = Position Switching (reference at 5\arcmin).}
}
\end{table*}

\begin{figure}[]
\centering
\resizebox{\hsize}{!}{
  \includegraphics[angle=0,width=0.9\textwidth]{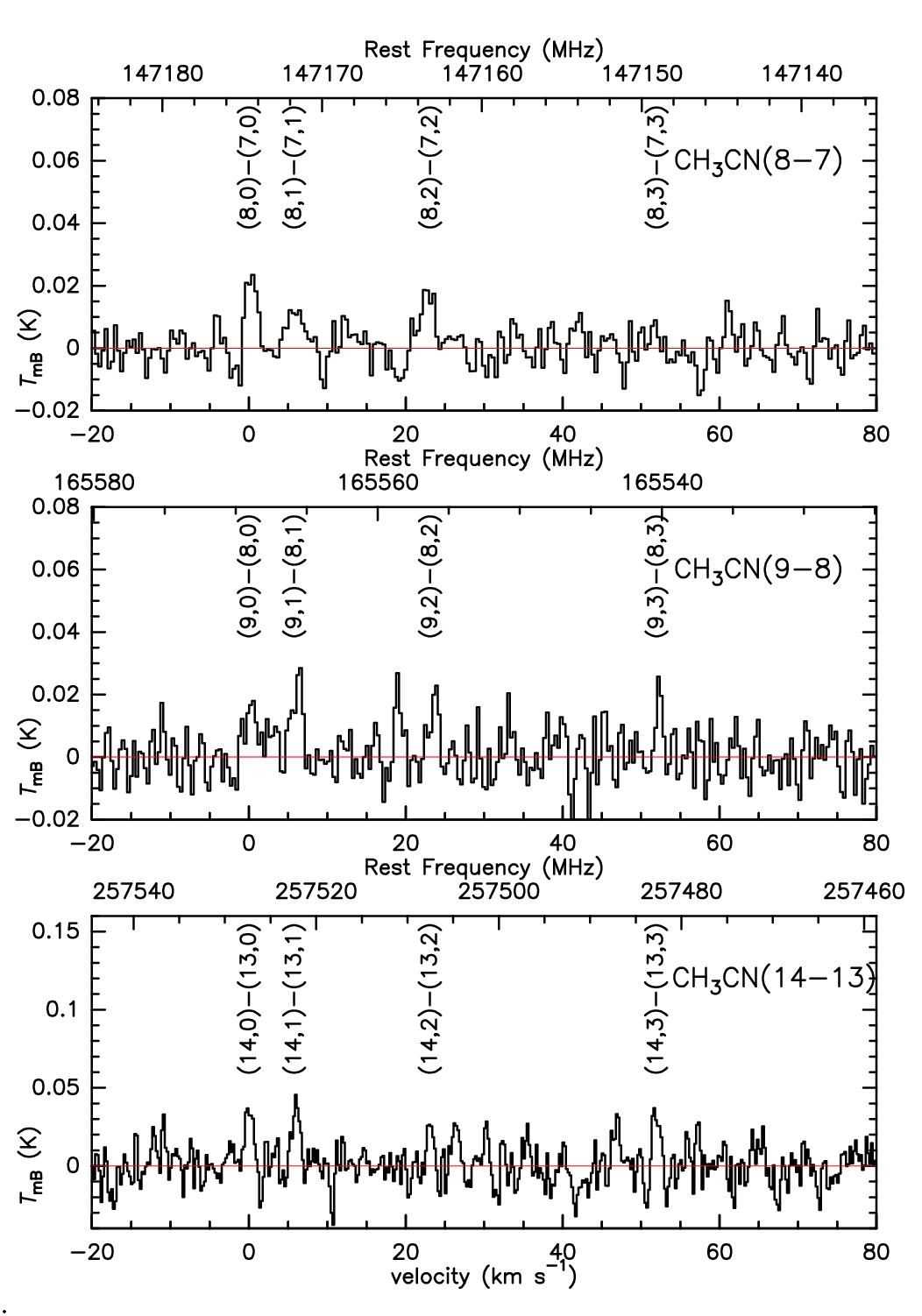}}
\caption{CH$_3$CN series of four lines observed with IRAM-30m in comet
  C/2021~A1 on 8--13 December 2021 (average intensity). 
  The vertical axis is main beam brightness temperature in K,
  horizontal axis is Doppler velocity in the rest frame of the comet
  with respect to the $(J,0)-(J-1,0)$ line, with frequencies in the rest
  frame of the comet given on the upper axis.}
\label{figspch3cnleonard}
\end{figure}

\begin{figure*}[]
\centering
\resizebox{0.85\hsize}{!}{
  \includegraphics[angle=270,width=0.85\textwidth]{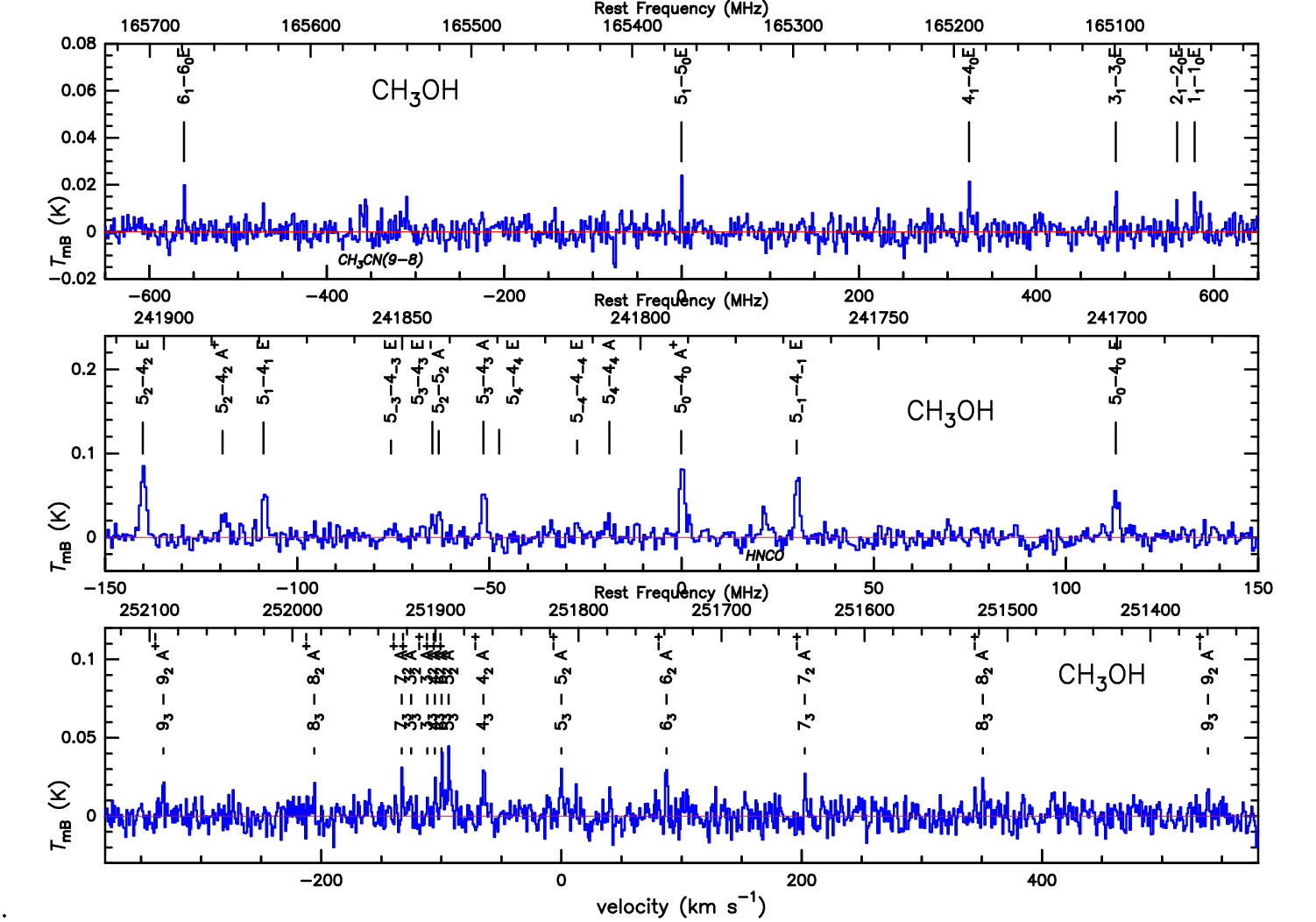}}
\caption{CH$_3$OH series of lines around 166, 242 and 252~GHz observed with
  IRAM-30m in comet C/2021~A1 on 8--13 December 2021 (average intensity). 
  The vertical axis is main beam brightness temperature in K,
  horizontal axis is Doppler velocity in the rest frame of the comet
  with respect to the main line, with frequencies in the rest
  frame of the comet given on the upper axis.}
\label{figspch3ohleonard}
\end{figure*}

Representative spectra of comet C/2021~A1 (Leonard) are provided in
Figs.~\ref{figspnovleonard}--\ref{figspcomsleonard}. They show individual lines
or averages of several lines for the complex organics from the observations
centred on the nucleus. Detailed line intensities are provided in
Tables~\ref{tabobsleonard} and \ref{tabobssumleonard} and for some molecules
with useful spatial information in Table~\ref{tabqdistrileonard} of
section~\ref{sect-distri}. Spectra showing the full spectral coverage with
the wide band FTS are shown in
Figs.~\ref{figsurveyleonard2mm2} and \ref{figsurveyleonard1mm6}.

\subsection{Observations of comet C/2022~E3 (ZTF) with {\it Odin}}\label{sect-odin}

\begin{figure}[]
\centering
\resizebox{\hsize}{!}{
  \includegraphics[angle=270,width=0.4\textwidth]{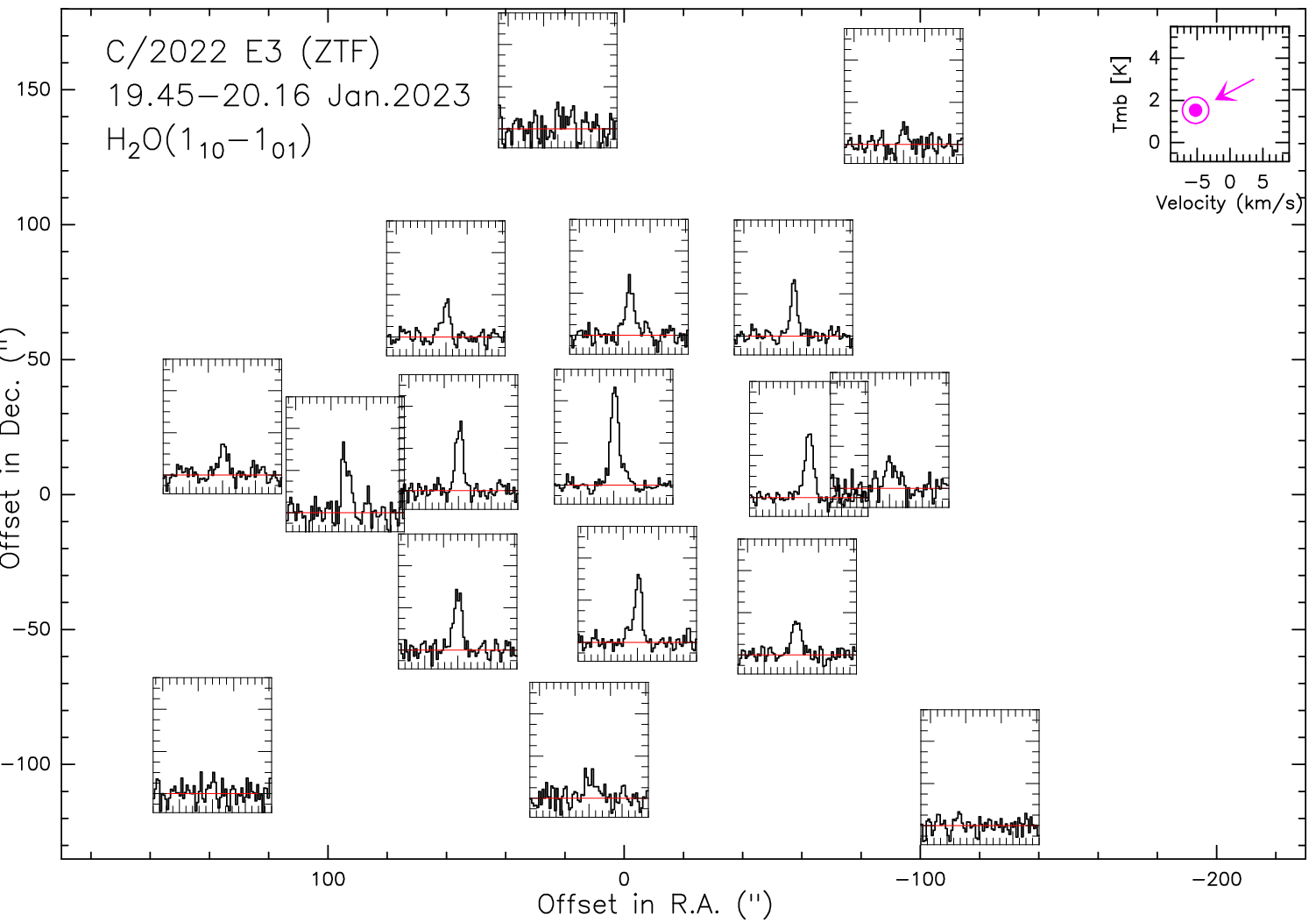}}
\caption{Map of the average spectra of the H$_2$O($1_{10}-1_{01}$) line obtained
  with {\it Odin} between 19.45 and 20.16 January 2023 in comet C/2022~E3 (ZTF),
  as a function of pointing offset (in arcsec). A baseline in red is plotted
  on each spectrum.
  Solar phase angle was 61.6\deg, and the direction
  of the Sun (PA=120.2\deg) is indicated at the upper right with scales
  for individual AC2 spectra.}
\label{figmaph2oztf}
\end{figure}

Comet C/2022~E3 (ZTF) was favourably placed (solar elongation between 60\deg
and 120\deg) for {\it Odin} during one of its yearly astronomy science operations.
15 orbits ($\sim24$h) were dedicated to observe the H$_2$O($1_{10}-1_{01}$)
line at 556.9~GHz in this comet.
{\it Odin} \citep{Fri03} is a small satellite in a
polar orbit (period 95~min) equipped with a 1.1~m sub-millimetre radiometer.
{\it Odin} houses four sub-mm receivers covering the 486--504 GHz and 541-581~GHz range with
3 backends: one acousto-optical-spectrometer (AOS) with 1~GHz band width
and 1.2~MHz spectral resolution and two autocorrelators (AC1 and AC2), covering
120~MHz with 140~kHz resolution in their highest resolution mode.
Two of the receivers ``555B2''
and ``549A1'' were supposed to observe the 556.9~GHz line, but the 549A1
receiver did not lock, and we only obtained data from the 555B2 receiver
connected to the AOS and AC2.
We aimed at alternating between pointing on the comet position and
coarse $3\times3$ points maps spaced by 60 or 120\arcsec. The beam size was 127\arcsec.
{\it Odin} achieved successful tracking of the comet on most of 14 of the 15 orbits
scheduled from 19.3 to 20.3 January 2023 UTC (Table~\ref{tablogztf}).
Part of the data from the 13th and 14th orbits was lost also due to
memory overload. Reconstruction of the attitude of {\it Odin} did not always converge
with good accuracy due to the lack of reference stars in the
star-trackers field of view, but most of the pointings were within 15\arcsec
($\approx12$\% of the beam width) of the comet position.
Figure~\ref{figmaph2oztf} shows the result from the combination of all the maps.

\subsection{Observations of comet C/2022~E3 (ZTF) with the IRAM-30m}\label{sect-iramztf}

Comet C/2022~E3 (ZTF) was the target of the observing proposal 097-22
scheduled at the IRAM 30-m telescope between 3 and 8 February 2023.
Weather conditions were very good to good during the first four days
(but too windy followed by a technical issue with the antenna
temperature control during 2/3 of the first night).
The last night (February 7/8) was lost due to bad weather (snow + wind).
Pointing and focus stability was not very good at the beginning of
the nights. Very low opacity on 5.0 February (0.2 to 1~mm $pwv$)
enabled a short observation at 177--185~GHz, but weather conditions degraded
and observing time was too limited to get a useful result on the H$_2$O
line at 183.3~GHz. Table~\ref{tablogztf} provides a log of the observations.

Wobbler switching mode with the secondary (wobbling) mirror alternating
pointing between the ON and OFF positions separated by 120\arcsec~
every 2 seconds could only be used on the last night as it had again issues
as in November 2021. We had to rely on position switching mode (PSW) with
reference OFF positions at 240\arcsec, alternating ON and OFF every 15 seconds.
Even using symmetric mode (ON OFF OFF ON sequence), alternating the reference
position in +RA and -RA, leaves some residual at the position of the strong
ozone atmospheric lines, where the spectra are noisier with poorer baselines,
especially around 231.13, 237.17, 242.34, 249.80, 249.97 and 267.28~GHz
(Fig.~\ref{figsurveyztf1mm6}). As for comet C/2021~A1, we used the EMIR
receivers and the FTS and VESPA spectrometers (Sect.~\ref{sect-iramleonard}).

Comet C/2022~E3 was tracked with the IRAM-30m control software (NCS)
using the latest JPL Horizons 
orbital elements available at the beginning of the observations, JPL\#41,
adding up offsets ($\sim$5\arcsec~ in declination) from the latest astrometric
measurements available and confirmed by coarse mapping of the strongest lines in the setup.
Final offset in the reduced data take into account a more recent (JPL\#45)
orbit solution and reconstructed pointing corrections of the antenna to cope
with potential approximations during real time observations.

\begin{figure}[]
\centering
\resizebox{\hsize}{!}{
  \includegraphics[angle=270,width=0.4\textwidth]{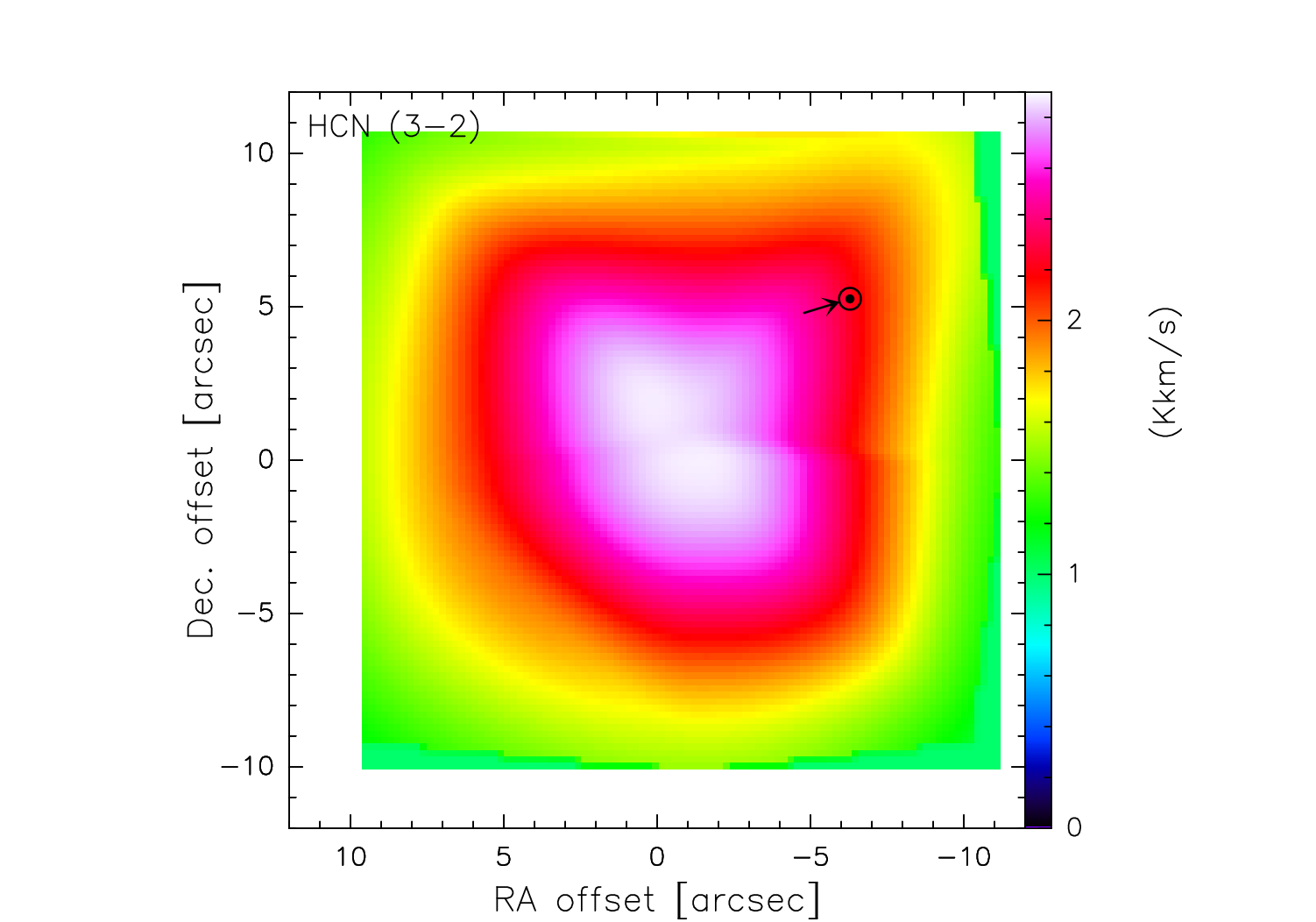}}
\caption{Coarse (3$\times$3 points) map of the HCN(3-2) line integrated
  intensity on 4.67 February 2023 in comet C/2022~E3.
  Solar phase angle was 45.5\deg. The direction
  of the Sun (PA=287\deg) is indicated at the upper right.}
\label{figmaphcn32ztf}
\end{figure}

\begin{table*}
\caption[]{Log of observations of comet C/2022~E3 (ZTF) in 2023.}\label{tablogztf}\vspace{-0.2cm}
\begin{center}
\begin{tabular}{rllcccccc}
\hline\hline
UT date & $<r_{h}>$ & $<\Delta>$ & Phase\tablefootmark{a} & Tel. & Integ. time & pwv\tablefootmark{b} & Mode\tablefootmark{d} & Freq. range \\
$($mm/dd.d--dd.d) & (au)  & (au) &  &  & (min)\tablefootmark{c} & (mm) & & (GHz)  \\
\hline
 01/19.33--19.42 & 1.117 & 0.523 & 61.7\deg & {\it Odin} &  81 &  0  & DSW & 556.43-557.44 \\
 01/19.45--19.75 & 1.118 & 0.516 & 61.7\deg & {\it Odin} & 233 &  0  & DSW & 556.43-557.44, map 3x3, step 1\arcmin \\
 01/19.78--19.88 & 1.118 & 0.511 & 61.6\deg & {\it Odin} &  85 &  0  & DSW & 556.43-557.44 \\
 01/19.98--20.14 & 1.119 & 0.504 & 61.6\deg & {\it Odin} & 108 &  0  & DSW & 556.43-557.44, map 3x3, step 2\arcmin  \\
 01/20.19--20.28 & 1.119 & 0.501 & 61.5\deg & {\it Odin} &  72 &  0  & DSW & 556.43-557.44 \\
\hline
 02/03.67--03.78 & 1.168 & 0.291 & 45.4\deg & IRAM &  93 &  0.8 &  PSW & 248.3-256.4, 264.3-272.4 \\
 02/04.65--04.69 & 1.173 & 0.300 & 45.5\deg & IRAM &  39 &  0.5 &  PSW & 248.3-256.4, 264.3-272.4 \\
    04.71--04.78 & 1.173 & 0.301 & 45.5\deg & IRAM &  65 &  0.2 &  PSW & 240.1-248.2, 256.1-264.2 \\
    04.81--04.86 & 1.174 & 0.302 & 45.5\deg & IRAM &  56 &  0.3 &  PSW & 240.1-248.2, 256.1-264.2 \\
    04.88--04.95 & 1.174 & 0.303 & 45.5\deg & IRAM &  76 &  0.2 &  PSW & 146.8-154.9, 162.8-170.9 \\
    04.97--05.02 & 1.174 & 0.304 & 45.6\deg & IRAM &  59 &  0.2 &  PSW & 209.4-217.5, 225.4-233.5 \\
    05.03--05.04 & 1.175 & 0.304 & 45.6\deg & IRAM &  16 &  1.1 &  PSW & 161.0-169.1, 177.0-185.4 \\
 02/05.65--05.69 & 1.178 & 0.312 & 45.8\deg & IRAM &  40 &  1.8 &  PSW & 248.3-256.4, 264.3-272.4 \\
    05.71--05.81 & 1.178 & 0.314 & 45.8\deg & IRAM &  94 &  0.9 &  PSW & 217.5-225.5, 233.5-241.5 \\
    05.84--05.89 & 1.179 & 0.315 & 45.9\deg & IRAM &  58 &  0.5 &  PSW & 146.8-154.9, 162.8-170.9 \\
    05.91--06.00 & 1.179 & 0.317 & 45.9\deg & IRAM &  95 &  0.4 &  PSW & 209.4-217.5, 225.4-233.5 \\
    06.03--06.04 & 1.180 & 0.318 & 46.0\deg & IRAM &  17 &  0.8 &  PSW & 248.3-256.4, 264.3-272.4 \\
 02/06.66--06.69 & 1.183 & 0.328 & 46.0\deg & IRAM &  31 &  1.7 &  WSW & 248.3-256.4, 264.3-272.4 \\ 
    06.69--06.70 & 1.183 & 0.328 & 46.0\deg & IRAM &   8 &  1.6 &  PSW & 248.3-256.4, 264.3-272.4 \\ 
    06.71--06.73 & 1.183 & 0.329 & 46.0\deg & IRAM &  23 &  2.0 &  WSW & 248.3-256.4, 264.3-272.4 \\
    06.75--06.80 & 1.184 & 0.329 & 46.0\deg & IRAM &  48 &  3.5 &  WSW & 240.1-248.2, 256.1-264.2 \\
    06.88--06.90 & 1.184 & 0.331 & 46.0\deg & IRAM &  23 &  1.7 &  WSW & 209.4-217.5, 225.4-233.5 \\ 
    06.92--07.00 & 1.185 & 0.333 & 46.0\deg & IRAM &  91 &  2.5 &  WSW & 209.4-217.5, 225.4-233.5 \\ 
    07.02--07.04 & 1.185 & 0.334 & 46.0\deg & IRAM &  25 &  1.9 &  WSW & 146.8-154.9, 162.8-170.9 \\
\hline
\end{tabular}
\end{center}
\tablefoot{
  \tablefoottext{a}{Phase angle.}
  \tablefoottext{b}{Mean precipitable water vapour in the atmosphere above the telescope.}\\
  \tablefoottext{c}{Total (offset positions included) integration time (ON+OFF) on the source.}
  \tablefoottext{d}{Observing mode: WSW = Wobbler Switching (reference at $\pm2$\arcmin);
    PSW = Position Switching (reference at 4\arcmin), DSW = Dicke Switching (reference at $\sim$42\deg).}
}
\end{table*}

\begin{figure*}[]
\centering
\resizebox{0.85\hsize}{!}{
  \includegraphics[angle=270,width=0.85\textwidth]{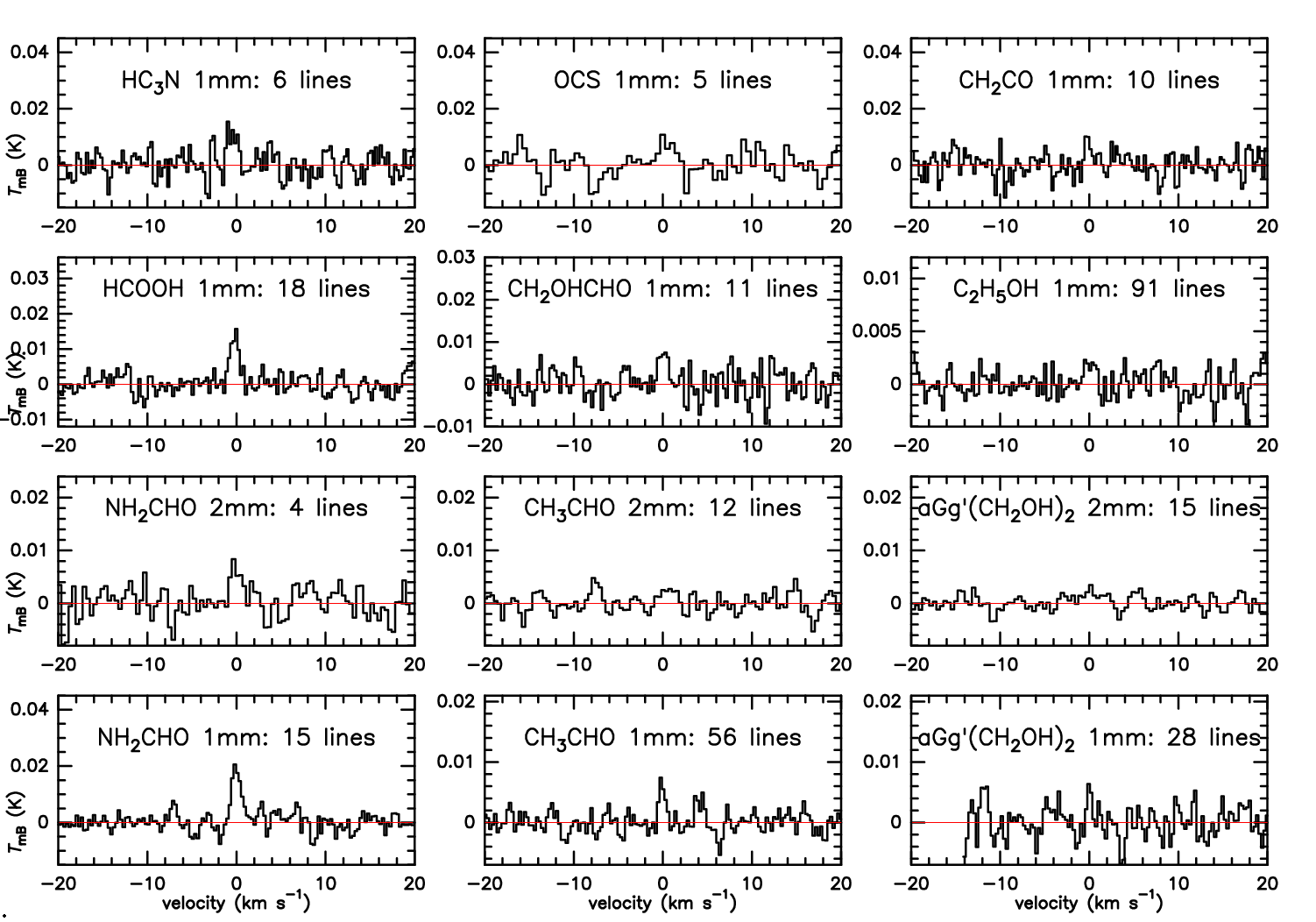}}
\caption{Spectra of molecules obtained by averaging several lines observed
  with IRAM-30m in comet C/2021~A1 between 8 and 13 December 2021. 
  The vertical axis is main beam brightness temperature in K. The
  horizontal axis is the Doppler velocity in the rest frame of the comet.
  The number of lines averaged is provided for each molecule, either in the
  2~mm band (147--153 and 163--171~GHz) or 1~mm band (209--272~GHz).}
\label{figspcomsleonard}
\end{figure*}

\begin{figure*}[]
\centering
\resizebox{0.85\hsize}{!}{
  \includegraphics[angle=270,width=0.85\textwidth]{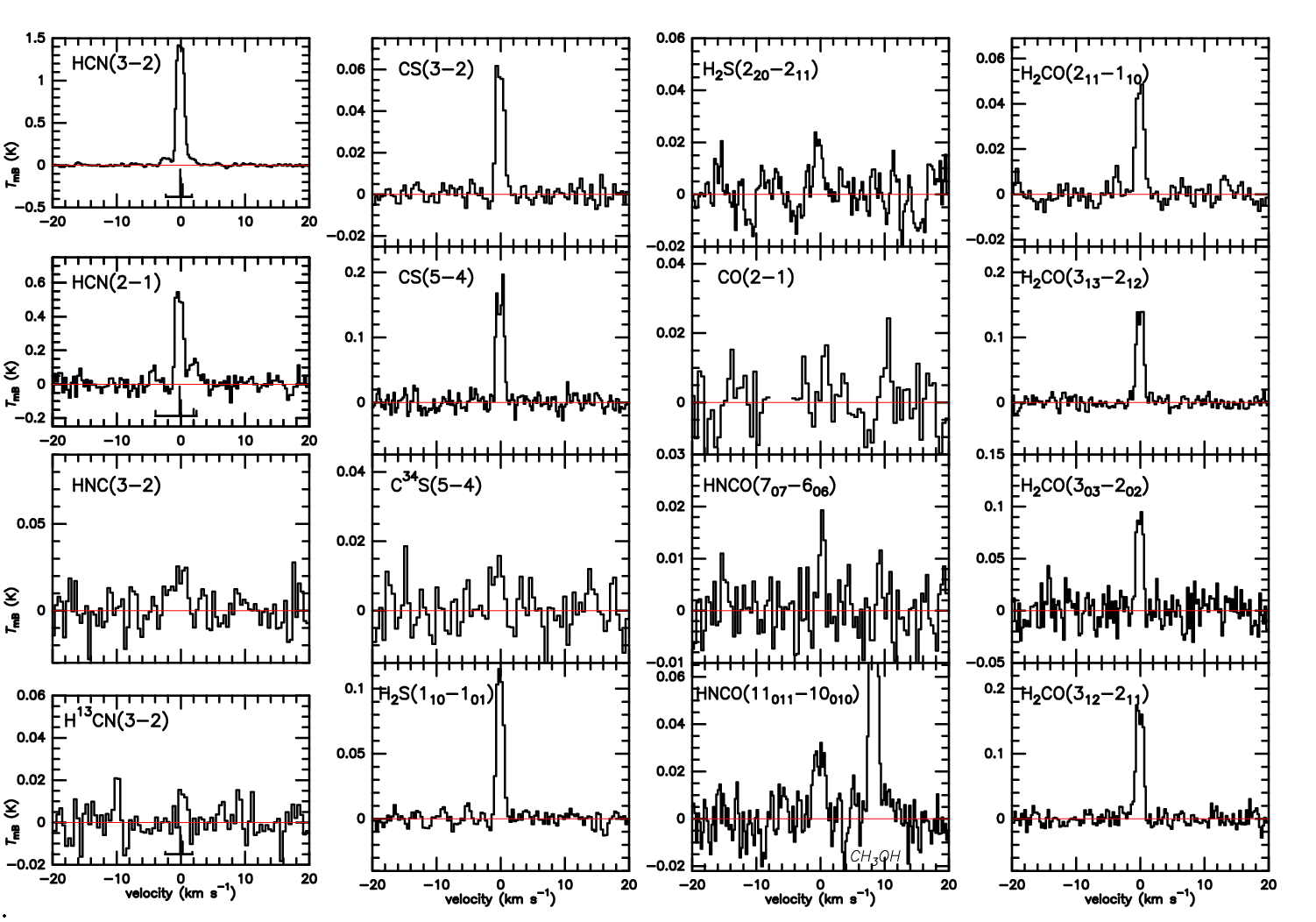}}
\caption{Individual lines observed with IRAM-30m in comet C/2022~E3 during 3--6
  February 2023 (average intensity, excluding offset positions). 
  The vertical axis is main beam brightness temperature in K. The
  horizontal axis is the Doppler velocity in the rest frame of the comet
  with respect to the line. Galactic contamination of CO has been blanked out.
  Position and relative intensity of the
    hyperfine components of the HCN(3-2), HCN(2-1) and H$^{13}$CN(3-2) lines
    have been drawn below the lines.}
\label{figspecztf}
\end{figure*}

\begin{figure*}[]
\centering
\resizebox{0.80\hsize}{!}{
  \includegraphics[angle=270]{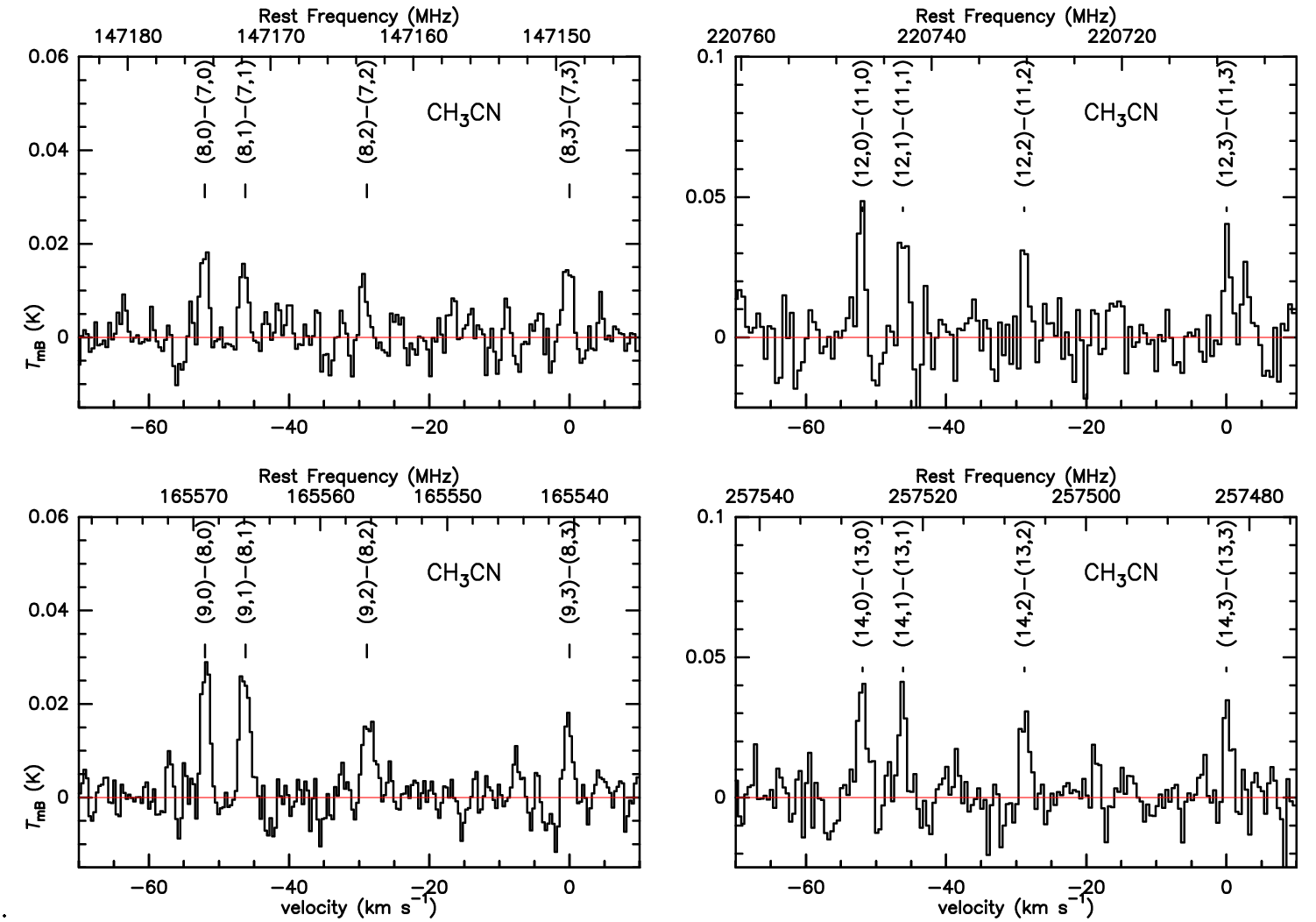}}
\caption{CH$_3$CN series of lines at 147, 165, 220 and 257~GHz observed with
  IRAM-30m in comet C/2022~E3 on 3--6 February 2023 (average intensity). 
  The vertical axis is main beam brightness temperature in K. The
  horizontal axis is the Doppler velocity in the rest frame of the comet
  with respect to the main line, with frequencies in the rest
  frame of the comet given on the upper axis.}
\label{figspch3cnztf}
\end{figure*}

\begin{figure*}[]
\centering
\resizebox{0.85\hsize}{!}{
  \includegraphics[angle=270,width=0.85\textwidth]{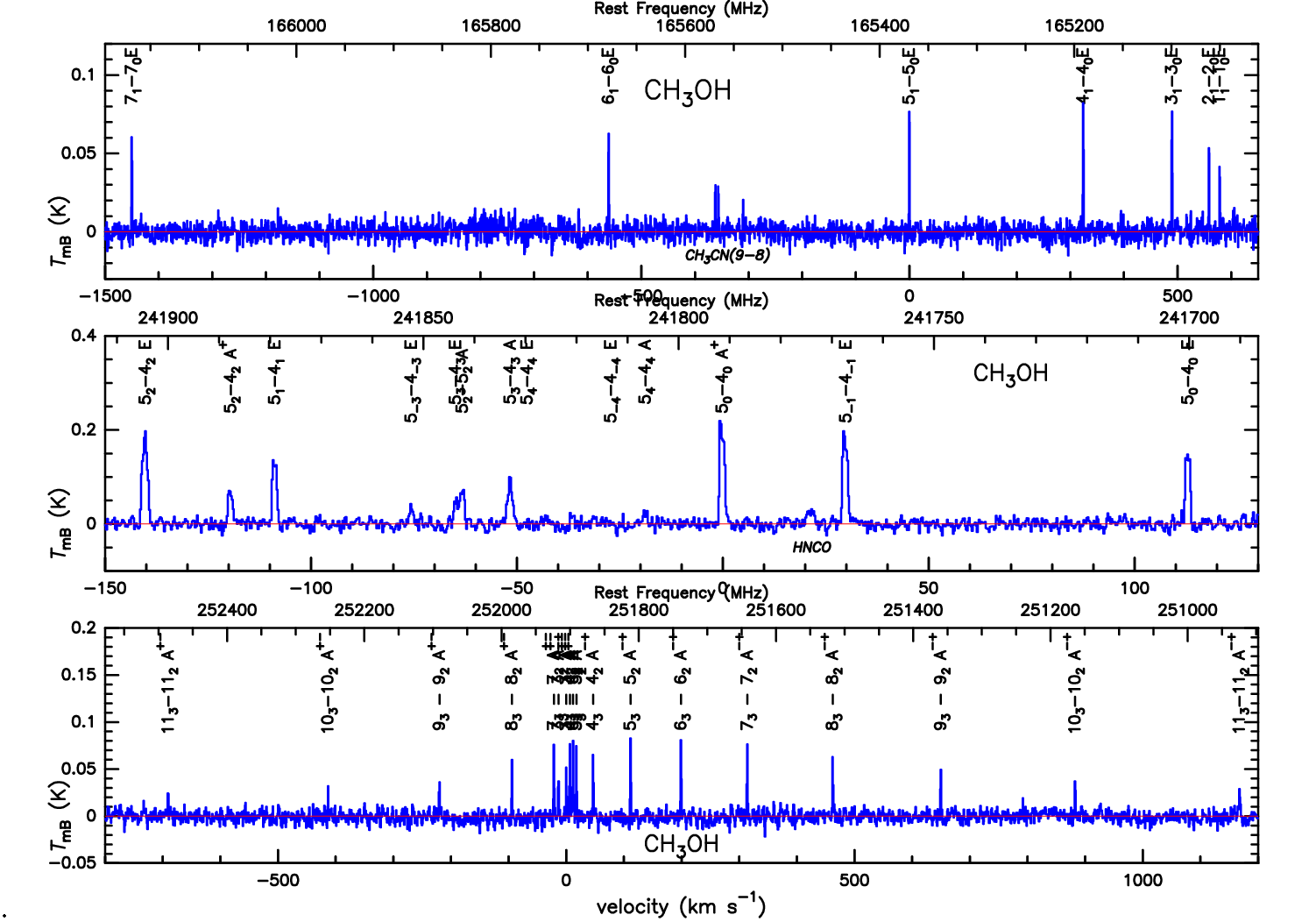}}
\caption{CH$_3$OH series of lines around 165, 242, and 252~GHz observed with
  IRAM-30m in comet C/2022~E3 on 3--6 February 2023 (average intensity). 
  The vertical axis is main beam brightness temperature in K. The
  horizontal axis is the Doppler velocity in the rest frame of the comet
  with respect to the main line, with frequencies in the rest
  frame of the comet given on the upper axis.}
\label{figspch3ohztf}
\end{figure*}

\begin{figure*}[]
\centering
\resizebox{0.85\hsize}{!}{
  \includegraphics[angle=270,width=0.85\textwidth]{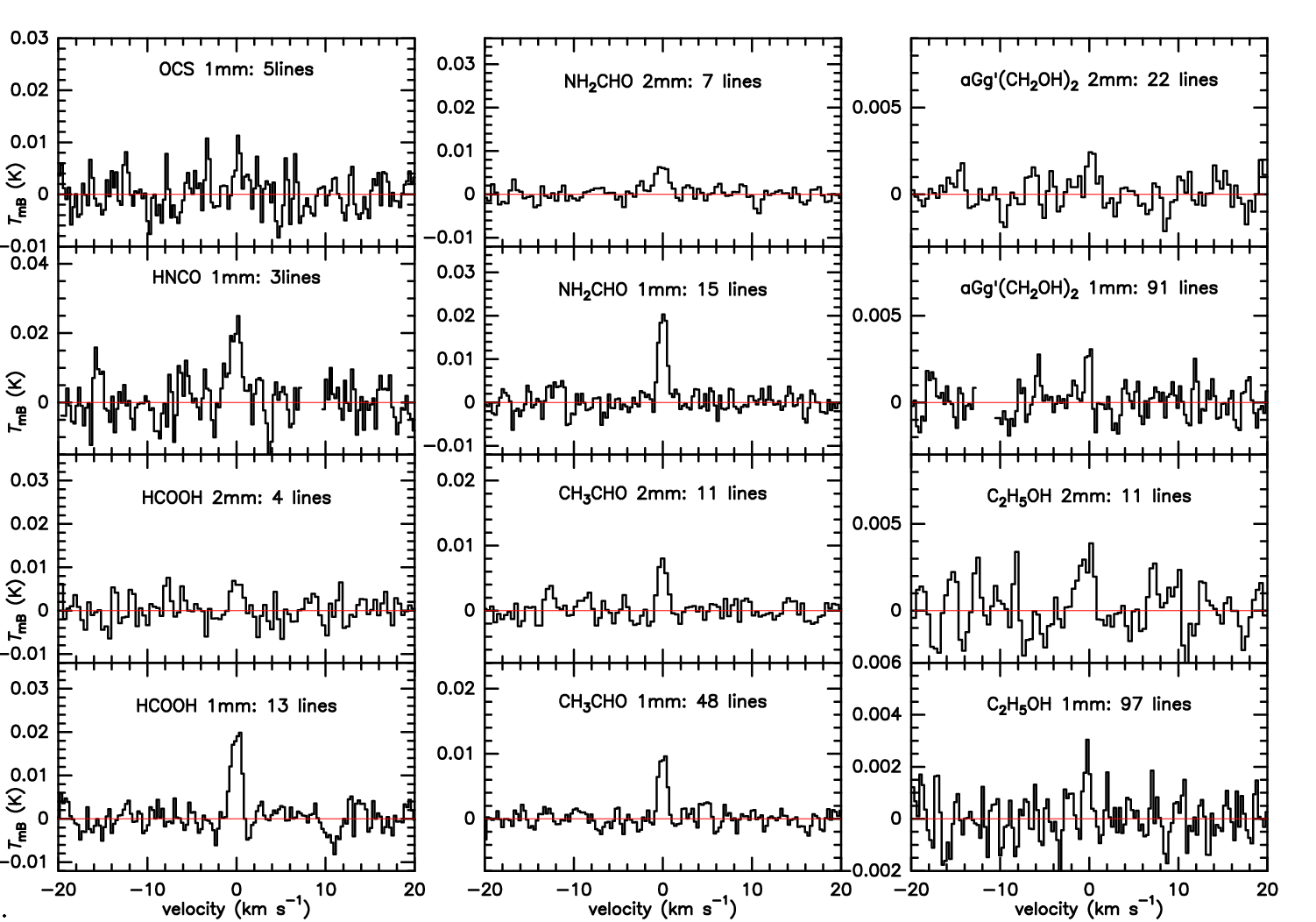}}
\caption{Spectra of molecules obtained by averaging several lines observed
  with IRAM-30m in comet C/2022~E3 between 3 and 6 February 2023. 
  The vertical axis is main beam brightness temperature in K. The
  horizontal axis is the Doppler velocity in the rest frame of the comet.
  The number of lines averaged is provided for each molecule, either in the
  2~mm band (147--155, 163--171 and 176--184~GHz) or 1~mm band (209--272~GHz).}
\label{figspcomsztf}
\end{figure*}

Representative spectra of comet C/2022~E3 (ZTF) are provided in
Figs.~\ref{figspecztf}--\ref{figspcomsztf}. They show individual lines
or averages of several lines for the complex organics from the observations
centred on the nucleus. Detailed line intensities are provided in
Tables~\ref{tabobsztf} and \ref{tabobssumztf} and for some molecules
with useful spatial information in Table~\ref{tabqdistriztf} of
Section~\ref{sect-distri}. Spectra showing the full spectral coverage with
the FTS are shown in Figs.~\ref{figsurveyztf2mm3} and
\ref{figsurveyztf1mm6}.
Figs.\ref{figspcomsleonard} and \ref{figspcomsztf} show the average of 3
to 97 lines, weighted according to the noise in each spectrum.
We have selected the strongest lines with expected similar S/N (within a
factor around four) which are not blended with other lines. This is only for the
purpose of highlighting detection and line shapes.

\subsection{Observations with the Nan\c{c}ay Radio Telescope}
\label{sect-nancay}

The NRT performances and the observing procedure used for comets were
described in \citet{Cro02}.  The integration time is
usually about one hour per day.  However, from 12 September to 14
December 2021, it was limited to 30 min due to work on the focal
track. The beam size at 18-cm wavelength is $3.5\times18$\arcmin.

\subsubsection{Comet C/2021~A1 (Leonard)}

C/2021~A1 (Leonard) was observed at the NRT from 1 October 2021 to
14 February 2022 at the NRT almost every day or every other
day\footnote{https://lesia.obspm.fr/planeto/cometes/basecom/LD/indexld.html}.

A succession of outbursts was observed, beginning on 13 December just
at the end of the IRAM observations \citep{Cro21,Jeh21}.
The OH production rate, which was $2.5\pm0.1\times10^{28}$~\mols 
on average for 8--12 December, rose to $6.3\pm0.3\times10^{28}$~\mols
on 13.5, up to $22.1\pm0.2\times10^{28}$~\mols on 15.5 December.

\begin{figure}[ht]
\centering
{\includegraphics[width=\hsize,angle=0.]{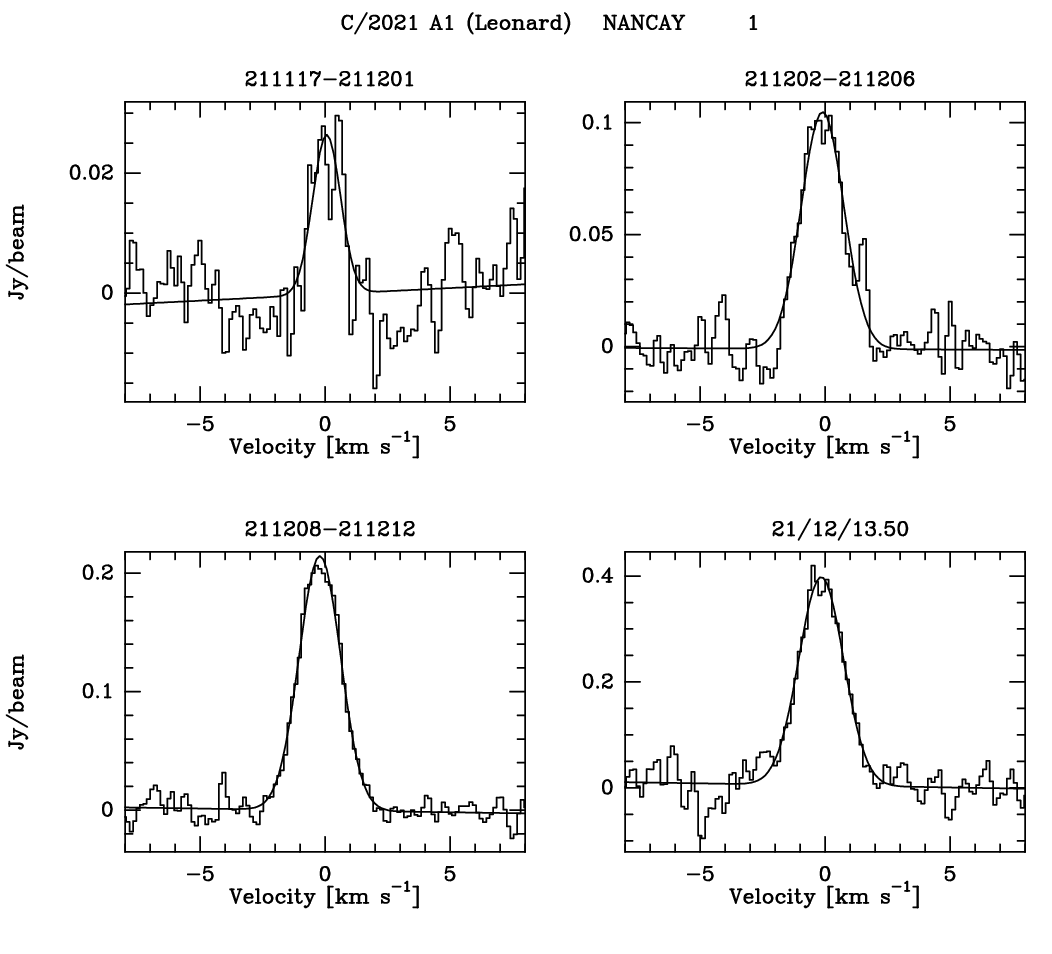}}
\caption{Selected averages of the OH lines observed at the NRT in
comet C/2021~A1 (Leonard) in November--December 2021  (averages of the
1667 and 1665~MHz lines scaled to 1667~MHz).}
\label{fig:leonard-OH-spectra}
\end{figure}

\begin{figure}[ht]
\centering
{\includegraphics[height=1.07\hsize,angle=270, trim= 30 0 20 10, clip]{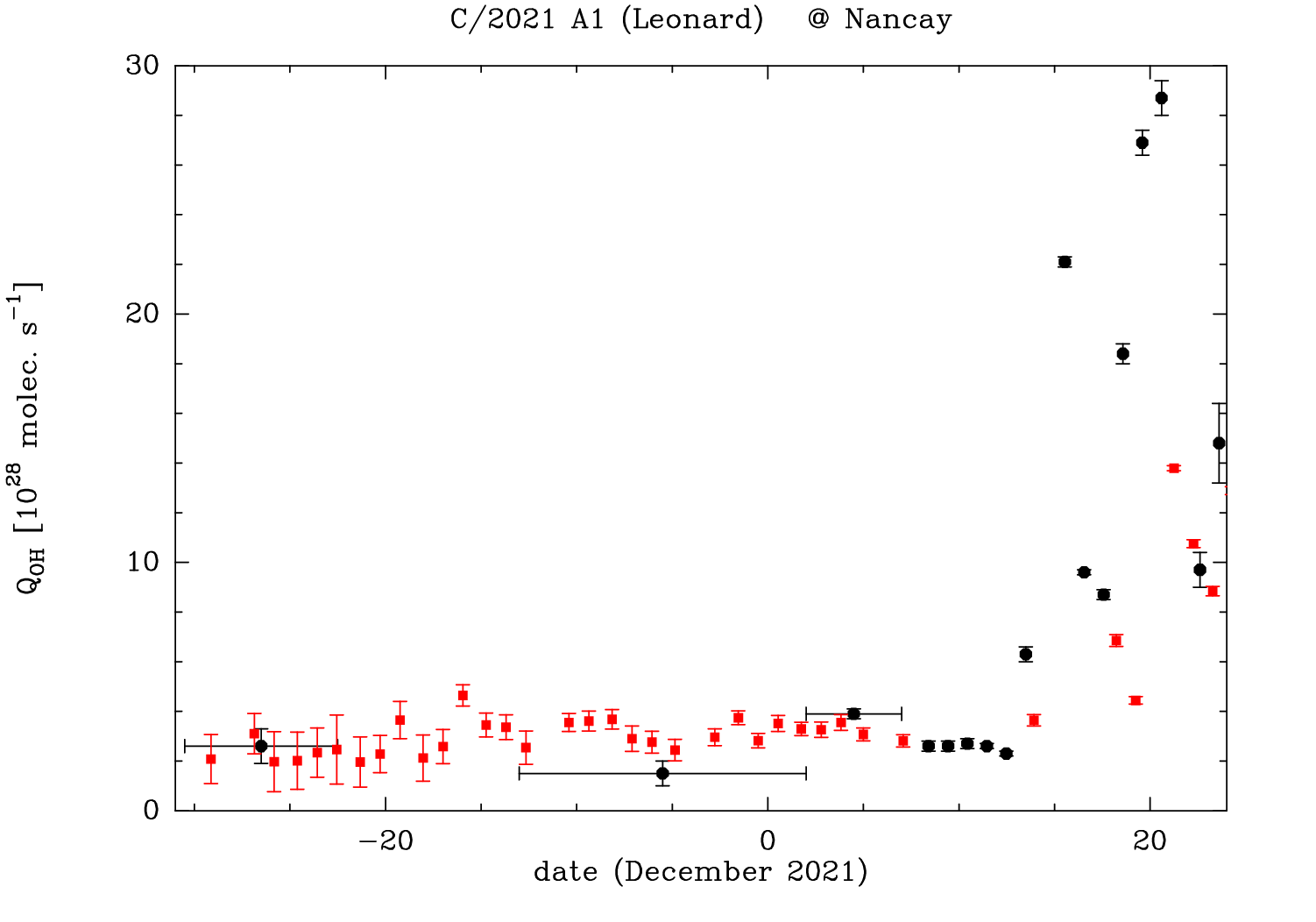}}
\caption{Time evolution of comet C/2021~A1 (Leonard), single days and
averages of selected periods in November--December 2021.  In black: OH
production from the NRT. In red: water production from SOHO/SWAN observations
of the wide H Lyman-$\alpha$ coma which tends to smooth out short term
variations.
\citep{Com23a}.}
\label{fig:leonard-OH-evolution} 
\end{figure}

Representative spectra for selected dates are shown in
Fig.~\ref{fig:leonard-OH-spectra}.  The evolution of the retrieved OH
production rate is plotted in Fig.~\ref{fig:leonard-OH-evolution}.

\subsubsection{Comet C/2022~E3 (ZTF)}

C/2022~E3 (ZTF) was observed at the NRT from 17 October to 23 December
2022\footnote{https://lesia.obspm.fr/planeto/cometes/basecom/ZT/indexzt.html}
The retrieved OH production rate rose
from 3.3 to $8.1\times10^{28}$\mols.  Then the
observations were interrupted due to a technical failure on 24 December 2022.
They were resumed from 17 to 24 February 2023 in a degraded mode,
resulting in an upper limit $Q(\rm{OH}) \leq 6 \times 10^{28}$~\mols.

\begin{figure}[]
\centering
\resizebox{\hsize}{!}{
  \includegraphics[angle=270,width=0.9\columnwidth]{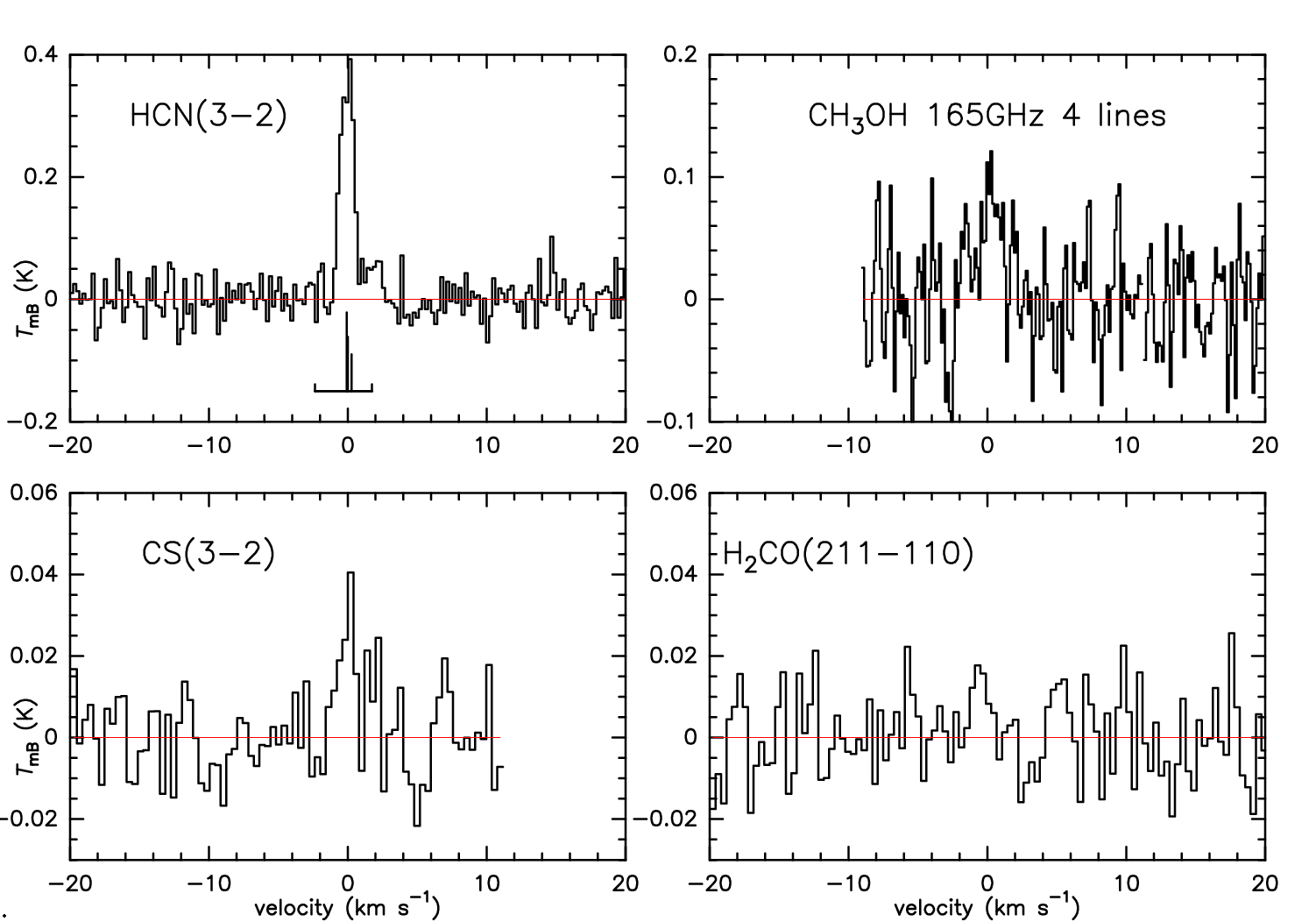}}
\caption{Millimetre lines observed with IRAM-30m in comet C/2021~A1 during
  12--16 November 2021 (average intensity). The CH$_3$OH line is the sum of the
  $J$=1 to 4 $J_1-J_0$ E lines at 165050.175, 165061.130, 165099.240, and
  165190.475~MHz observed with the high resolution backend VESPA.
  The vertical axis is main beam brightness temperature in K,
  horizontal axis is Doppler velocity in the rest frame of the comet
  with respect to the main line.}
\label{figspnovleonard}
\end{figure}

\begin{figure*}[]
\centering
\resizebox{0.85\hsize}{!}{
  \includegraphics[angle=270,width=0.85\textwidth]{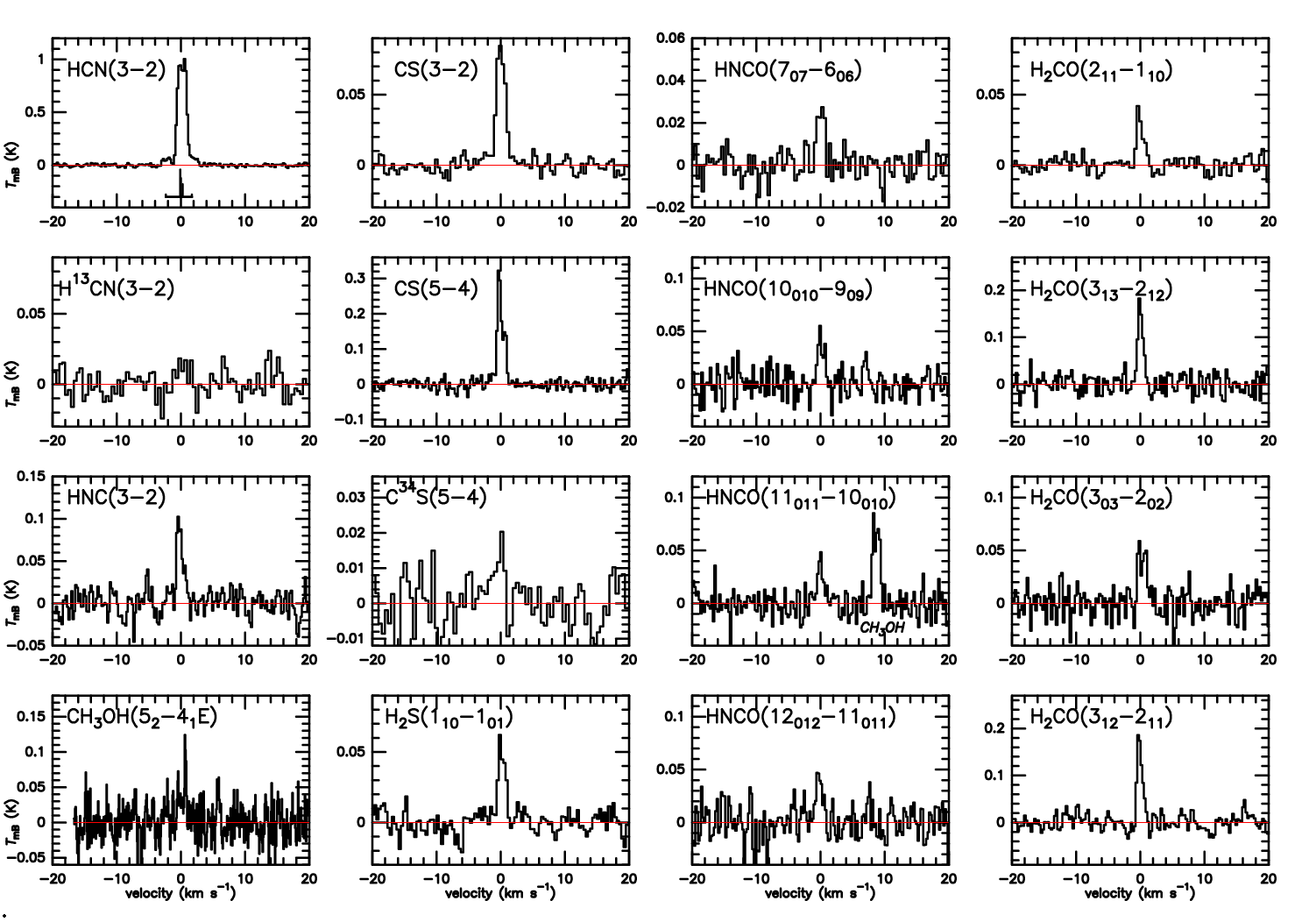}}
\caption{Individual lines observed with IRAM-30m in comet C/2021~A1 during 8--13
  December 2021 (weighted averages). 
  The vertical axis is main beam brightness temperature in K,
  horizontal axis is Doppler velocity in the rest frame of the comet
  with respect to the line. Position and relative intensity of the
  hyperfine components of the HCN(3-2) line have been drawn below the line.}
\label{figspdecleonard}
\end{figure*}

\section{Data analysis}
\label{sect-analysis}

\subsection{Expansion velocity and outgassing pattern}\label{sect-vexp}
\begin{figure}[]
\centering
\resizebox{\hsize}{!}{
  \includegraphics[angle=270,width=0.9\textwidth]{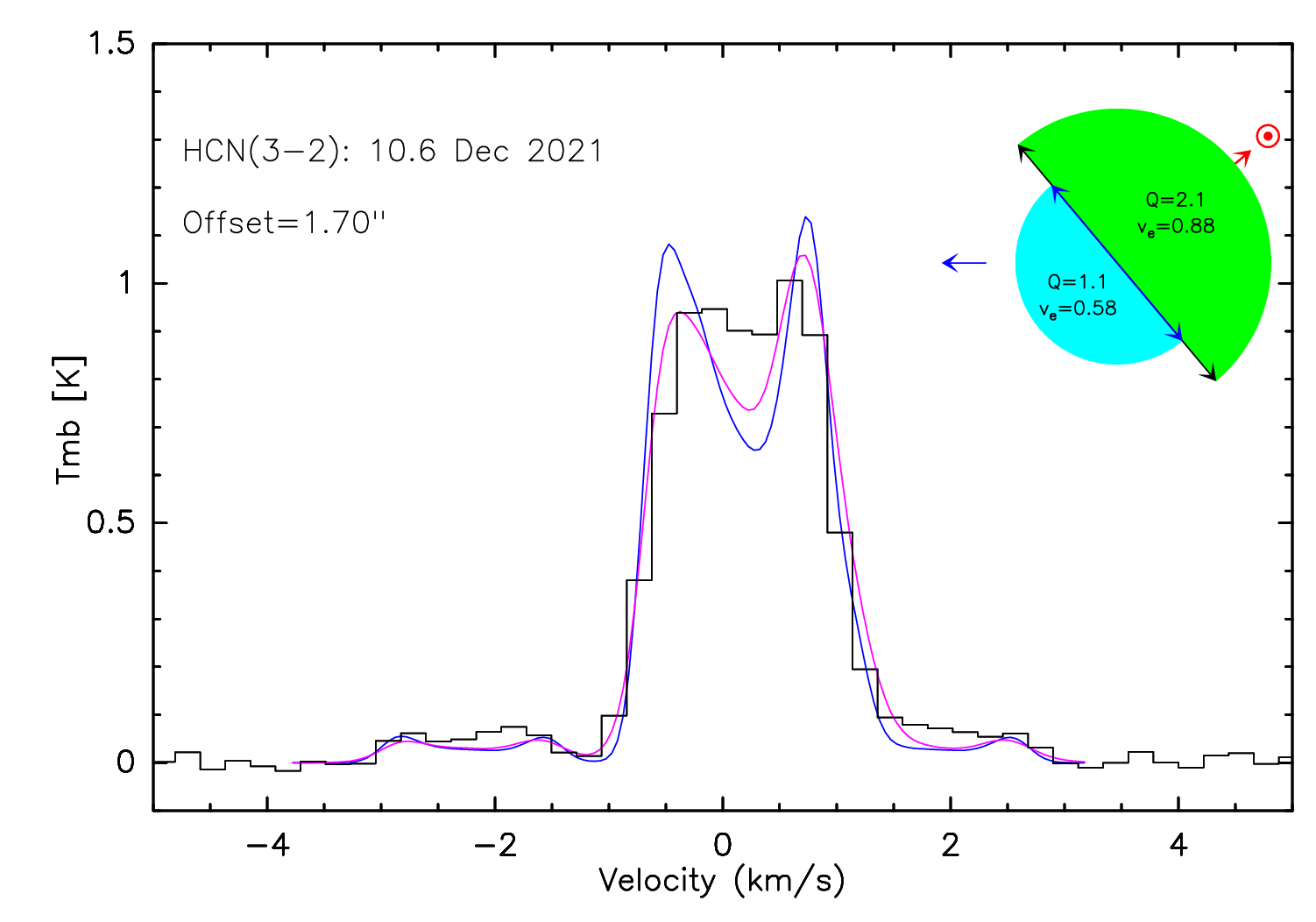}}
\caption{Average spectrum of the HCN(3-2) line observed from 8.5 to 13.4
  December in comet C/2021~A1, with simulated profile.
  The model assumes a production rate of
  $2.1\times10^{25}$~\mols~ at 0.88~\kms~ on the sunward hemisphere and
  $1.1\times10^{25}$~\mols~ at 0.58~\kms~ on the opposite hemisphere mostly
  facing the observer. The mean phase angle is $\sim140$\deg, and
  the tilt of 40\deg (or 140\deg) with respect to the comet-observer line of
  sight is taken into account in this 3D simulation, as depicted in the
  upper right. The model in purple uses variable velocities in both
  hemispheres (see text), with production rates of $2.3\times10^{25}$~\mols~
  and $0.9\times10^{25}$~\mols~, respectively.
  The vertical axis is main beam brightness temperature in K. The
  horizontal axis is the Doppler velocity in the rest frame of the comet.}
\label{figsimhcn32-2pop}
\end{figure}

The lines profiles (Figs.~\ref{figspnovleonard}-\ref{figspdecleonard}) of
comet C/2021~A1 do not show systematic asymmetry, excepted for small shifts
with respect to the rest velocity in the comet frame.
In November the mean Doppler shift of the line is slightly
negative ($\delta$v=$-0.03\pm0.03$~\kms~ for HCN(3-2)) and positive for most
lines in December ($\delta$v(HCN(3-2))$=+0.18\pm0.01$~\kms).
The two-Gaussian fit \citep{Biv21a} used to estimate the expansion velocity from
the velocity at half maximum intensity $VHM$, yields $VHM = -0.75\pm0.04$~\kms~
and $VHM = +0.53\pm0.03$~\kms~ in November (from HCN) and
$VHM = -0.63\pm0.01$~\kms~ and $VHM = +0.97\pm0.01$~\kms~ in December (all
lines). Those asymmetries are expected for a preferential outgassing at a
higher rate and velocity on the sunward hemisphere, since in November the
phase angle was $\sim51$\deg~ (the sunlit hemisphere is mostly facing us) and
in December it was between 116\deg~ and 159\deg~ (Table~\ref{tablogleonard})
so that we were mostly facing the night side of the comet.
We could have simulated a two-component outgassing pattern with a higher
production rate and expansion velocity on the day side, but this would not
change significantly the retrieved production rates for the optically thin
lines from using an isotropic model with a constant expansion velocity equal
to the mean of the day and night sides.
Thus, we assumed isotropic outgassing with $v_{exp}=0.60$~\kms~ in November and
$v_{exp}=0.73$~\kms~ in December.
The actual expansion velocities needed to fit the observed profiles are 5-10\%
lower than the $VHM$ (that is -0.58 and 0.88~\kms~ in December) due to thermal
broadening. An example of simulated profile with asymmetric outgassing is shown
in Fig.~\ref{figsimhcn32-2pop}.

\begin{figure}[]
\centering
\resizebox{\hsize}{!}{
  \includegraphics[angle=270,width=0.9\textwidth]{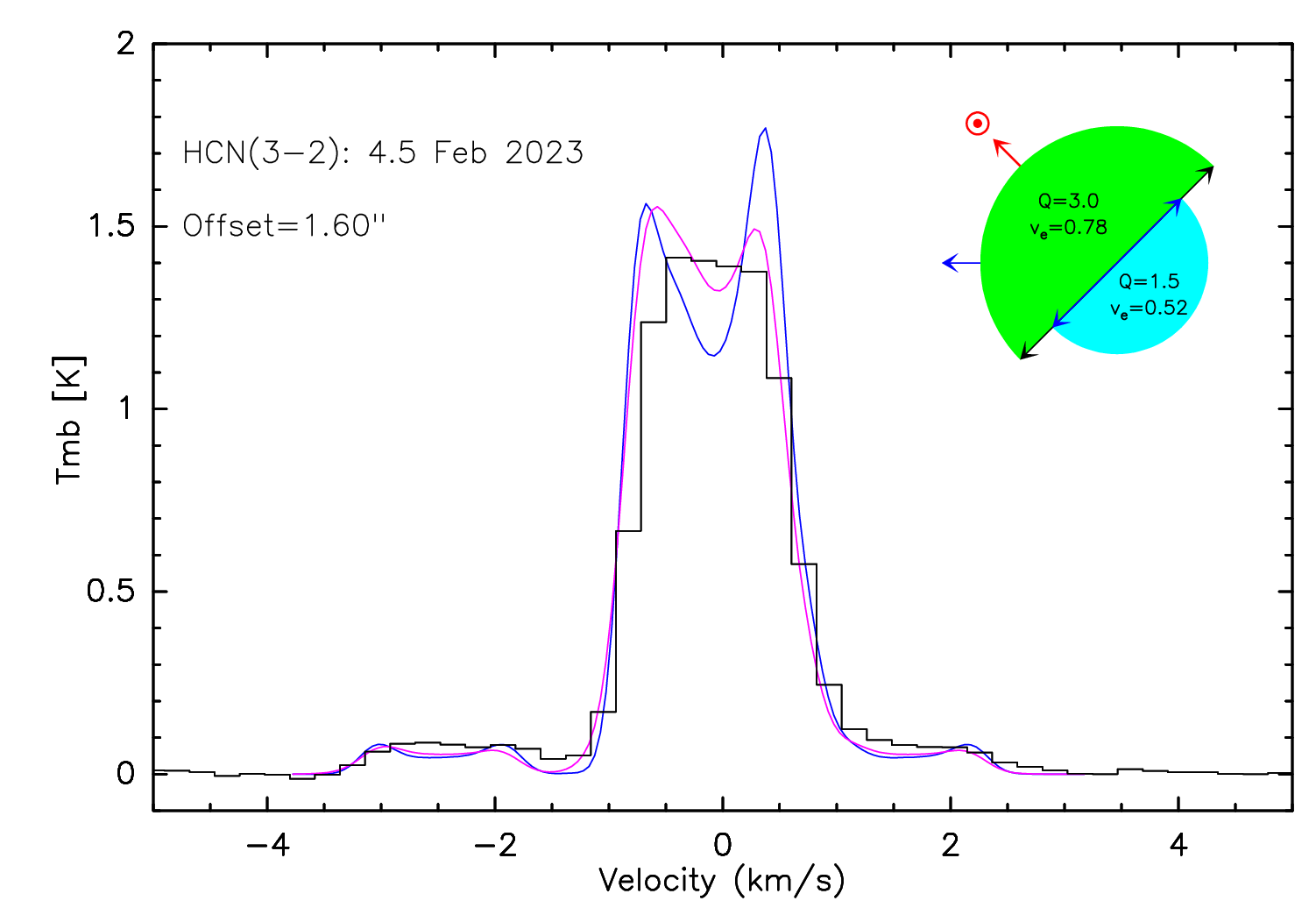}}
\caption{Average FTS spectrum of the HCN(3-2) line observed from 3.7 to 6.7
  February in comet C/2022~E3 with simulated profiles.
  The model in blue assumes a production rate of
  $3.0\times10^{25}$~\mols~ at 0.76~\kms~ on the sunward hemisphere and
  $1.5\times10^{25}$~\mols~ at 0.52~\kms~ on the other hemisphere, as depicted
  in the upper right. The model in purple uses variable velocities in both
  hemispheres (see text), with production rates of $3.0\times10^{25}$~\mols~
  and $1.0\times10^{25}$~\mols~, respectively.
  The mean phase angle is $\sim45$\deg.
  The vertical axis is main beam brightness temperature in K,
  horizontal axis is Doppler velocity in the rest frame of the comet.}
\label{figsimhcn32-2popztf}
\end{figure}

\begin{figure}[]
\centering
\resizebox{\hsize}{!}{
  \includegraphics[angle=270,width=0.9\textwidth]{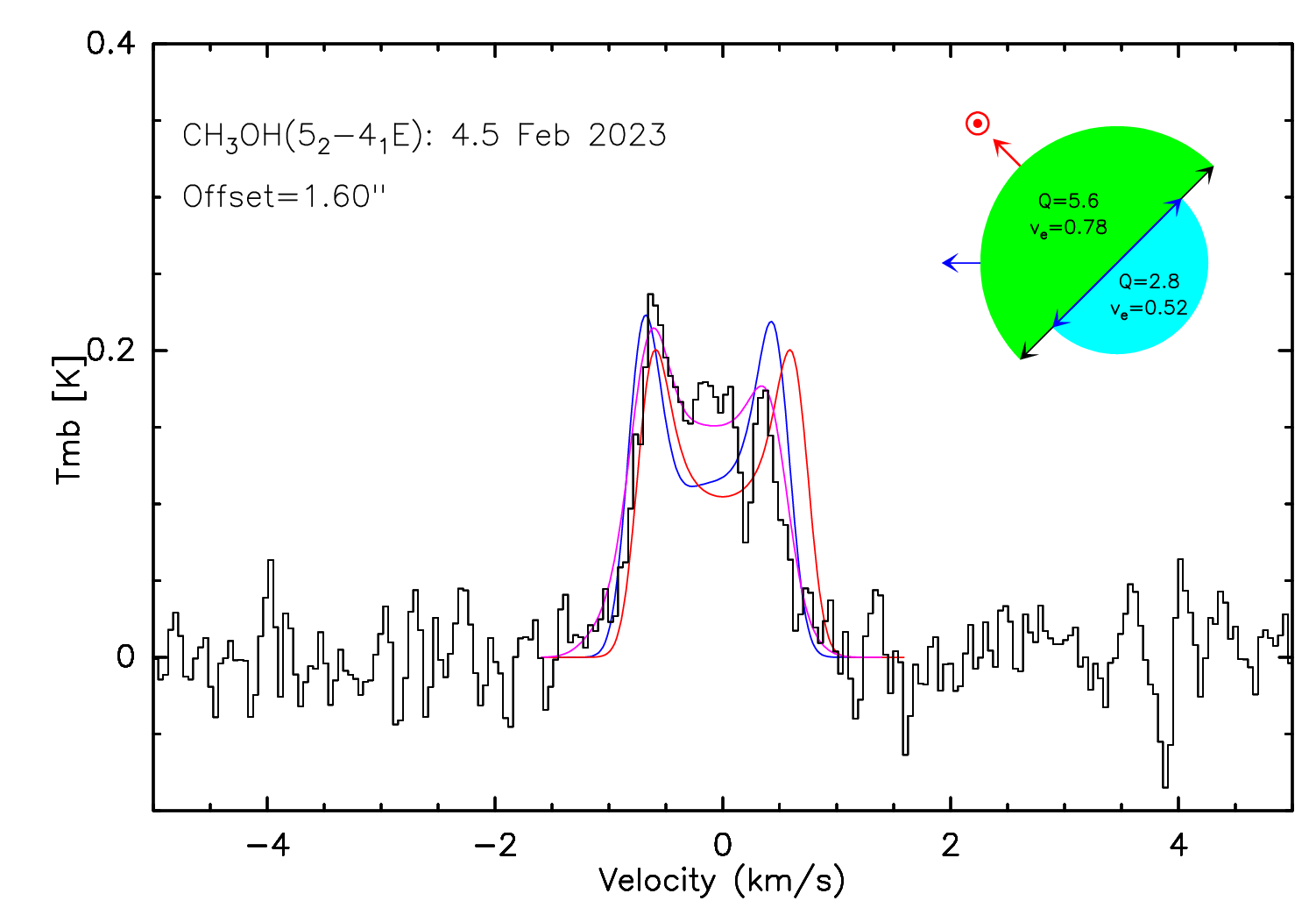}}
\caption{Average spectrum of the CH$_3$OH line at 266.838~GHz observed
  from 3.7 to 6.7 February 2023 in comet C/2022~E3 at the same time as HCN(3-2).
  The model in blue assumes a production rate of
  $5.6\times10^{26}$~\mols~ at 0.76~\kms~ on the sunward hemisphere and
  $2.8\times10^{26}$~\mols~ at 0.52~\kms~ on the other hemisphere, as depicted
  in the upper right, while the red profile comes from symmetric outgassing at
  0.68~\kms and same total production rate ($8.4\times10^{26}$~\mols~).
  The model in purple uses the same variable velocities in both hemispheres as
  in fig~\ref{figsimhcn32-2popztf}, with production rates of
  $6.4\times10^{25}$~\mols~ and $2.0\times10^{25}$~\mols~, in sunward and other
  hemisphere, respectively.
  The mean phase angle is $\sim45$\deg.
  The vertical axis is main beam brightness temperature in K,
  horizontal axis is Doppler velocity in the rest frame of the comet.}
\label{figsimmet-2popztf}
\end{figure}

In the case of comet C/2022~E3 (ZTF), the full width at half maximum ($FWHM$)
of the water line at 556.9~GHz is on the order of 1.60~\kms, but the line is
asymmetric and redshifted due to opacity effects
\citep[see for example][]{Lec03}. In addition the uncertainty on the
frequency calibration is on the order of 0.05~\kms, so it is difficult to
derive information on the gas velocity asymmetry from these data, but the
average expansion velocity should be on the order of 0.8~\kms.
IRAM data obtained two weeks later provide more accurate line profiles but
show also evidence of variation with time. On the basis of the lines
with the best signal-to-noise ratio (HCN, CH$_3$OH, H$_2$S, CS, H$_2$CO,
CH$_3$CN), the weighted average $VHM$s are $-0.83\pm0.01$ and
$+0.64\pm0.01$~\kms, with a
mean Doppler shift of the lines of $-0.12\pm0.01$~\kms. Since the phase angle
was not too large (45\deg, Table~\ref{tablogztf}), we deduce that outgassing
was larger on the day-side with a larger expansion velocity. From the $VHM$s and
fits to line profiles (Figs.~\ref{figsimhcn32-2popztf},\ref{figsimmet-2popztf})
we estimate that the expansion velocity was on the order of 0.76~\kms on the
day side and 0.52~\kms on the night side, with a production rate two times
higher on the day side to get the average measured Doppler shift. Nevertheless,
assuming isotropic outgassing with the average $v_{exp}=0.68$~\kms~ yields very
similar production rates and we will assume isotropic outgassing at this
velocity to compute and compare the production rates.
Model fits with constant velocities in Figs.~\ref{figsimhcn32-2pop},
 \ref{figsimhcn32-2popztf}, and \ref{figsimmet-2popztf} are underestimating the
 signal around zero velocity. Better fits to the line shapes are obtained
 when simulating radial acceleration of the expansion velocity, using the
 parameter $x_{acc}=4$ from the formula in \citet{Biv11, Biv22} and
 $v_{exp,0} = 0.9\times v_{exp}$. This illustrate a possible way to improve
 the fit, with a slightly lower ($\approx3-10$\%) production rate for most
 molecules. But there are other possible explanations, like
 variation of the gas temperature and different azimutal gas distribution.
 We keep the simpler constant velocity, isotropic model for computation of
 production rates and abundances of the numerous molecules observed.

\subsection{Gas temperature}\label{sect-temp}

Several species such as methanol are detected through multiple transitions
coming from different rotational upper energy levels $E_u$. In the case of local
thermal equilibrium (LTE) we expect the relative population levels ($p_u$) to
follow the Boltzmann law ($p_u \propto \exp(-E_u/kT)$). When we plot the
logarithm of the populations $p_u$ versus the upper energy levels $E_u$
(the so-called rotational diagram)
the mean slope of the linear fit of the data is $1/T_{rot}$, where $T_{rot}$
is the rotational temperature, equal to the gas kinetic temperature $T$ in LTE.
Due to radiative decay and infrared pumping, deviations from LTE can be observed
in some series of lines, especially the CH$_3$OH lines at 242~GHz, for
which $T_{rot} < T$. In such a case the full non-LTE modelling of the evolution
of the population of the rotational level throughout the cometary atmosphere
is required to estimate the value of $T$ resulting in the measured $T_{rot}$
within the radio telescope beam. We provide in Figs.~\ref{diagrot2021a1-met166}
--~\ref{diagrotztf-ch3cho} the rotational diagrams for species for which we
observed several transitions with sufficient signal-to-noise ratio (S/N$>$3) and
in Table~\ref{tabtemp} the measured  $T_{rot}$ and inferred $T$.

The average gas kinetic temperature $T$ for comet C/2021~A1 inferred
from CH$_3$OH and CH$_3$CN
rotational temperatures (Table~\ref{tabtemp}) is $T=25\pm5$~K in November and
$T=61\pm3$~K in December. In December the temperature inferred from HNCO lines
(Figs.~\ref{diagrot2021a1-hnco}) suggests a lower value of $T$,
but we have not taken into account the effect of infrared pumping that could
modify the value expected for $T_{rot}$(HNCO) for a given $T$. The
spatial distribution of HNCO is also not well constrained: if it comes from a
distributed source then the observed  rotational temperature will be compatible
with a higher value of $T$.
We have adopted $T=30$~K in November and $T=60$~K in December.

In the case of comet C/2022~E3 (ZTF), due to higher signal-to-noise ratios,
we have numerous measurements of rotational temperatures (rotational diagrams in
Figs~\ref{diagrotztf-met166}-\ref{diagrotztf-ch3cho}). Inferred gas temperatures
$T$ are provided in Table~\ref{tabtemp}. From methanol data, the weighted
average and dispersion is $T=58\pm4$~K, from CH$_3$CN we get $T=68\pm9$~K and
$T=53\pm16$~K from the four other species in Table~\ref{tabtemp}.
We adopt the average $T=60$~K to derive production rates of comet
C/2022~E3 for all species. All measurements, given their uncertainties,
are within 10 K of this value (Table~\ref{tabtemp}) but due to likely different
collision rates for the different molecules, small deviations are not
surprising. Also for NH$_2$CHO, HCOOH, HNCO and CH$_3$CHO we have not taken
into account infrared pumping that could introduce more differences between
$T_{rot}$ and $T$.
We have also investigated possible day versus night differences of the
coma temperature. The positive parts of the 166~GHz methanol lines only suggest
marginally higher temperature on the night side than on the day-side
(blueshifted parts of the lines, Table~\ref{tabtemp}).

\begin{table}
\renewcommand{\tabcolsep}{0.08cm}
\caption[]{Rotational temperatures and inferred gas kinetic temperatures.}\label{tabtemp}
\begin{center}
\begin{tabular}{lccrrcc}
\hline\hline\noalign{\smallskip}
UT   & Molecule & Freq. range  & lines & off.\tablefootmark{a} & $T_{rot}$\tablefootmark{b} & $T_{gas}$ \\
\multicolumn{2}{l}{(mm/dd.d)} &  (GHz)        &\tablefootmark{c}  & (\arcsec)   & (K)  & (K) \\
\hline\noalign{\smallskip}
\multicolumn{7}{l}{C/2021~A1 (Leonard)} \\
\hline
11/15.9 & CH$_3$OH & 250-254   & 12 &  1.4 &   $23\pm5$    & $26\pm6$  \\
11/14.8 & CH$_3$OH & 165-169   &  6 &  1.5 &   $22\pm9$    & $21\pm9$  \\

12/12.8 & CH$_3$OH & 241.8     & 13 &  1.6 & $54.1\pm7.1$  & $63\pm8$  \\
12/10.6 & CH$_3$OH & 250-254   & 20 &  1.7 & $65.3\pm5.2$  & $70\pm6$  \\
12/11.1 & CH$_3$OH & 165-169   & 10 &  1.9 & $56.0\pm6.9$  & $56\pm7$  \\
12/10.6 & CH$_3$OH & 250-267   &  5 &  1.7 & $54.4\pm7.3$  & $55\pm7$  \\
12/12.5 & HNCO     & 153-264   &  9 &  1.5 & $30.4\pm3.6$  & $37\pm6$  \\
12/12.5 & HNCO$_d$ &           &    &      & $37.0\pm5.4$  & $48\pm7$  \\
12/11.1 & CH$_3$CN & 147-165   & 11 &  1.9 &  $53\pm11$    & $53\pm10$  \\ 
12/12.8 & CH$_3$CN & 220-257   & 13 &  1.7 & $62.9\pm9.3$  & $63\pm9$  \\
12/12.5 & NH$_2$CHO & 213-267 & 25 &  1.5 &  $116\pm59$   &$102\pm52$  \\
\hline
\multicolumn{7}{l}{C/2022~E3 (ZTF)} \\
\hline
02/05.2 & CH$_3$OH & 165-169   & 10 &  2.0 & $53.0\pm1.7$  & $55\pm2$  \\
        &          &           &    &  9.1 & $51.5\pm6.3$  & $57\pm7$  \\
        &          &           &    & 13.0 & $53.5\pm7.8$  & $63\pm10$  \\
02/05.2 & CH$_3$OH & 165-169   & v-\tablefootmark{e} & 2.0 & $52.3\pm1.9$ & $54\pm2$  \\
        &          &           & v+ & 2.0 &  $55.0\pm3.5$  & $58\pm4$ \\
02/04.6 & CH$_3$OH & 250-254   & 28 &  1.6 & $58.4\pm1.6$  & $63\pm2$  \\
        &          &           & 22 &  9.8 & $48.7\pm4.6$  & $59\pm6$  \\
02/04.6 & CH$_3$OH & 250-267   &  5 &  1.6 & $54.3\pm1.9$  & $56\pm2$  \\
02/05.1 & CH$_3$OH & 241.8     & 14 &  1.8 & $42.2\pm1.4$  & $54\pm3$  \\
02/05.1 & CH$_3$CN & 257       &  4 &  1.8 &   $68\pm16$   & $68\pm16$  \\ 
02/05.2 & CH$_3$CN & 147-165   & 12 &  2.1 & $52.3\pm4.0$  & $73\pm8$  \\
02/05.8 & CH$_3$CN & 220-239   &  9 &  2.7 & $47.4\pm9.2$  & $54\pm13$  \\
02/05.2 & NH$_2$CHO & 149-265 & 40 &  2.1 &    $79\pm14$   & $84\pm15$  \\
02/05.2 & HNCO       & 153-264 & 12 &  2.1 &   $50\pm 9$   & $76\pm13$  \\
02/05.2 & HCOOH      & 151-270 & 29 &  2.1 &   $49\pm 9$   & $49\pm9$  \\
02/05.2 & CH$_3$CHO  & 151-271 & 79 &  2.1 &   $43\pm 6$   & $44\pm6$  \\
\hline
\end{tabular}
\end{center}
\tablefoot{Subscript ``$_d$'' has been added to the molecules for which
  a daughter Haser density profile is assumed with the parent scale length
  provided below.\\
  \tablefoottext{a}{Mean pointing offset.}\\
  \tablefoottext{b}{Result of non-linear fit with $\chi^2$ minimisation.}\\
  \tablefoottext{c}{Number of lines used for the determination of $T_{rot}$.
    In some cases lines have been averaged by groups of two lines having
    similar spectral characteristics
    (same J level, close energy levels and Einstein A coefficients).}\\
  \tablefoottext{d}{Assuming a distributed source with a parent scale length
    of $L_p$=5000~km.}\\
  \tablefoottext{e}{Temperatures deduced from the negative (blueshifted)
    part of the line (respectively positive or redshifted side for v+)
    sampling mostly the day-side of the coma (respectively night-side).}\\
}
\end{table}


\section{Production rates and abundances}
\label{sect-results}
Production rates are computed using our excitation and radiative
transfer codes, and parameters as in previous papers \citep{Biv21a}.
We assumed isotropic outgassing and a constant velocity and temperature
(Sect.~\ref{sect-vexp} and \ref{sect-temp}).
We provide both daily production rates when the molecules are detected
with a sufficiently high signal-to-noise (Table~\ref{tabqpdayleonard}
and~\ref{tabqpdayztf})
and averages over the 13--16 November and 8--13 December 2021 periods for
comet C/2021~A1 (Leonard) and 3--7 February 2023 for comet C/2022~E3 (ZTF)
(Tables~\ref{tabqpleonard} and~\ref{tabqpztf}).
For the molecules for which we observed several lines '$i$', either
detected individually with a S/N $>$5 or not, the final production rate is the
weighted average of all production rates $Q_i$. They are computed on each
considered line (even when $Q_i$ is negative) and averaged with weighting
according to $1/\sigma(Q_i)^2$ were $\sigma(Q_i)$ is the
uncertainty in production rate deduced from the line '$i$'.

\subsection{Reference water production rate}\label{sect-qh2o}
Water production rates are inferred from the monitoring of OH
lines at 18-cm with the  NRT (Sect.~\ref{sect-nancay}) 
and H$_2$O observations at 557~GHz with {\it Odin} (Sect.~\ref{sect-odin}).
Water productions rates for these comets were also obtained from
SOHO/SWAN \citep{Com23a,Com23b}.
For the period of observation with the IRAM-30m radio telescope, and to compute
relative abundances, we estimate $Q_{\rm H_2O}$ = 2, 3 and 4$\times10^{28}$~\mols
for the 12, 13--16 November and 8--13 December 2021, respectively, for
comet C/2021~A1 (Leonard). For comet C/2022~E3 (ZTF), extrapolation from
{\it Odin} production rates in Table~\ref{tabqperztf}, following $1/r_h^2$, yields
$Q_{\rm H_2O} = 5\times10^{28}$~\mols~ for the 3--7 February period.
These values also follow the longer term trend of brightness 
evolution. We use these values to determine collisional rates and abundances
relative to water.

\subsection{Temporal variations in the production rates}\label{sect-qvar}
\begin{figure}[]
\centering
\resizebox{\hsize}{!}{
  \includegraphics[angle=270,width=0.9\textwidth,trim= 30 0 0 0, clip]{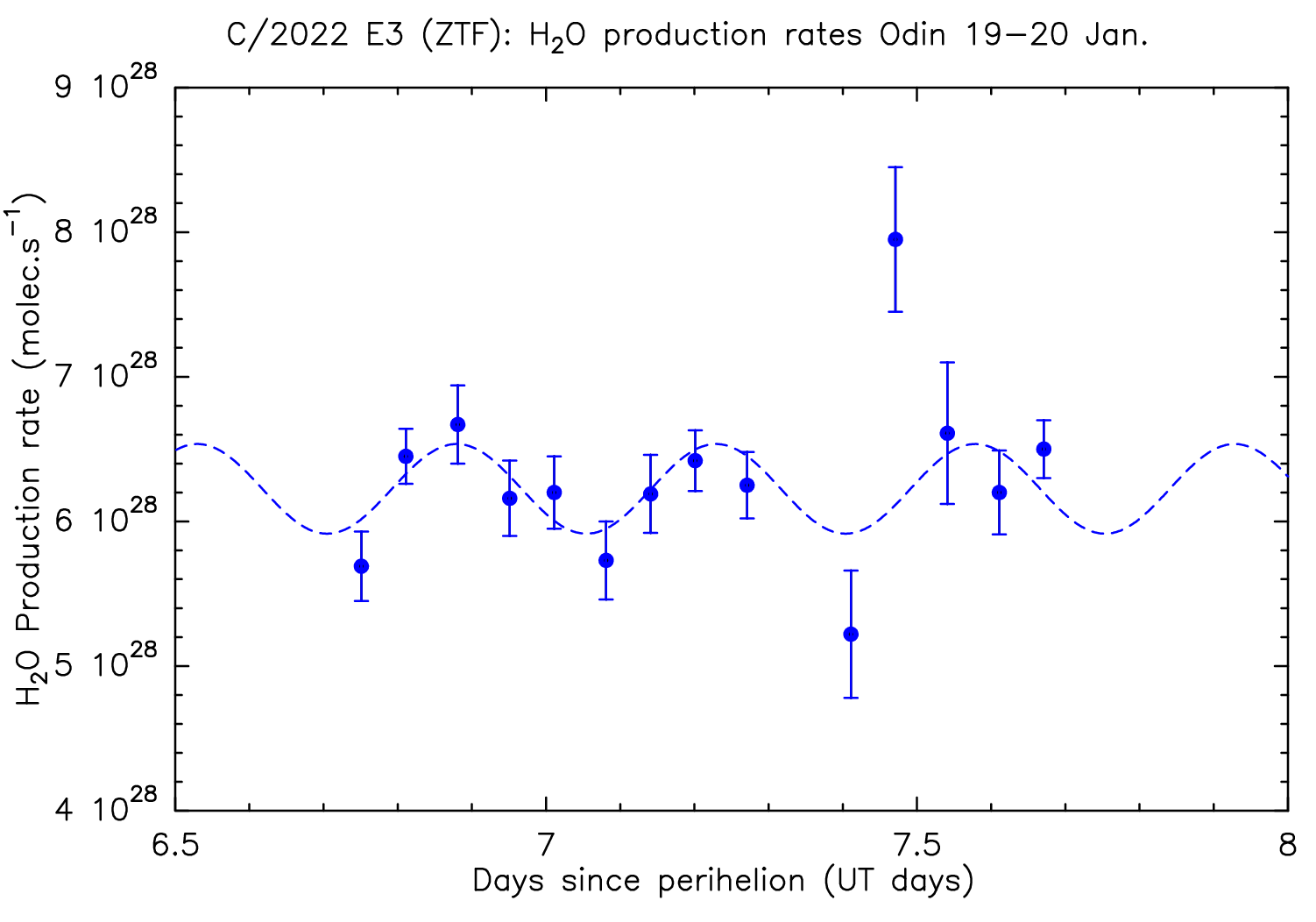}}
\caption{Water production rates (in~\mols) of comet C/2022~E3 (ZTF) between 19.3
  and 20.3 January 2023 derived from observations
  with {\it Odin} (central pointings). A best fit sinusoidal variation is plotted
  in blue.}
\label{figqperh2oztf}
\end{figure}

\begin{figure}[]
\centering
\resizebox{\hsize}{!}{
  \includegraphics[angle=270,width=0.9\textwidth,trim= 30 0 0 0, clip]{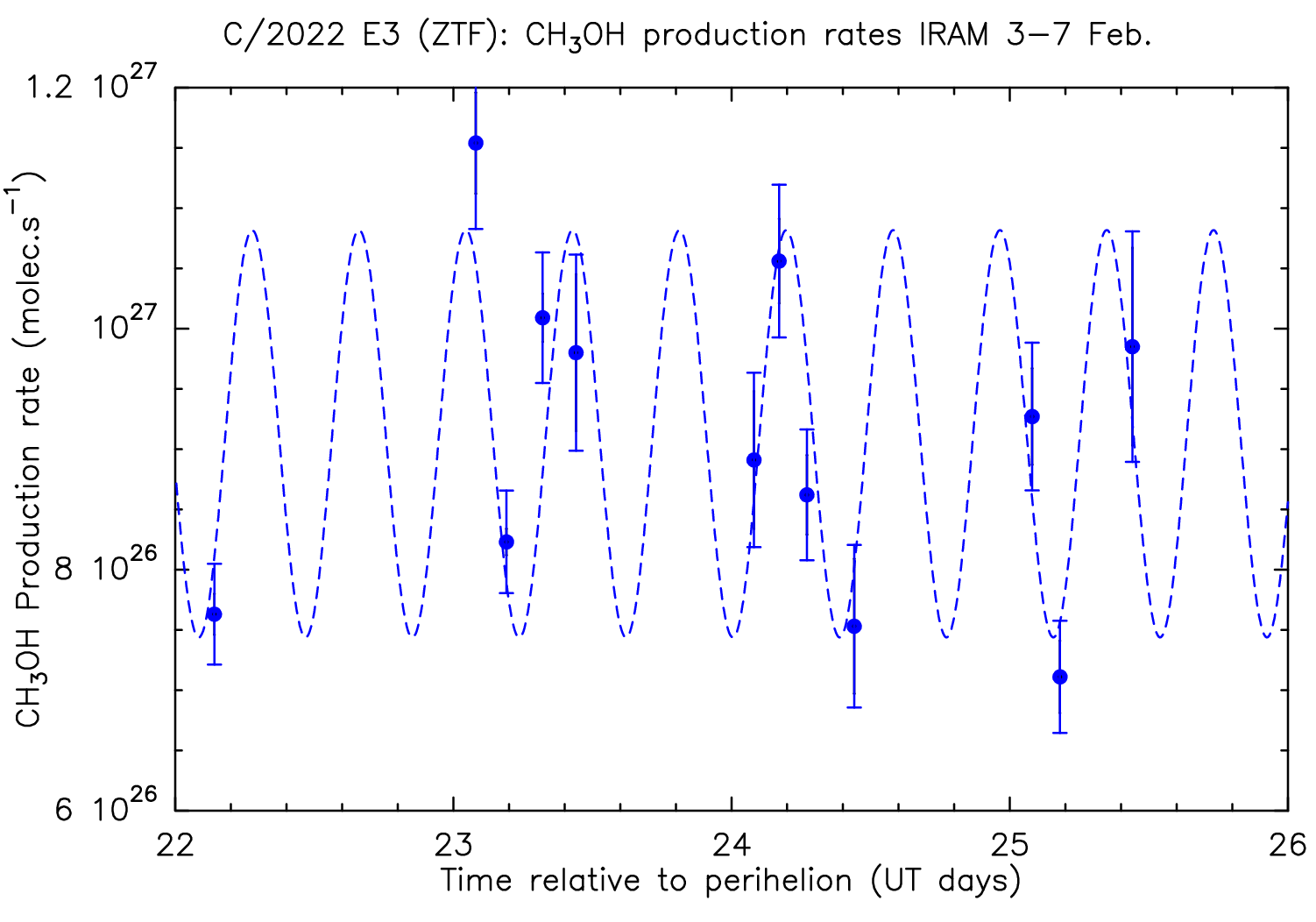}}
\caption{Methanol production rates (in~\mols) of comet C/2022~E3 (ZTF) between
  3.7 and 7.1 February 2023 derived from observations with IRAM~30m.
  An average 5\% calibration uncertainty has been added to the formal
  uncertainty due to system temperature noise.
  A best fit sinusoidal variation is plotted in blue.}
\label{figqperch3ohztf}
\end{figure}

Evidence for short-term variability in the outgassing is present both
in {\it Odin} and IRAM data for comet C/2022~E3 (ZTF), but not for comet C/2021~A1.
Visual inspection of the production rates folded on a single period seems to
indicate a periodic
pattern of about 0.35 days in $Q_{\rm H_2O}$ and 0.38 days in $Q_{\rm CH_3OH}$,
while optical observations have reported a periodic pattern in the CN structures
of $0.363\pm0.004$ days \citep{Kni23} and $0.354\pm0.004$ days \citep{Man23}.
We fitted (Figs.~\ref{figqperh2oztf} and ~\ref{figqperch3ohztf}) a sine-wave
variation (for simplicity - a more complex profile would require more
parameters) to our water and methanol production rates and found similar
periods.
Table~\ref{tabqperztf} provides the fitted parameters and their uncertainty
based on $\chi^2$ minimisation for a simple sinusoidal pattern.
The relative amplitudes are 18\% for methanol and 5\% for H$_2$O. H$_2$O being
observed with a beam about ten times larger (the {\it Odin} beam radius of 33000~km
covers molecules emitted over a full period of 0.38 days), and having optically
thick lines, it is expected to display shallower variations. The amplitude of
the variations for production rate of methanol are more likely to represent the
full extent of variations of outgassing of the nucleus and are similar to the
variations ($\pm20$\%) observed for HCN. This has to be taken into account when
deriving precise ratios (e.g. isotopic ratios sect.~\ref{sect-isotopicratio}),
but for general abundances
averaged over the four days the impact is small, especially as on the 4, 5 and 6
February evenings (Table~\ref{tablogztf}) the observations covered more than a
full period of $\sim$0.36 days.

\begin{table*}[ht]
\caption[]{Time periods found to fit comet C/2022~E3 (ZTF) production rates}\label{tabqperztf}
\begin{center}
\begin{tabular}{cccccc}
\hline\hline
Molecule  & $Q$\tablefootmark{a}  & $\Delta Q$\tablefootmark{b} & $T_p$ & Confidence & Method \\[0cm]
  & (\mols) & (\mols) & (Days) &  Level or $\chi^2$  &        \\[0cm]
\hline
H$_2$O  & $6.27\pm0.07\times10^{28}$ & 0 & 0 & $\chi^2_{10}=2.51$ & Constant \\
H$_2$O  & $6.23\pm0.08\times10^{28}$ & $0.31\pm0.11\times10^{28}$ & 
               $0.349\pm0.022$ & $\chi^2_{10}=2.44$ & Sine first order \\
CH$_3$OH & $9.12\pm0.18\times10^{26}$ & $1.69\pm0.28\times10^{26}$ & 
               $0.3841\pm0.0029$ & $\chi^2_{8}=2.31$ & Sine first order \\
\hline
\end{tabular}
\end{center}
\tablefoot{
  \tablefoottext{a}{Mean production rate.}
  \tablefoottext{b}{Amplitude in production rate of the sine fitting.}\\
  }
\end{table*}

\begin{figure}[]
\centering
\resizebox{\hsize}{!}{
  \includegraphics[angle=0,width=0.9\textwidth]{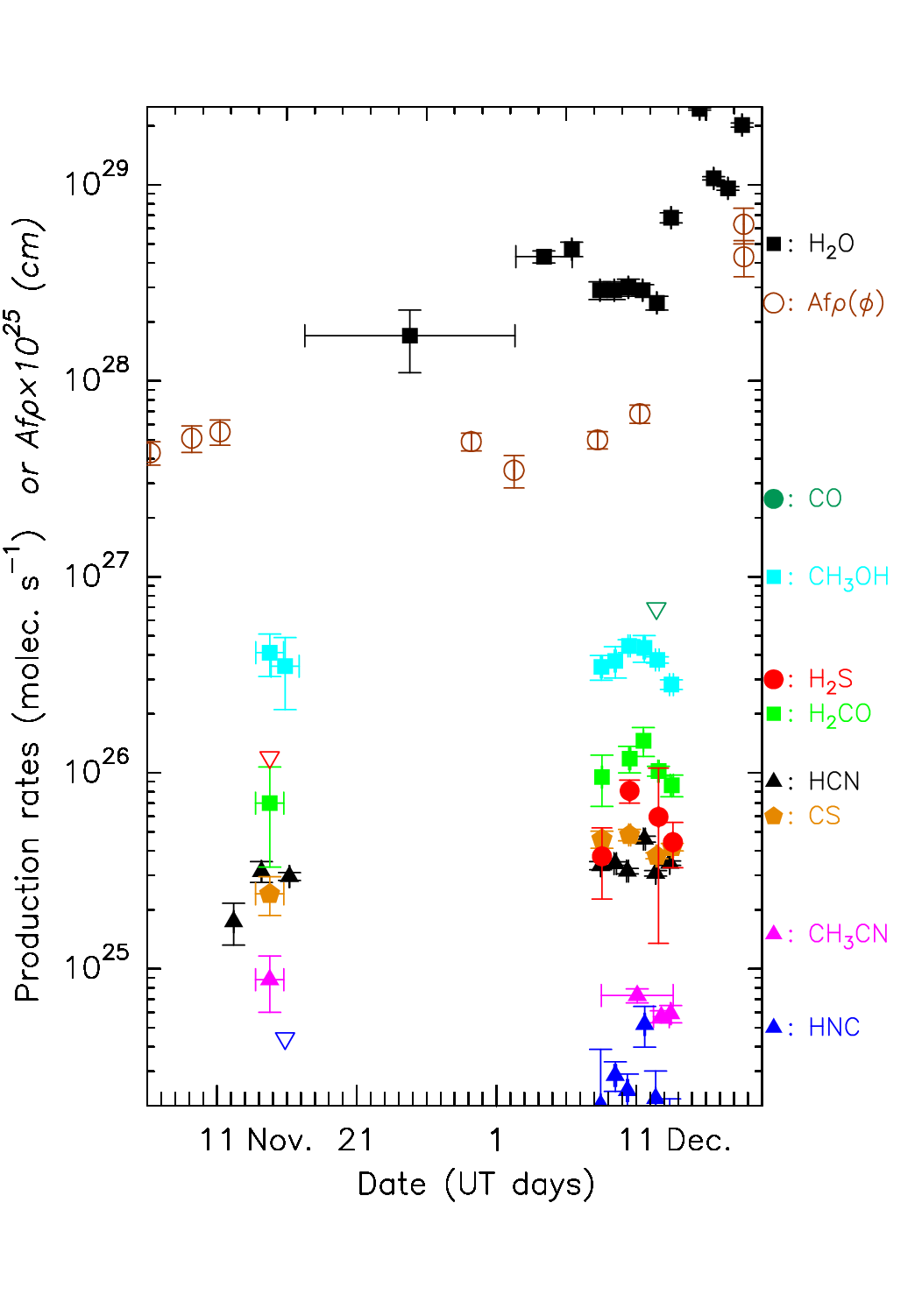}}
\caption{Production rates of comet C/2021~A1 (Leonard) in November and December
  2021. Water production rates are $1.1\times Q_{OH}$ from
  Sect.\ref{sect-nancay}. $Af\rho$ were measured from images of the comet
  by N. Biver, not corrected for the phase angle.}
\label{figqpleonard}
\end{figure}

\begin{figure}[]
\centering
\resizebox{\hsize}{!}{
  \includegraphics[angle=0,width=0.9\textwidth]{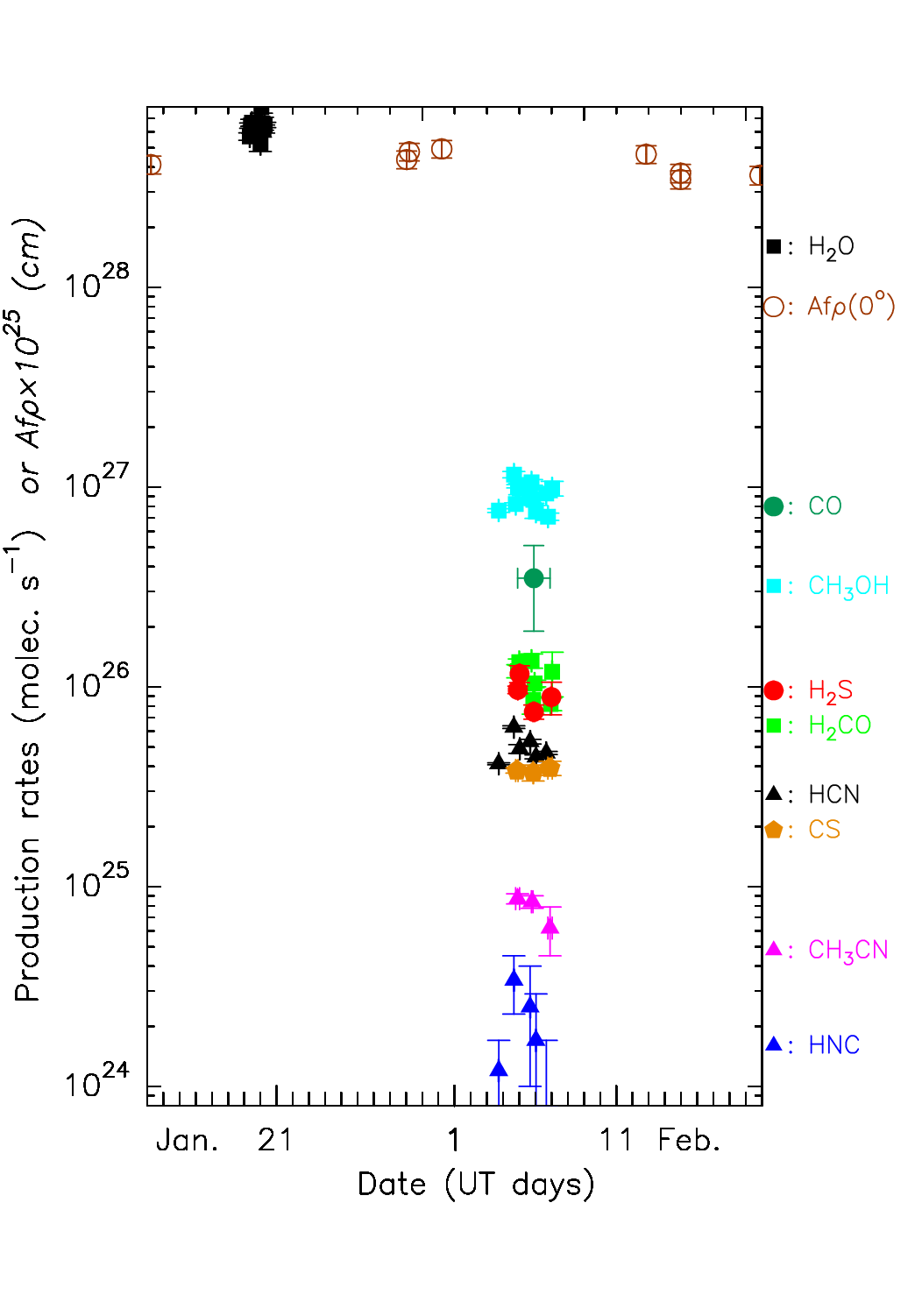}}
\caption{Production rates of comet C/2022~E3 (ZTF) in January and February
  2023. Water production rates are from {\it Odin} observations 
  (Sect.\ref{sect-odin}, Fig.~\ref{figqperh2oztf}).
  $Af\rho$(0\deg) were measured from images of the
  comet by N. Biver, corrected for the phase angle.}
\label{figqpztf}
\end{figure}

For other species day-to-day variations of the production rates are provided in
Tables~\ref{tabqpdayleonard} and ~\ref{tabqpdayztf}, and plotted in
Figs.~\ref{figqpleonard} and ~\ref{figqpztf}, together with some estimates of
the $Af\rho$ quantity\footnote{https://www.lesia.obspm.fr/comets}
in order to have a rough idea of the relative evolution
of the dust production rate.

\subsection{Molecules coming from a distributed source}
\label{sect-distri}

The coarse mapping done during the observations provides some information
on the spatial extent of the emission for some molecules, within 14\arcsec~
from the nucleus (2400~km at the distance of comet C/2021~A1 in December 2021)
to $\sim20$\arcsec (4500~km) for comet C/2022~E3 (ZTF).
Tables~\ref{tabqdistrileonard} and~\ref{tabqdistriztf}
provide information on the spatial extent of the
emission of some molecules (average line intensities as a function of pointing
offset) and constraints on the scale length and production from a distributed
source. We adjusted a Haser daughter species density profile to the data
and provide
the result of the $\chi^2$ minimisation with $1\sigma$ uncertainty on the
scale length ($L_p$) and production rate ($Q_p$). Values for other fixed scale
lengths ($Q_p$ and corresponding $\chi^2$) are provided.
These results have to be taken with caution since:
(i) the excitation of the rotational levels is not
perfectly known due to variable temperature in the coma, unknown collisional
cross-sections and other poorly estimated processes that can mimic extended
emission, (ii) the comet was variable in activity (it underwent a huge
outburst on 14 December just after the last observation),
(iii) the observations are
mostly probing the 1600--5000~km spatial scales (beam size - extent of the
mapping), so information on very different scale lengths is not well
constrained.

{\bf HCN:} it has been shown to be mostly released from nucleus ices
\citep{Cor14}, but the IR observations \citep{Del16,Lip21} that probe the
inner part of the coma also generally find higher production rates of HCN than
the radio ones \citep{Biv24}.
The observations of both comets (data were treated day by day to limit time
variability effects) suggest that a fraction of HCN is produced in the coma.
The best fit to the 12.5 and 13.4 December observations of comet C/2021~A1
yield a parent scale length on the order of 350~km with $Q_{p}(\rm HCN)$
larger by 15 to 35\% than with $L_p=0$. HCN(3-2) data from comet
C/2022~E3 (ZTF) yield a
scale length of $600\pm500$~km while HCN(2-1) suggest $L_p$(HCN)$<780$km,
implying also $Q_{p}(\rm HCN)$ larger by $\approx20$\%
(Tables~\ref{tabqdistrileonard} and~\ref{tabqdistriztf}).
In both comets the retrieved parent scale length is smaller than the beam size,
so the extended production of HCN needs to be further investigated at higher
spatial resolution.
The retrieved production rate would only be increased by $\approx20$\%, but
other issues show that the excitation of HCN is not fully understood.
We also cannot exclude that an excitation process, 
for example a larger gas temperature or electron collision rate that would
decrease the $J$=3 rotational level population close to the nucleus, is
responsible for this apparent distributed production.
Also the imperfect modelling of the line shape
(underestimation of the $J(3-2)$ $F=2-2$ and $F=3-3$ hyperfine
satellite lines and a signal strength around zero velocity in
Figs.~\ref{figsimhcn32-2pop},\ref{figsimhcn32-2popztf} being too high),
suggest that there are other issues. A much larger opacity could provide
a better fit to the line shape, although requiring unrealistic production rate
and line intensity.

{\bf HNC:} The mean of offset observations of HNC(3--2) on the whole period
does not yield a significant detection (signal-to-noise is below 2) but
$L_p=0\pm700$ km from $\chi^2$ minimisation is the best fit to comet
C/2021~A1 data.
In comet C/2022~E3 we find $L_p>2800$~km at 1-$\sigma$, also poorly constrained
with signal below 3-$\sigma$ at offsets larger than 5\arcsec.
Interferometric observations
of HNC \citep{Cor17, Rot21} suggest a parent scale length on the order of
$\sim2000$~km at 1 au (scaled as $r_h^{-2}$), somewhat compatible with our
data. Therefore, we provide in the tables $Q_{p}(\rm HNC)$ for nuclear source
$L_p=0$ and $L_p$(HNC) = 1000~km and 2000~km for comets C/2021~A1 and C/2022~E3,
respectively.

{\bf CS:} in comet Leonard we obtained data at offset positions on three
different dates: 10.5, 12.5 and 13.6 December.
The only two points obtained on the first date do not yield significant
constraints ($L_p(\rm CS)>800$~km at $1\sigma$) and are not listed in Table~\ref{tabqdistrileonard}.
The larger dataset of $J$=5--4 line observations of 12.48 December
(Table~\ref{tabqdistrileonard}) yields a reliable value
(reduced $\chi^2_\nu = 0.94$): $L_p(\rm CS)=500_{-390}^{+580}$~km.
We tried to combine the $J$=5--4 and $J$=3--2 observations of 13.6 December
but they yield a large value for $L_p(C\rm S)>2700$~km (mostly driven by the
CS(5-4) data that yield $L_p(\rm CS)>2400$~km at $1\sigma$). This implies
a too large value $Q_{p}(\rm CS) = 12.6\times10^{25}$~\mols~ for the nominal
fitted value of $L_p(\rm CS)=7000$~km. Since CS(3-2) and CS(5-4) were not
observed simultaneously, the constraint may not be reliable.
CS(3-2) data taken alone suggests $L_p(\rm CS)\sim1400$~km, with
$Q_{p}(\rm CS) = 5.2\times10^{25}$~\mols~ although compatible with any
value of $L_p$ at $1\sigma$.

In comet C/2022~E3, taken separately, CS(3-2) and CS(5-4) maps yield parent
scale lengths of $L_p(\rm CS)=1400^{+2000}_{-930}$ and $700^{+1450}_{-620}$km,
or $L_p(\rm CS)=1900^{+900}_{-570}$~km combining all data (larger since
constrained by a unique value for $Q_p(\rm CS)$, Table~\ref{tabqdistriztf}).

Around these heliocentric distances (0.8 and 1.2~au), the CS$_2$
photo-dissociation scale lengths are expected to be $\sim$260 and 560~km.
The average of all measurements obtained, yield $L_p(\rm CS)$ $>800$ and
$1300-2200$~km, respectively. Most other observations also suggest
a parent scale length for CS that is about 3--5 times the CS$_2$ dissociation
scale length \citep{Biv22,Rot21}. Hence, we use $L_p(\rm CS)$=1000 and 2000~km
for C/2021~A1 and C/2022~E3, respectively. This is equivalent to assuming that CS
comes from a parent molecule X-CS with a photo-dissociation rate at 1~au
$\beta_0({\rm X-CS})\sim4.5\times10^{-4}$~s$^{-1}$,
which is comparable to the revised photo-destruction rate of H$_2$CS
\citep[$\beta_0(\rm H_2CS)=4.9\times10^{-4}$~s$^{-1}$,][]{Hro23}.
However upper limits on the H$_2$CS abundance are much lower than for CS parent.
Note that assuming that CS comes
directly from the nucleus reduces the production rate by only $\sim$14\%
when comparing to the case of production by dissociation of CS$_2$ with its
known lifetime.

{\bf H$_2$CO:} Combining the two $J_{K_aK_c}=3_{1K_c}$ lines at 211.211 and
225.698~GHz observed on 12.6 December provides some constraints on the spatial
distribution of H$_2$CO in the coma of comet C/2021~A1. The retrieved parent
scale length is $1500_{-970}^{+3700}$~km (Table~\ref{tabqdistrileonard})
within the range of values
($0.2-1.6\times$ the photo-destruction scale length $L_d$ of H$_2$CO,
which is 2200~km around 12 December), found from previous observations
\citep{Biv22,Biv99,Rot21,Cor14}. The $+1\sigma$ uncertainty is large
and the upper limit not very precise ($\chi^2$ does not increase as steeply
as for the lower limit). In addition, if there is some nucleus contribution,
we can mostly say that the H$_2$CO coming from a distributed source must have
a parent scale length larger than 1400~km.
We use  $L_p(\rm H_2CO)=1500$ and 2800~km ($0.68\times L_d(\rm H_2CO)$) to
retrieve the (parent) production rate of H$_2$CO in December and November,
respectively.

\begin{figure}[]
\centering
\resizebox{\hsize}{!}{
  \includegraphics[angle=270,width=0.9\textwidth]{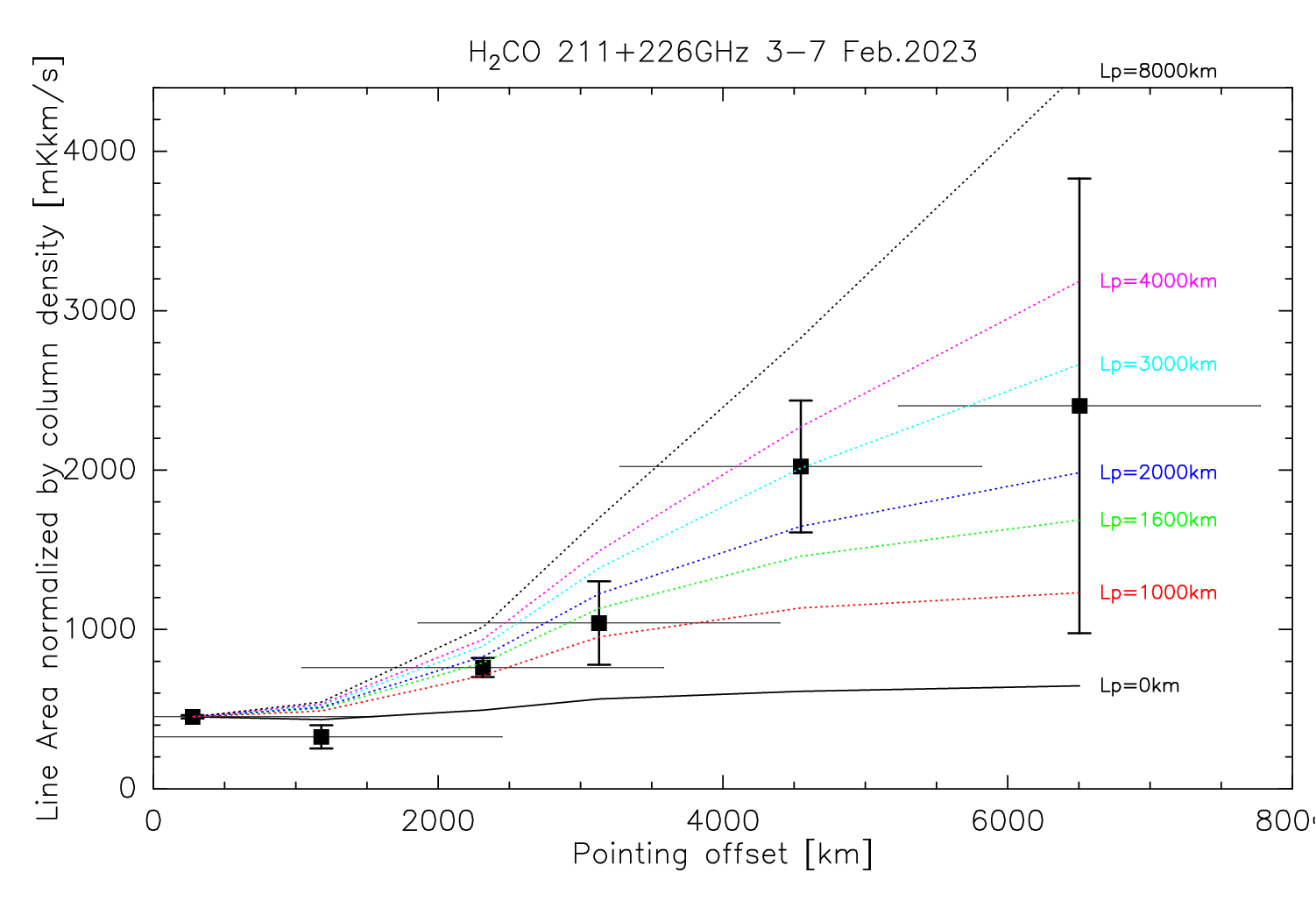}}
\caption{Spatial distribution of formaldehyde in comet C/2022~E3 (ZTF).
  This plot shows for the average of all observations of the twin 211~GHz
  and 226~GHz lines observed together in comet C/2022~E3 (ZTF) from 3 to 7
  February 2023: (i) in black squares (with error-bars) the observed line
  integrated intensity, (ii) in solid and doted lines the expected intensity
  for parent scale length of 0 or 1000 to 8000~km. For better visibility a
  normalisation has been applied: at each offset the line intensity
  is multiplied by the column density at the first offset divided by the one
  expected at the considered offset point, for a parent distribution.
  Production rate of the models are adjusted to get the same integrated
  intensity at the first point. Deviation from a flat line for $L_p=0$
  (black continuous line) is due to excitation effects.
  Vertical axis is normalised integrated intensity in mK km/s and
  horizontal axis is pointing offset converted into km at the distance of the
  comet.}
\label{figprofileh2coztf}
\end{figure}

For comet C/2022~E3 (Table~\ref{tabqdistriztf}), we made observations at offset
positions up to 20\arcsec~ on 5.00 February and 28\arcsec~ on 5.95 February
(4400--6500~km with a 2400~km beam) that provide more stringent constraints on
the spatial distribution of H$_2$CO.
For the assumed Haser daughter distribution profile, we
found $L_p(\rm H_2CO)=1000\pm430$ and $1700\pm1100$~km, respectively.
The observations of 6.94 February as well as those obtained for other lines
($2_{11}-1_{10}$ line at 150.498~GHz) provide looser constraints that fully
encompass those values. Fig.~\ref{figprofileh2coztf} shows the combination of
all H$_2$CO observations at 211 and 226~GHz in comet C/2022~E3 versus expected
evolution of line intensity as a function of pointing offset for various parent
scale lengths - the retrieved $L_p(\rm H_2CO)$ is $1700\pm500$~km which
corresponds to $0.36\pm0.11\times L_d(\rm H_2CO)$ at 1.18~au from the sun,
with $v_{exp}=0.68$~\kms.

{\bf NH$_2$CHO} and {\bf CH$_3$CHO}: We have also studied the spatial
distribution of these two species (only NH$_2$CHO in C/2021~A1): results are
not conclusive and may be affected by excitation effects
(especially for the high $K_a$ transitions of formamide).
Slightly distributed production (with $L_p\sim1500$~km) could be favoured for
comet C/2022~E3, but a nuclear source is not clearly excluded (within the
$2-\sigma$ uncertainty) as for formaldehyde.

In addition, if a significant production of the molecules comes from the
sublimation of icy grains in the coma, all species could also show some
distributed production. But the least distributed species like HCN, could put
an upper limit on the spatial extent of the distributed production.
The production of CS and H$_2$CO remains clearly more distributed and they are
thus daughter products.

\begin{table*}
\renewcommand{\tabcolsep}{0.08cm}
\caption[]{Production from a distributed source based on offset pointings on comet C/2021~A1 (Leonard).}\label{tabqdistrileonard}
\begin{center}
\begin{tabular}{llccrclccc}
\hline\hline\noalign{\smallskip}
UT   & Molecule & Freq. range  & lines\tablefootmark{a} & off.\tablefootmark{b} & Line intensity & & $L_{p}$\tablefootmark{c} & $Q_{p}$\tablefootmark{d} & $\chi^2$ \\
(mm/dd.d) & & (GHz) &         & (\arcsec) & (K~\kms) & & (km) & ($\times10^{25}$~\mols) & \\
\hline\noalign{\smallskip}
12/12.40 & HCN   & 265.886   & 1 &  1.1 &   $1.714\pm0.050$   & \vline &      0        & $2.9\pm0.1$ & 15.2 \\
         &       &           &   &  4.8 &   $1.274\pm0.091$   & \vline & $560_{-220}^{+330}$ & $3.9\pm0.1$ & 7.5 \\
         &       &           &   &  8.4 &   $0.977\pm0.092$   & \vline &              & & \\
         &       &           &   & 11.0 &   $1.113\pm0.125$   & \vline\vspace{0.2cm}&  & & \\
12/13.40 & HCN   & 265.886   & 1 &  0.4 &   $2.022\pm0.045$   & \vline &     0        & $3.4\pm0.1$ & 4.5  \\
         &       &           &   & 10.0 &   $0.939\pm0.072$   & \vline &  $240\pm100$ & $3.9\pm0.1$ & 0.1  \\
         &       &           &   & 14.1 &   $0.686\pm0.067$   & \vline\vspace{0.2cm}& & & \\
\hline
12/10.6  & HNC   & 271.981   & 1 &  1.7 &   $0.121\pm0.012$   & \vline &    $0\pm700$ & $0.19\pm0.02$ & 2.7 \\
12/11.0  &       &           &   &  4.7 &   $0.011\pm0.049$   & \vline &    1000      & $0.30\pm0.03$ & 4.1 \\
12/11.0  &       &           &   & 10.3 &   $0.050\pm0.027$   & \vline\vspace{0.2cm}& & & \\
\hline
12/12.5  & CS    & 244.936   & 1 &  1.8 &   $0.344\pm0.013$   & \vline &   0              & $2.2\pm0.1$  & 3.7 \\
         &       &           &   &  3.8 &   $0.274\pm0.079$   & \vline & 260              & $2.6\pm0.1$  & 2.2 \\
         &       &           &   &  8.2 &   $0.242\pm0.038$   & \vline & $500_{-390}^{+580}$ & $3.0\pm0.1$ & 1.9 \\
         &       &           &   & 11.8 &   $0.105\pm0.043$   & \vline\vspace{0.2cm}& 1000 & $3.8\pm0.1$ & 2.7 \\
12/13.6  & CS    & 244.936   & 1 &  1.1 &   $0.357\pm0.020$   & \vline &   0              & $2.5\pm0.1$  & 27.6 \\
         &       &           &   &  8.7 &   $0.170\pm0.079$   & \vline & 260              & $2.9\pm0.1$  & 20.0 \\
         &       &           &   & 10.4 &   $0.271\pm0.039$   & \vline &1000              & $4.2\pm0.2$  & 8.3 \\
         &       & 146.969   & 1 &  1.7 &   $0.143\pm0.011$   &\vline & $7000_{-430}^{+36000}$& $12.6\pm0.5$ &2.4\\
         &       &           &   & 10.3 &   $0.109\pm0.031$   & \vline\vspace{0.2cm}& & &  \\
\hline
12/12.6  & H$_2$CO & 211,226 & 2 &  0.9 &   $0.185\pm0.012$   & \vline & 0  & $2.2\pm0.1$  & 4.8 \\
         &       &           &   & 10.2 &   $0.078\pm0.022$   & \vline & $1500_{-970}^{+3700}$& $10.2\pm0.6$ & 1.4 \\
         &       &           &   & 14.7 &   $0.098\pm0.043$   & \vline\vspace{0.2cm} & 4000 & $18.8\pm1.2$& 2.0 \\
\hline
12/11.1  & HNCO  & 153.865   & 1 &  1.9 &   $0.040\pm0.006$   & \vline &   0 & $3.2\pm0.3$ & 13.5  \\
12/11.8  &       &           &   & 10.3 &   $0.036\pm0.021$   & \vline & 2000 & $7.1\pm0.6$ & 4.1 \\
12/12.8  &       & 219--264  & 3 &  1.4 &   $0.050\pm0.005$   & \vline & $10000$ & $23.0\pm1.9$ & 1.8 \\
12/12.9  &       &  242,264  & 2 &  9.9 &   $0.030\pm0.014$   & \vline\vspace{0.2cm}& & & \\
\hline
12/12.7  & HCOOH & 215--268  & 13\tablefootmark{e} & 1.5 & $0.0166\pm0.0024$ & \vline & 0 & $7.3\pm0.9$  & 4.9 \\
         &       &           & 10 & 10.1 & $0.0168\pm0.0074$ &\vline\vspace{0.2cm} & 2000 & $19.4\pm2.5$ & 2.7 \\
\hline
12/12.2  & NH$_2$CHO & 211--267 & 11 & 1.4 & $0.0321\pm0.0031$ & \vline & $0\pm400$       & $10.8\pm1.1$  &   \\
12/12.7  &       &  $K_a<3$  & 11 & 10.2 &   $0.0081\pm0.0079$ &\vline\vspace{0.2cm} &  &        &   \\
\hline
\hline
\end{tabular}
\end{center}
\tablefoot{
  \tablefoottext{a}{Number of lines used or averaged.}
  \tablefoottext{b}{Mean pointing offset.}
  \tablefoottext{c}{Parent scale length in km. The value with uncertainty
    (or lower limit) is the one obtained from the $\chi^2$ minimisation.}
  \tablefoottext{d}{Production rates from nuclear or distributed source with given parent scale length.}\\
  \tablefoottext{e}{The $\chi^2$ fitting was computed using all
    individual lines intensities separately, not their average.}
  \tablefoottext{f}{This $\chi^2$ for HCN(3-2) includes an additional
    10\% calibration uncertainty. Without, $\chi^2$ is about two orders of
    magnitude but leads to the same $L_p$.}
}
\end{table*}

\begin{table*}
\renewcommand{\arraystretch}{0.90}
\renewcommand{\tabcolsep}{0.08cm}
\caption[]{Production from a distributed source based on offset pointings on comet C/2022~E3 (ZTF).}\label{tabqdistriztf}\vspace{-0.6cm}
\begin{center}
\begin{tabular}{llccrclccc}
\hline\hline\noalign{\smallskip}
UT   & Molecule & Freq. range  & lines\tablefootmark{a} & off.\tablefootmark{b} & Line intensity & & $L_{p}$\tablefootmark{c} & $Q_{p}$\tablefootmark{d} & $\chi^2$\\
(mm/dd.d) & & (GHz) & & (\arcsec) & (K \kms) & & (km) & ($\times10^{25}$~\mols) & \\
\hline\noalign{\smallskip}
02/04.55 & HCN   & 265.886   & 1 &  1.6 &   $2.292\pm0.011$   & \vline &  0  & $4.6\pm0.3$ & 2.2\tablefootmark{e} \\
02/05.13 &       &           &   &  5.4 &   $1.716\pm0.029$   & \vline & $600\pm500$ & $5.4\pm0.3$ & 0.9 \\
02/04.47 &       &           &   &  9.8 &   $1.106\pm0.017$   & \vline &              & & \\
02/04.38 &       &           &   & 14.0 &   $0.953\pm0.122$   & \vline\vspace{0.1cm} & & & \\  
02/05.03 & HCN   & 177.261   & 1 &  0.9 &   $1.122\pm0.061$   & \vline &  $0\pm800$      & $4.9\pm0.2$ & 0.01  \\
         &       &           &   & 10.1 &   $0.731\pm0.079$   & \vline\vspace{0.1cm} &    &  &   \\
\hline
02/04.55 & HNC   & 271.981   & 1 &  1.6 &   $0.038\pm0.009$   & \vline & 0         & $0.74\pm0.14$ & 9.2 \\
02/05.13 &       &           &   &  5.4 &   $0.077\pm0.026$   & \vline & $2000$    & $1.59\pm0.29$ & 6.3 \\
02/04.47 &       &           &   &  9.8 &   $0.034\pm0.017$   & \vline & $>2800$   & $>1.92\pm0.34$ & 6.0 \\
02/04.38 &       &           &   & 14.0 &   $0.084\pm0.033$   & \vline\vspace{0.1cm} & & & \\
\hline
02/05.23 & CS    & 146.969   & 1 &  2.1 &   $0.096\pm0.004$   & \vline &    &   &  \\
02/05.26 &       &           &   &  8.8 &   $0.087\pm0.010$   & \vline &    0              & $1.95\pm0.05$ & 30.1 \\
02/05.41 &       &           &   & 13.0 &   $0.059\pm0.009$   & \vline &  550              & $2.44\pm0.06$ & 11.8 \\
02/05.08 & CS    & 244.936   & 1 &  1.8 &   $0.247\pm0.008$   & \vline & 1400              & $3.23\pm0.08$ &  3.3 \\
02/04.78 &       &           &   &  5.1 &   $0.196\pm0.018$   & \vline & $1900_{-470}^{+900}$ & $3.65\pm0.08$ &  2.6 \\
02/05.18 &       &           &   &  8.4 &   $0.166\pm0.034$   & \vline\vspace{0.1cm} & 2000  & $3.74\pm0.09$  &  2.7 \\
\hline
02/05.00 & H$_2$CO & 211,226 & 2 &  1.2 &   $0.282\pm0.012$   & \vline & &  &  \\
         &       &           &   & 10.1 &   $0.138\pm0.016$   & \vline &  0               & $7.29\pm0.28$  & 12.4 \\
         &       &           &   & 14.2 &   $0.078\pm0.019$   & \vline & $1000_{-330}^{+500}$ & $10.89\pm0.41$ & 1.7 \\
         &       &           &   & 20.0 &   $0.080\pm0.027$   & \vline\vspace{0.1cm} & 1700 & $13.25\pm0.51$  & 3.4 \\
02/05.95 & H$_2$CO & 211,226 & 2 &  1.3 &   $0.209\pm0.008$   & \vline &  &  &  \\
         &       &           &   & 10.1 &   $0.107\pm0.018$   & \vline &   0               & $5.59\pm0.21$  & 8.5 \\
         &       &           &   & 20.1 &   $0.053\pm0.020$   & \vline & $1700_{-640}^{+1700}$& $10.37\pm0.39$ & 0.5 \\
         &       &           &   & 28.3 &   $0.032\pm0.019$   & \vline\vspace{0.1cm} &  &  & \\
02/06.94 & H$_2$CO & 211,226 & 2 &  1.0 &   $0.168\pm0.013$   & \vline &  &  &  \\
         &       &           &   &  4.9 &   $0.117\pm0.026$   & \vline &   0               & $4.65\pm0.33$ & 5.0 \\
         &       &           &   & 10.0 &   $0.100\pm0.018$   & \vline\vspace{0.1cm} & $1700_{-1100}^{+3700}$&$8.21\pm0.57$ & 1.5 \\
02/05.2  & H$_2$CO & 150.498 & 1 &  2.1 &   $0.075\pm0.004$   & \vline &   &  &  \\
         &       &           &   &  8.8 &   $0.045\pm0.011$   & \vline & $0\pm2200$         & $6.97\pm0.36$  & 0.5 \\
         &       &           &   & 13.0 &   $0.037\pm0.012$   & \vline\vspace{0.1cm} & 1700 & $11.02\pm0.57$ & 1.1 \\
\hline
02/05.25  & HNCO  & 220-264  & 3 &  2.1 &   $0.036\pm0.005$   & \vline &    0    &  $2.06\pm0.25$ & 6.2  \\
02/04.92  &       & $K_a=0$  &   &  5.1 &   $0.048\pm0.011$   & \vline &  $>2000$ & $>3.97\pm0.46$ & 3.4 \\
02/05.42  &       &          &   &  9.1 &   $0.043\pm0.018$   & \vline\vspace{0.1cm} & 6000   &  $7.95\pm0.92$ & 2.7 \\
02/05.25  & HNCO  & 219-265  & 6 &  2.0 &   $0.013\pm0.003$   & \vline &  0    &  $2.99\pm0.57$ & 8.4  \\
02/04.92  &       & $K_a=1$  &   &  5.2 &   $0.021\pm0.007$   & \vline & $>1000$ & $>6.44\pm1.20$ & 7.3 \\
02/05.42  &       &          &   &  9.3 &   $0.023\pm0.009$   & \vline\vspace{0.1cm} & 6000   &  $24.9\pm4.6$  & 6.5 \\
\hline
02/05.34 & HCOOH & 215--268  & 15 & 1.7 & $0.0276\pm0.0017$  & \vline &  0       & $10.39\pm0.71$ & 0.9 \\
02/05.25 &       & $K_a=0-2$ & 15 & 5.1 & $0.0189\pm0.0057$  & \vline & $1100_{-1100}^{+3100}$ & $17.11\pm1.03$ & 0.1 \\
02/05.31 &       &           & 15 & 9.9 & $0.0102\pm0.0046$  & \vline\vspace{0.1cm} & &  &  \\
\hline
02/05.23 & NH$_2$CHO & 147--169 & 7 & 2.1 & $0.0120\pm0.0014$ & \vline &  &  &  \\
02/05.23 &        & $K_a=0-2$ & 7 &  8.9 & $0.0040\pm0.0039$  & \vline & $0\pm2500$         & $0.84\pm0.10$ & 1.1 \\
02/05.41 &        & $J=7-8$    & 7 & 13.0 & $0.0068\pm0.0036$  & \vline\vspace{0.1cm} & $1300$  & $1.18\pm0.14$ & 1.7 \\
02/05.44 & NH$_2$CHO & 211--267 & 15 & 1.7 & $0.0235\pm0.0016$ & \vline &  &  &  \\
02/05.26 &        & $K_a=0-2$ & 15 &  5.1 & $0.0329\pm0.0052$  & \vline &  0                 & $0.93\pm0.06$ & 8.6 \\
02/05.49 &        & $J=10-13$ & 15 & 9.9 & $0.0106\pm0.0038$  & \vline & $1300_{-900}^{+2500}$ & $1.50\pm0.10$ & 6.6 \\
02/04.83 &        &          &  6 & 14.1 & $0.0062\pm0.0078$  & \vline\vspace{0.1cm} &  &  &  \\
02/05.43 & NH$_2$CHO & 212--256 & 15 & 1.6 & $0.0131\pm0.0018$ & \vline &  0               & $1.07\pm0.13$ & 13.5 \\
02/05.26 &        & $K_a=3-5$ & 15 &  5.2 & $0.0158\pm0.0066$  & \vline & 1300              & $1.98\pm0.23$ & 8.5 \\
02/05.49 &        & $J=10-12$ & 15 & 10.0 & $0.0155\pm0.0031$  & \vline\vspace{0.1cm} & $>3000$ & $>3.24\pm0.37$ & 6.6 \\
\hline
02/05.94 & CH$_3$CHO & 211--232 & 18 & 1.2 & $0.0115\pm0.0013$ & \vline &  &  &  \\
02/06.94 &        & $K_a=0-3$ & 18 &  4.9 & $0.0061\pm0.0069$  & \vline &  0                 & $3.3\pm0.4$ & 8.0 \\
02/05.85 &        & $J$=11,12 & 18 & 10.1 & $0.0043\pm0.0017$  & \vline & $1600_{-1000}^{+5000}$& $5.9\pm0.6$ & 6.5 \\
02/05.00 &        &          & 18 & 14.2 & $0.0139\pm0.0042$  & \vline &  &  &  \\
02/05.62 &        &          & 18 & 20.1 & $0.0022\pm0.0034$  & \vline\vspace{0.1cm} &  &  &  \\
02/04.55 & CH$_3$CHO & 251--271 & 16 & 1.6 & $0.0133\pm0.0017$ & \vline &  &  &  \\
02/05.13 &        & $K_a=1-3$ & 16 &  5.4 & $0.0125\pm0.0044$  & \vline & 0                 & $3.5\pm0.4$ & 1.8 \\
02/04.47 &        & $J$=13,14 & 16 &  9.8 & $0.0067\pm0.0029$  & \vline & $2000_{-1600}^{+4000}$& $7.7\pm0.9$ & 0.2 \\
02/04.38 &        &          & 16 & 14.0 & $0.0040\pm0.0049$  & \vline\vspace{0.1cm} & &  &  \\
\hline
\end{tabular}
\end{center}\vspace{-0.5cm}
\tablefoot{
  \tablefoottext{b}{Mean pointing offset.}
  \tablefoottext{a}{Number of lines used or averaged.}
  \tablefoottext{c}{Parent scale length in km. The value with uncertainty
    (or lower limit) is the one obtained from the $\chi^2$ minimisation.}
  \tablefoottext{d}{Production rates from nuclear or distributed source with given parent scale length.}
  \tablefoottext{e}{This $\chi^2$ for HCN(3-2) includes an additional
    10\% calibration uncertainty. Without, $\chi^2$ is about two orders of
    magnitude larger but minimisation leads to the same $L_p$.}
}
\end{table*}

\begin{table}\renewcommand{\tabcolsep}{0.07cm}
\caption[]{Production rates in C/2021~A1 in November-December 2021 (weekly average).}\label{tabqpleonard}
\begin{center}
\begin{tabular}{lcccc}
\hline\hline\noalign{\smallskip}
UT date  & Molecule & $r_h$  & Production rate  & Lines\tablefootmark{a} \\
(mm/dd.d) &      &  (au)  & ($\times10^{25}$~\mols) &  \\
\hline
\multicolumn{5}{c}{13-16 November average} \\
\hline
11/15.8     & HCN           & 1.17   & $3.0\pm0.1$    &   1  \\
11/15.8     & HNC$_{\rm d}$\tablefootmark{b}  & 1.17   & $<0.44$      &  1 \\ 
11/14.8     & CH$_3$CN      & 1.18   & $0.9\pm0.3$    &  (8) \\
11/14.8     & H$_2$S        & 1.18   & $<11.5$        &  (1) \\
11/14.8     & CS            & 1.18   & $2.2\pm0.5$    &   1  \\
11/14.8     & H$_2$CO$_d$   & 1.18   & $7.0\pm3.7$    &  (1) \\
11/15.3     & CH$_3$OH      & 1.17   & $39\pm9$       & 2+(23) \\
\hline
\multicolumn{5}{c}{08-13 December average} \\
\hline
12/10.8     & HCN              & 0.79   & $3.4\pm0.1$  &  2 \\
12/12.8     & H$^{13}$CN       & 0.77   & $<0.072$     &  1 \\
12/12.8     & HC$^{15}$N       & 0.77   & $<0.047$     &  1 \\
12/10.8     & HNC$_{\rm d}$\tablefootmark{b}  & 0.79   & $0.24\pm0.02$ &  1 \\ 
12/12.5     & CH$_3$CN         & 0.77   & $0.62\pm0.03$  & 19+(6) \\ 
12/12.4     & HC$_3$N          & 0.78   & $0.14\pm0.05$  & (8) \\
12/12.4     & NH$_2$CHO        & 0.78   & $0.91\pm0.08$  & 8+(27) \\ 
12/12.5     & HNCO             & 0.78   & $2.95\pm0.30$  & 6+(6) \\ 
12/11.2     & H$_2$S           & 0.79   & $5.9\pm0.7$    & 1+(1) \\ 
12/12.0     & CS               & 0.78   & $2.9\pm0.1$    &  2 \\ 
     & CS$_d$\tablefootmark{c} & 0.78   & $4.1\pm0.1$    &  2 \\ 
12/13.1     & C$^{34}$S         & 0.77   & $0.15\pm0.05$ &  1 \\ 
12/12.4     & SO               & 0.78    & $<0.6$        & (5) \\
     & SO$_d$\tablefootmark{d} & 0.78    & $<1.5$        & (5) \\ 
12/12.4     & SO$_2$            & 0.78   & $<1.6$        & (20) \\
12/12.4     & OCS               & 0.78   & $4.3\pm1.1$   & (6) \\
12/12.4     & H$_2$CS           & 0.78   & $<0.8$        & (5) \\
12/12.0     & H$_2$CO           & 0.78   & $5.0\pm0.2$   &  4+(2) \\
   & H$_2$CO$_d$\tablefootmark{e} & 0.78  & $10.2\pm0.5$  & 4+(2) \\
12/12.5     & CO                & 0.77   & $<78$         & (1) \\
12/12.4     & HCOOH             & 0.78   & $7.6\pm0.9$   & (28) \\
12/11.6     & CH$_3$OH          & 0.78   & $35\pm1$      & 42+(14) \\ 
12/12.4     & CH$_3$CHO         & 0.78   & $1.5\pm0.3$   & (53) \\ 
12/12.4     & (CH$_2$OH)$_2$    & 0.78   & $5.3\pm1.3$   & (39) \\ 
12/12.0     & HCOOCH$_3$        & 0.78   & $<5.8$        & (6) \\ 
12/12.4     & CH$_2$OHCHO       & 0.78   & $2.0\pm0.5$   & 1+(10) \\  
12/12.4     & C$_2$H$_5$OH      & 0.78   & $7.3\pm1.5$   & (84) \\  
12/12.4     & CH$_2$CO          & 0.78   & $1.3\pm0.6$   & (10) \\  
12/11.6     & CH$_3$SH          & 0.79   & $<3.2$        & (32) \\  
12/10.6     & PH$_3$            & 0.80   & $<6.4$        & (1) \\  
12/11.6     & c-C$_3$H$_2$      & 0.80   & $<0.9$        & (11) \\  
12/11.6     & l-C$_3$H$_2$      & 0.80   & $<0.2$        & (13) \\  
\hline
\end{tabular}
\end{center}
\tablefoot{Subscript ``$_d$'' has been added to the molecules for which
  a daughter Haser density profile is assumed with the parent scale length
  provided below.\\
  \tablefoottext{a}{Number of lines used for the determination of $Q$,
    in parentheses the number of lines that are not individually detected.}\\
  \tablefoottext{b}{Where we assume that HNC is produced in the
  coma with a Haser parent scale length of 1000~km and 500~km in December \citep{Cor17}.}\\
  \tablefoottext{c}{Assuming a parent scale length of 1000~km (about 4$\times$
    longer than the photo-dissociation scale length of CS$_2$, assumed parent otherwise).}\\
  \tablefoottext{d}{Assuming a parent scale length of 1700~km (SO$_2$).}\\
  \tablefoottext{e}{Where we assume that H$_2$CO is produced in the
  coma with a Haser parent scale length of 2800~km in November and 1500~km in December.}
}
\end{table}

\begin{table}\renewcommand{\tabcolsep}{0.07cm}
\caption[]{Production rates in C/2022~E3 (ZTF) in February 2023.}\label{tabqpztf}
\begin{center}
\begin{tabular}{lcccc}
\hline\hline\noalign{\smallskip}
UT date  & Molecule & $r_h$  & Production rate  & Lines\tablefootmark{a} \\
(mm/dd.d) &      &  (au)  & ($\times10^{25}$~\mols) &  \\
\hline
\multicolumn{5}{c}{3-7 February 2023 average} \\
\hline
02/04.6     & HCN              & 1.17   & $4.6\pm0.1$  &  2 \\
02/05.1     & H$^{13}$CN       & 1.18   & $<0.065$     &  (1) \\
02/05.1     & HC$^{15}$N       & 1.18   & $<0.044$     &  (1) \\
02/05.9     & DCN              & 1.18   & $<0.084$     &  (1) \\
02/04.6     & HNC$_{\rm d}$\tablefootmark{b}  & 1.17   & $0.16\pm0.03$ &  1 \\ 
02/05.1     & CH$_3$CN         & 1.18   & $0.84\pm0.03$  & 20+(4) \\ 
02/05.2     & HC$_3$N          & 1.18   & $<0.11$     & (8) \\
02/05.3     & NH$_2$CHO        & 1.18   & $0.94\pm0.05$  & 14+(25) \\ 
02/05.3     & HNCO             & 1.18   & $2.08\pm0.21$  & 3+(9) \\ 
02/05.5     & H$_2$S           & 1.18   & $9.0\pm0.2$    & 2 \\ 
02/05.2     & H$_2^{34}$S       & 1.18   & $<0.7$        & (1) \\ 
02/05.2     & CS               & 1.18   & $2.4\pm0.1$    &  2 \\ 
     & CS$_d$\tablefootmark{c} & 1.18   & $3.7\pm0.1$    &  2 \\ 
02/05.3     & C$^{34}$S$_d$    & 1.18   & $0.03\pm0.01$ &  1 \\ 
02/05.3     & SO               & 1.18    & $<0.5$        & (5) \\
     & SO$_d$\tablefootmark{d} & 1.18    & $<1.5$        & (5) \\ 
02/05.3     & SO$_2$            & 1.18   & $<1.5$        & (16) \\
02/05.3     & OCS               & 1.18   & $3.4\pm0.8$   & (6) \\
02/05.1     & H$_2$CS           & 1.18   & $<0.9$        & (7) \\
02/05.2     & H$_2$CO           & 1.18   & $6.1\pm0.1$   &  5+(1) \\ 
   & H$_2$CO$_d$\tablefootmark{e} & 1.18  & $10.9\pm0.2$  & 5+(1) \\
02/04.9     & HDCO$_d$          & 1.17   & $<0.69$       &  (5) \\ 
02/05.9     & CO                & 1.18   & $35\pm16$     & 1 \\
02/05.3     & HCOOH             & 1.18   & $9.0\pm0.7$   & 13+(17) \\
02/05.1     & CH$_3$OH          & 1.18   & $87.8\pm0.8$  & 62+(12) \\ 
02/05.3     & CH$_3$CHO         & 1.18   & $3.4\pm0.2$   & 13+(66) \\ 
02/05.3     & (CH$_2$OH)$_2$    & 1.18   & $5.5\pm0.7$   & 6+(108) \\ 
02/05.3     & HCOOCH$_3$        & 1.18   & $<4.8$        & (6) \\ 
02/05.3     & CH$_2$OHCHO       & 1.18   & $<1.2$        & (13) \\  
02/05.3     & C$_2$H$_5$OH      & 1.18   & $8.7\pm1.3$   & 4+(103) \\  
02/05.3     & CH$_2$CO          & 1.18   & $<1.5$        & (7) \\  
02/05.3     & CH$_3$SH          & 1.18   & $<3.1$        & (20) \\  
02/05.3     & CH$_3$NH$_2$      & 1.18   & $<4.8$        & (21) \\  
02/05.3     & CH$_3$COCH$_3$    & 1.18   & $<1.6$        & (29) \\  
02/04.6     & PH$_3$            & 1.17   & $<4.4$        & (1) \\
02/05.3     & c-C$_3$H$_2$      & 1.18   & $<0.6$        & (23) \\  
02/05.3     & l-C$_3$H$_2$      & 1.18   & $<0.1$        & (13) \\  
\hline
\end{tabular}
\end{center}
\tablefoot{Subscript ``$_d$'' has been added to the molecules for which
  a daughter Haser density profile is assumed with the parent scale length provided below.\\
  \tablefoottext{a}{Number of lines used for the determination of $Q$,
    in parentheses the number of lines that are not individually detected.}\\
  \tablefoottext{b}{Where we assume that HNC is produced in the
  coma with a Haser parent scale length of 2000~km \citep{Cor17}.}\\
  \tablefoottext{c}{Assuming a parent scale length of 2000~km (about 4$\times$
    longer than the photo-dissociation scale length of CS$_2$, assumed parent otherwise).}\\
  \tablefoottext{d}{Assuming a parent scale length of 3800~km (SO$_2$).}\\
  \tablefoottext{e}{Where we assume that H$_2$CO is produced in the
  coma with a Haser parent scale length of 1700~km.}
}
\end{table}

\subsection{Relative abundances}
\label{sect-abundances}
Abundances relative to water, assuming $Q_{\rm H_2O}=2-3\times10^{28}$~\mols~
in November and $Q_{\rm H_2O}=4\times10^{28}$~\mols~ in December 2021 for comet
C/2021~A1 and $Q_{\rm H_2O}=5\times10^{28}$~\mols~ in early February 2023 for
comet C/2022~E3 are provided in Table~\ref{tababund}.
Abundances have also been measured for several molecules in comet C/2021~A1
in the infrared by \citet{Fag23}, but during the outbursting period of the
comet in December-January. While these authors found very similar abundances
for HCN, OCS, H$_2$CO, and compatible with our upper limit on CO, they did not
detect CH$_3$OH. Their inferred CH$_3$OH/H$_2$O is very low, down to one order
of magnitude lower than our abundance (based on the secure detection of over
42 lines). Either the comet had exhausted its
methanol content during the outburst phase at the end of December, or the
infrared observations missed part of the methanol emission, following a trend
also seen in millimetre investigations \citep{Biv11} which show an apparent
decrease in methanol abundance as comets get closer to the Sun.
The methanol observations of \citet{Fag23} were obtained at 0.62~au from the
Sun, where the sampled gas coma gets warmer (assumed to be 120~K) and the
fraction of methanol in an excited torsional state may also depopulate the
ground vibrational levels.

\subsection{Upper limits}
\label{sect-upperlimits}

The surveys cover part of the 2~mm wavelength range and most of the 1~mm
(210--272 GHz) as shown in spectra in Figs.~\ref{figsurveyleonard2mm2}--
\ref{figsurveyleonard1mm6} and \ref{figsurveyztf2mm3}--\ref{figsurveyztf1mm6}.
Most of the known organic
molecules have many transitions in these wavelength ranges. We have not noticed
clearly unidentified lines (above the $5-\sigma$ level). We checked
the sensitivity of the observations for the following molecules: CH$_3$SH,
CH$_3$OCH$_3$, CH$_3$COCH$_3$, c-C$_2$H$_4$O, c-C$_3$H$_2$, CH$_3$NH$_2$,
C$_2$H$_3$CN, CH$_3$Cl, and propanal.
Averaging the lines expected to be the strongest did not
yield any significant signal. $3-\sigma$ upper limits on production rates are
provided in Tables~\ref{tababund} or \ref{tabobssumztf} and for the species
not listed the upper limits on abundances are not expected to be better than
in previous comets such as 46P/Wirtanen \citep{Biv21a}.

\begin{table*}
\caption[]{Molecular abundances.}\label{tababund}
\begin{center}
\begin{tabular}{llllll}
\hline\hline
Molecule & Name &\multicolumn{2}{c}{Abundance relative to water in \%} \\  
         &      & C/2021~A1 (Leonard) & C/2022~E3 (ZTF) & all comets \\
\hline
HCN         & hydrogen cyanide    & $0.09\pm0.01$    & $0.09\pm0.01$     & 0.08--0.25 \\
HNC         & hydrogen isocyanide & $0.005\pm0.0005$ & $0.0015\pm0.0003$ & 0.0015-0.035  \\
HNC$_{\rm d}$\tablefootmark{a}  &  & $0.006\pm0.0007$ & $0.0032\pm0.0006$      \\
CH$_3$CN     & methyl cyanide     & $0.016\pm0.001$  & $0.017\pm0.001$  & 0.008-0.054 \\
HC$_3$N      & cyanoacetylene     & $0.004\pm0.001$  & $<0.0022$        & 0.002-0.068 \\
HNCO         & isocyanic acid     & $0.073\pm0.008$  & $0.042\pm0.004$  & $<$0.01-0.62 \\
NH$_2$CHO    & formamide          & $0.023\pm0.002$  & $0.019\pm0.001$  & 0.015-0.022 \\
\hline
CO           & carbon monoxide    & $<2.0$           & $0.7\pm0.3$      & 0.4- 35    \\ 
H$_2$CO      & formaldehyde       & $0.12\pm0.01$    & $0.12\pm0.01$    &  \\
H$_2$CO$_{\rm d}$\tablefootmark{b} && $0.26\pm0.01$   & $0.22\pm0.01$    & 0.13- 1.4 \\
CH$_3$OH     & methanol           & $0.88\pm0.02$    & $1.76\pm0.01$    & 0.7 - 6.1   \\
HCOOH        & formic acid        & $0.19\pm0.02$    & $0.19\pm0.01$    & $<$0.04--0.58  \\
CH$_3$CHO    & acetaldehyde       & $0.036\pm0.009$  & $0.070\pm0.004$  & 0.05--0.08  \\
(CH$_2$OH)$_2$ & ethylene-glycol  & $0.13\pm0.03$    & $0.13\pm0.02$    & 0.07--0.35  \\
C$_2$H$_5$OH & ethanol            & $0.18\pm0.04$    & $0.17\pm0.03$    & 0.11-0.19  \\
CH$_2$OHCHO  & glycolaldehyde     & $0.051\pm0.012$  & $<0.024$         & 0.016--0.039  \\
CH$_2$CO     & ketene             & $0.033\pm0.015$  & $<0.03$          & $\leq0.0078$  \\
HCOOCH$_3$   & methyl formate     & $<0.15$          & $<0.10$          & 0.06--0.08  \\
c-C$_2$H$_4$O & ethylene-oxide    & $<0.024$         & $<0.017$         & $<0.029$ \\
\hline
H$_2$S      & hydrogen sulphide   & $0.15\pm0.02$    & $0.18\pm0.04 $   & 0.09- 1.5   \\
CS          & carbon monosulphide & $0.06\pm0.01$    & $0.049\pm0.004$  & 0.05--0.20  \\
CS$_{\rm d}$\tablefootmark{c}   &  & $0.10\pm0.01$    & $0.075\pm0.003$  & 0.05--0.20  \\
OCS         & carbonyl sulphide   & $0.11\pm0.03$    & $0.068\pm0.016$  & 0.05--0.40  \\
SO          & sulphur monoxide    & $<0.04$          & $<0.03$          & 0.04--0.30  \\
SO$_2$      & sulphur dioxide     & $<0.04$          & $<0.03$          & 0.03--0.23  \\
H$_2$CS     & thioformaldehyde    & $<0.03$          & $<0.02$          & 0.009--0.090 \\
CH$_3$SH    & methyl mercaptan    & $<0.08$          & $<0.06$          & $<0.06$   \\
\hline
c-C$_3$H$_2$ & cyclopropenylidene  & $<0.024$         & $<0.011$         & $<0.009$      \\
l-C$_3$H$_2$ & propadienylidene    & $<0.004$         & $<0.002$         & $<0.0026$     \\
PH$_3$      & phosphine           & $<0.16$          & $<0.09$          & $<0.07$       \\
\hline
\end{tabular}
\end{center}
\tablefoot{Molecules with a "$_{\rm d}$" are modelled with a daughter distribution with the following parent scale length:\\
  \tablefoottext{a}{Assuming a daughter distribution with $L_p=1000$ and 2000 km \citep{Cor14,Cor17}.}\\
  \tablefoottext{b}{Assuming a daughter distribution with $L_p=1500$ and 1700 km (fitted values for C/2021~A1 and C/2022~E3).}\\
  \tablefoottext{c}{Assuming a daughter distribution with $L_p=4\times L(\rm CS_2)$.}\\
}
\end{table*}

\subsection{Isotopic ratios}\label{sect-isotopicratio}
Isotopic ratios of H, C, N, and S and upper limits on their abundances in some
molecules are provided in Table~\ref{tabisotop}.
Regarding HCN isotopologues, none are clearly detected but upper limits or
marginal signal at the 2--3-$\sigma$ level are compatible with values observed
in other comets \citep{Biv24}. Note that for $^{15}$N/$^{14}$N, the Earth value
of 272 corresponds already to an enrichment compared to the estimated solar
value of 450\citep{Mar11}, showing that fractionation was important already
for Earth nitrogen but even more for the material that was incorporated
into cometary ices. 

The $^{32}$S/$^{34}$S could be measured in CS in both comets,
but in comet C/2022~E3 it seems lower than the Earth value while
compatible with terrestrial value for H$_2$S: this comet seems enriched in
C$^{34}$S, at the 3-$\sigma$ level. However the PSW
observing mode used for comet C/2022~E3 is more prone to produce ripples that
affect the signal of marginal lines.
We also provide upper limits on abundances of deuterated species. None yields
a D/H below the Earth VSMOW (Vienna Standard Mean Ocean Water) value, but as
some enrichment has been observed in the interstellar medium and in some
molecules ejected by comet 67P/Churyumov-Gerasimenko observed by the ROSINA
mass spectrometer on board
the Rosetta spacecraft \citep{Dro21,Mul22}, we provide those upper limits.

\begin{table*}
\caption[]{Isotopic ratios.}\label{tabisotop}
\begin{center}
\begin{tabular}{llcccc}
\hline\hline
Ratio & Molecule & C/2021~A1 & C/2022~E3\tablefootmark{a} & other comets\tablefootmark{b} & Earth\tablefootmark{c} \\  
\hline
$^{12}$C/$^{13}$C & HCN    & $>47$        & $135\pm75$   & 88--114   & 89.4 \\
$^{14}$N/$^{15}$N & HCN    & $131\pm81$   & $223\pm138$  & 139--205  & 272  \\
$^{32}$S/$^{34}$S & CS     & $17.8\pm5.6$ & $11.2\pm3.7$ & 16--23    & 22.7 \\
                 & H$_2$S & $>3.4$       & $16.5\pm7.3$ & 16--23    & 22.7 \\
D/H              & H$_2$O &              & $<6.4\times10^{-4}$ & $[1.4-6.4]\times10^{-4}$ & $1.6\times10^{-4}$\tablefootmark{d} \\
                 & HCN    &              & $<1.8$\%     & 0.23\%    &      \\
                 & H$_2$CO &             & $<3.1$\%     & $<0.7$\%  &      \\
                 & CH$_2$DOH &           & $<1.7$\%     & $<0.35$\% &      \\
                 & CH$_3$OD  &           & $<2.5$\%     & $<1.0$\%  &      \\
\hline
\end{tabular}
\end{center}
\tablefoot{
  \tablefoottext{a}{The lines observed in C/2022 E3 are only present at the 1.5--3$\sigma$ level.}\\
  \tablefoottext{b}{\citet{Boc15,Biv16,Cor19}}\\
  \tablefoottext{c}{\citet{Sim06}}\\
  \tablefoottext{d}{VSMOW value}\\
}
\end{table*}

\section{The abundance of HNCO and HCOOH in comets}\label{sect-hncohcooh}

The molecules HNCO and HCOOH were first detected in comets C/1996~B2 (Hyakutake)
and C/1995~O1 (Hale-Bopp) in 1996-1997 \citep{Lis97,Boc00}.
They have been detected in several comets since \citep{Biv23} but there were 
large uncertainties on their destruction rates and
lifetimes and a simple excitation model was used to derive their abundances.
\citet{Hro23} have recently published photo-dissociation and photo-ionisation
rates for HNCO and HCOOH that significantly differ from those assumed in the
past \citep[][ and references therein]{Biv21a}. The new values
($\beta_0(\rm HNCO)= 38\times10^{-5}$~s$^{-1}$,
$\beta_0(\rm HCOOH)= 54\times10^{-5}$~s$^{-1}$, in the solar radiation field at 1~au)
are 5 to 10 times higher than previously assumed, with a stated 2-$\sigma$
uncertainty on the order of 20 to 30\% \citep{Hro23}.
We have decided to revisit all millimetre observations of comets reporting
a detection of at least one of these molecules and recomputed production rates
and abundances relative to water assuming values of:
\begin{itemize}
  \item $\beta_0(\rm HNCO)= 3.5\times10^{-4}$s$^{-1}$;
  \item $\beta_0(\rm HCOOH)= 5.0\times10^{-4}$s$^{-1}$;
\end{itemize}
The contribution of Solar Lyman-$\alpha$ to their photo-dissociation is minor
and their photo-dissociation cross-section extends well into the 130-200~nm
range where the solar flux is less affected by solar activity\citep{Hue92}.
Therefore,
we do not expect a large variation with solar activity.
An overview of the observations analysed in this work, partly published in
previous papers is given in Table~\ref{tabobs-hnco-hcooh}.

Table~\ref{hnco-hcooh} provides the abundances relative to water of
HNCO and HCOOH in these comets.
In these calculations infrared pumping via the rotational bands is not taken
into account, but with those reduced lifetimes, the photo-dissociation of the
molecules takes place in a region where collisional excitation
dominates (within $\sim$2000~km from the nucleus) and the photo-dissociation
process is as efficient as vibrational excitation (g-factors comparable
to $\beta$). So the new reduced
lifetimes (compared to previous calculations) should also reduce the
uncertainties from the neglected infrared pumping.

\begin{table*}
\renewcommand{\tabcolsep}{0.08cm}
\caption[]{Observations of HNCO and HCOOH in comets.}\label{tabobs-hnco-hcooh}
\begin{center}
\begin{tabular}{llclclclc}
\hline\hline\noalign{\smallskip}
Comet                & \multicolumn{4}{c}{HNCO} & \multicolumn{4}{c}{HCOOH} \\
  & (yyyy/mm/dd-dd) & $r_h$ (au) & telescopes\tablefootmark{a} & lines\tablefootmark{b} & (yyyy/mm/dd-dd) & $r_h$ (au) & telescopes\tablefootmark{a} & lines\tablefootmark{b} \\
\hline\noalign{\smallskip}
C/1996~B2           & 1996/03--04  & 1.10--0.65 & CSO, & 1(+2) & 1996/04/11 & 0.65 & IRAM & (2) \\
\hspace{1cm} (Hyakutake) &         &        & PdB,IRAM &       &            &      &      &   \\
C/1995~O1           & 1997/02--05  & 0.92--1.28 & CSO, & 4(+1) & 1997/02--05 & 0.92-1.28 & CSO, & 7(+1) \\
\hspace{1cm} (Hale-Bopp) &         &        & PdB,IRAM &       &            &      &  PdB,IRAM  &   \\
153P/Ikeya-Zhang    & 2002/05/12   & 1.255 & IRAM   & 1     & 2002/05/10--12 & 1.23 & IRAM  & (2) \\ 
C/2004~Q2 (Machholz)& 2005/01/16--18 & 1.21 & IRAM   & 1     & 2005/01/15--18 & 1.21 & IRAM  & (3) \\
73P-C/S.-W.3& 2006/05/09--18& 1.00 & CSO, IRAM & 1(+1) & 2006/05/13--18 & 1.00  & IRAM    & (3)  \\
73P-B/S.-W.3& 2006/05/09--18& 1.00 & CSO, IRAM & 2     & 2006/05/13--18 & 1.00  & IRAM    & (3)  \\
8P/Tuttle           & 2007/12/29--31 & 1.10 & IRAM   & 1     & 2007/12/29--31 & 1.11 & IRAM  & 1+(2) \\
C/2007 N3 (Lulin)   & 2009/02/26--27 & 1.41 & IRAM   & 1     & 2009/02/25--27 & 1.41 & IRAM  & (3)  \\
103P/Hartley 2      & 2010/10/25--29 & 1.06 & IRAM, CSO & 1(+3) & 2010/10/25--39 & 1.06  & IRAM, CSO & (9)\\
C/2009~P1 (Garradd) & 2012/02/15--19 & 1.73 & IRAM   & 1(+6) & 2012/02/15--19 & 1.73 & IRAM  & (10) \\
C/2013~R1 (Lovejoy) & 2013/11/08--16 & 1.13 & IRAM   & (5)   & 2013/11/08--16 & 1.13 & IRAM  & (21)  \\
                    & 2013/11/27--31 & 0.92 & IRAM   & 1(+5) & 2013/11/27--31 & 0.92 & IRAM  & 7(+13)   \\
                    & 2013/12/09--16 & 0.83 & IRAM   & 1(+9) & 2013/12/09--16 & 0.82 & IRAM  & 2(+24)   \\
C/2012~S1 (ISON)    & 2013/11/14--15 & 0.62 & IRAM   &  1    & 2013/11/14--15 & 0.62 & IRAM  & 2(+6)   \\
C/2012~F6 (Lemmon)  & 2013/03/14--18 & 0.75 & IRAM   & (6)   & 2013/03/14--39 & 0.75 & IRAM  & (16) \\
                    & 2013/04/06--08 & 0.75 & IRAM   & 1(+8) & 2013/04/06--08 & 0.78 & IRAM  & (22) \\
C/2011~L4           & 2013/03/14--18 & 0.36 & IRAM   & 2(+4) & 2013/03/14--18 & 0.36 & IRAM  & 1(+21)  \\
\hspace{1cm}(PanSTARRS)&2013/04/05--08 & 0.81 & IRAM   & (9)   & 2013/04/05--08 & 0.81 & IRAM  & (28)  \\
C/2014~Q2 (Lovejoy) & 2015/01/13--16 & 1.31 & IRAM   & 4(+2) & 2015/01/13--16 & 1.31 & IRAM  & 6+(14) \\
                    & 2015/01/23--26 & 1.30 & IRAM   & 4(+1) & 2015/01/23--26 & 1.29 & IRAM  & 3(+9) \\
C/2013~US$_{10}$ (Catalina)
                    & 2015/12/30--34 & 1.18 & IRAM   & (6)   & 2015/12/30--34 & 1.18 & IRAM  & (10)    \\
46P/Wirtanen        & 2018/12/11--18 & 1.06 & IRAM   & (10)  & 2018/12/11--18 & 1.06 & IRAM  & (17) \\
67P/Churyumov-G.    & 2021/11/12--16 & 1.22 & IRAM   & (10)  & 2021/11/12--16 & 1.22 & IRAM  & (14)  \\
C/2021~A1 (Leonard) & 2021/12/08--13 & 0.78 & IRAM   & 6(+6) & 2021/12/08--13 & 0.78 & IRAM  & 2(+26)   \\
C/2022~E3 (ZTF)     & 2023/02/03--07 & 1.18 & IRAM   & 3(+9) & 2023/02/03--07 & 1.18 & IRAM  & 13(+17) \\
\hline
\end{tabular}
\end{center}
\tablefoot{
  \tablefoottext{a}{Radio telescopes used: IRAM 30-m,
    CSO = Caltech Submillimeter Observatory 10.4-m,
    PdB = Plateau de Bure interferometer in single-dish ON-OFF mode.}\\
  \tablefoottext{b}{The number of lines in parenthesis have a S/N < 3, 
    but are used to infer abundances.}\\
}
\end{table*}

\begin{table*}
\caption[]{Revised abundances of HNCO and HCOOH in comets.}\label{hnco-hcooh}
\begin{center}
\begin{tabular}{lllcc}
\hline\hline\noalign{\smallskip}
Comet   & Dates & $r_h$ range  & $Q_{\rm HNCO}/Q_{\rm H_2O}$ & $Q_{\rm HCOOH}/Q_{\rm H_2O}$ \\
        & (yyyy/mm/dd-dd) & (au) & & \\
\hline\noalign{\smallskip}
C/1996~B2 (Hyakutake)& 1996/03--04   & 1.10--0.65 & $0.11\pm0.02$\%   & $<1.4$\%  \\
C/1995~O1 (Hale-Bopp) & 1997/02--05  & 0.92--1.28 & $0.22\pm0.01$\%   & $0.44\pm0.03$\% \\
153P/Ikeya-Zhang    & 2002/05/08--12 & 1.25       & $0.049\pm0.008$\% & $ < 0.12$\%   \\ 
C/2004~Q2 (Machholz)& 2005/01/13--17 & 1.21       & $0.025\pm0.003$\% & $0.09\pm0.02$\% \\
73P-C/Schwassmann-W.3& 2006/05/10--18& 1.00       & $0.066\pm0.014$\% & $ < 0.18$\%   \\
73P-B/Schwassmann-W.3& 2006/05/10--18& 1.00       & $0.075\pm0.011$\% & $ < 0.15$\%   \\
8P/Tuttle           & 2007/12/29--31 & 1.11       & $<0.06$\%         & $0.42\pm0.11$\% \\
C/2007 N3 (Lulin)   & 2009/02/26--27 & 1.41       & $0.025\pm0.009$\% & $ < 0.13$\%   \\
103P/Hartley 2      & 2010/10/25--29 & 1.06       & $0.034\pm0.012$\% & $ < 0.17$\% \\
C/2009~P1 (Garradd) & 2012/02/15--19 & 1.73       & $0.031\pm0.010$\% & $ < 0.14$\% \\
C/2013~R1 (Lovejoy) & 2013/11/08--46 & 0.83--1.13 & $0.050\pm0.007$\% & $0.17\pm0.03$\% \\
C/2012~S1 (ISON)    & 2013/11/14--15 & 0.61--0.64 & $0.21\pm0.06$\%   & $0.58\pm0.09$\% \\
C/2012~F6 (Lemmon)  & 2013/03/14--39 & 0.75--0.78 & $0.10\pm0.02$\%   & $0.25\pm0.06$\% \\
C/2011~L4 (PanSTARRS) &2013/03/14--39& 0.33--0.83 & $0.62\pm0.12$\%   & $\leq0.93$\%  \\
C/2014~Q2 (Lovejoy) & 2015/01/13--26 & 1.30       & $0.016\pm0.001$\% & $0.055\pm0.004$\% \\
C/2013~US$_{10}$ (Catalina)
                    & 2015/12/30--34 & 1.18       & $0.051\pm0.019$\% & $ < 0.22$\%  \\
46P/Wirtanen        & 2018/12/11--18 & 1.06       & $<0.011$\%        & $ <0.043$\% \\
67P/Churyumov-G.    & 2021/11/12--16 & 1.22       & $0.085\pm0.024$\% & $ < 0.33$\% \\
C/2021~A1 (Leonard) & 2021/12/08--13 & 0.76--0.81 & $0.073\pm0.008$\% & $0.19\pm0.02$\% \\
C/2022~E3 (ZTF)     & 2023/02/03--07 & 1.18       & $0.042\pm0.004$\% & $0.19\pm0.02$\% \\
\hline
\end{tabular}
\end{center}
\end{table*}

\begin{figure}[]
\centering
\resizebox{\hsize}{!}{
  \includegraphics[angle=270,width=0.9\textwidth]{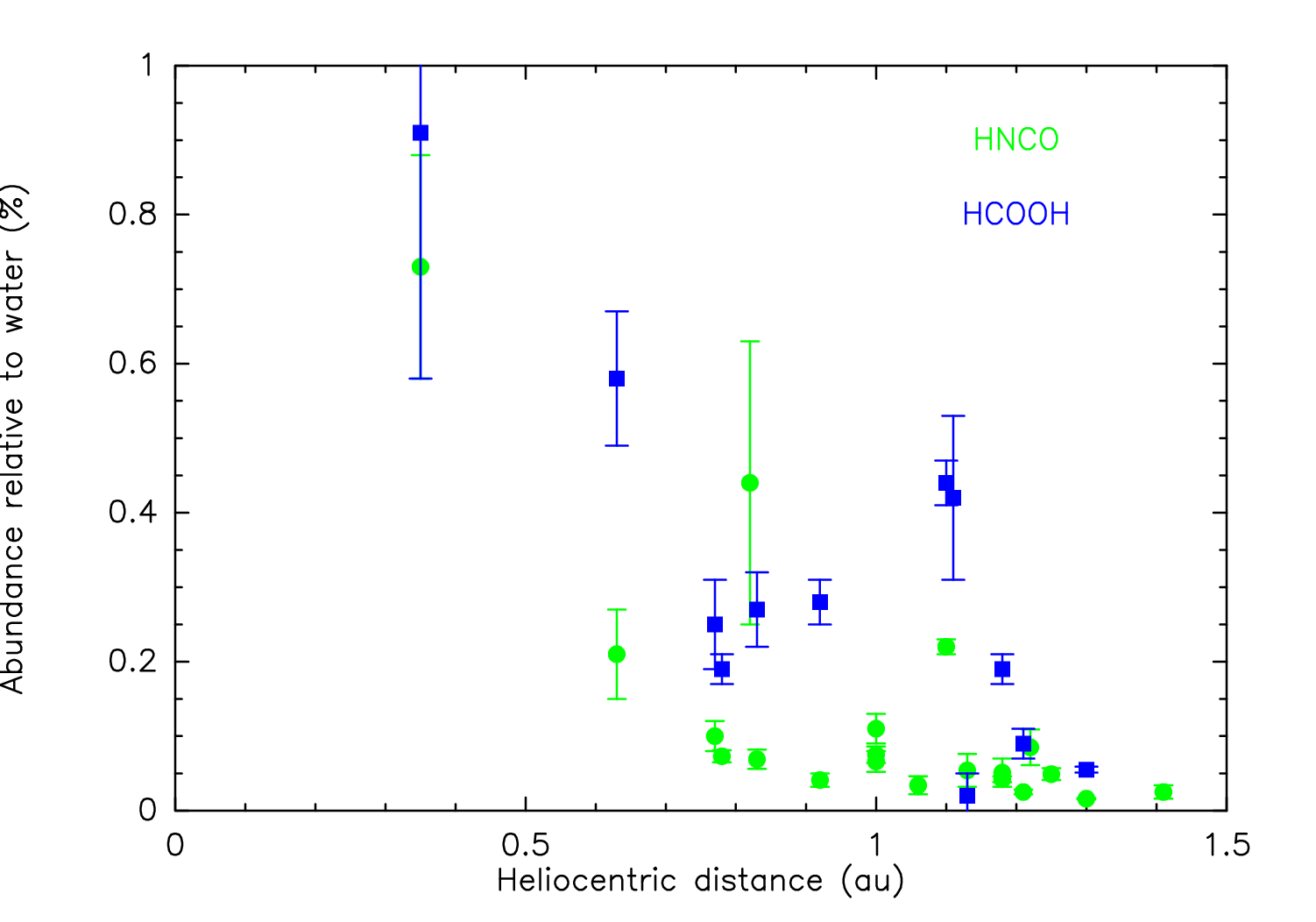}}
\caption{Abundances relative to water of HNCO (green) and HCOOH (blue) in
  comets as a function of heliocentric distance. Albeit dispersion between
  comets, the trend appears to be that these species are
  more abundant in the coma close to the Sun.}
\label{fighncohcoohrh}
\end{figure}

{\bf HNCO} is relatively well detected in the two comets
via its $K_a=0$ lines at 2~mm and
1~mm wavelengths, and marginal signals at an offset of 10\arcsec~
(Tables~\ref{tabqdistrileonard},~\ref{tabqdistriztf})
suggest that it may be produced by a distributed
source with a scale length $L_P$(HNCO) larger than 2000~km.
Indeed, assuming a parent scale length of 5000~km also provides a better
agreement between the observed rotational temperature and the predicted one
for $T=60$~K (Table~\ref{tabtemp}).

{\bf HCOOH:} In Tables~\ref{tabqdistrileonard} and \ref{tabqdistriztf}
we have selected the lower energy lines that are expected to
be the strongest, that is those with $K_a=0$ to 2. For comet C/2021~A1,
since at the offset position the weighted average intensity is similar
to the one at the central position, we do not find a solution for a
reasonably small parent scale length, but data are compatible within
$1\sigma$ with $L_p$(HCOOH) = 2000~km, and
$L_p$(HCOOH) = 0~km cannot be fully excluded. Similarly, for comet
C/2022~E3, we derive $L_p$(HCOOH) = 0--5200~km in the $\pm1\sigma$ interval,
so that distributed source as well as nuclear source are not excluded.
Hence, if HCOOH is produced by the sublimation and dissociation of ammonium
formate, then this must take place within a few thousands of km from the
nucleus at 0.78--1.18 au from the Sun.

The revised measured abundances in comets (increased by a factor 1.2 to 5)
of these two molecules are
plotted in Fig.~\ref{fighncohcoohrh} as a function of heliocentric distance
at which they were detected. This plots suggest that their abundance
relative to water increases for comets observed closer to the Sun. This needs
to be confirmed, but can suggest that the molecules could be at least in part
produced by the degradation of parents in the coma such as ammonium salts.
Their combined abundance relative to water in comets ($\sim0.1-1$\%) is comparable
to that of NH$_3$\citep[0.2-5\%,][]{Biv24}, the other decomposition
product of ammonium salts. Interestingly, a similar trend with heliocentric distance
has been observed for the abundance of ammonia \citep[see, e.g.][]{Del16}.

\section{Conclusions}
\label{sect-discussion}

We have undertaken a comprehensive and sensitive spectroscopic survey
at 2~mm and 1~mm of two long period comets, C/2021~A1 (Leonard) and
C/2022~E3 (ZTF) when they were relatively close to the Earth (0.2--0.3 au).
We collected in addition some useful spatial information from coarse
mapping, to probe the spatial distribution of selected molecules.
The main results can be summarised as follows.

\begin{itemize}

\item Both comets exhibited relatively stable (within $\pm$20\%) production
  rates during the observational periods, although C/2021~A1 underwent a series of
  disintegration outbursts just after our IRAM-30m observing run.

\item We determined the abundance relative to water or a significant upper
  limit for 26 molecules. Comets C/2021~A1 and C/2022~E3 share
  similar compositions: they are depleted in hypervolatiles (low CO and
  H$_2$S abundances relative to water) and have a relatively low methanol
  abundance relative to water (0.9 and 1.8\% respectively) compared to the mean of other comets.

\item Constraints on the presence of distributed sources have been obtained.
  A slightly distributed source (scale length smaller than 1/3 beam) for HCN
  fits better the data than direct release from the nucleus.
  The parent source of formaldehyde was found to have a scale length
  $L_p(\rm H_2CO) = 0.68$ and $0.36\pm0.11$ times the photo-dissociation length
  of formaldehyde ($L_d(\rm H_2CO)$) in C/2021~A1 and C/2022~E3 respectively.
  We measured $L_p(\rm CS)\approx4\times L_d(\rm CS_2)$, 
  suggesting that the dissociation rate of the parent producing CS is
  $\beta_0=4.5\times10^{-4}$~s$^{-1}$ at 1 au from the Sun. This is comparable to
  the photo-dissociation rate of H$_2$CS, but the
  H$_2$CS upper limit is more than 3 times lower than the CS abundance.

\item Both comets are relatively depleted in sulphur-bearing species compared to
  other comets, with H$_2$S/H$_2$O in the low range (0.15--0.18\%) and the sum
  of sulphur-bearing molecule abundances below $\sim$0.5\%. SO and SO$_2$ are not
  detected with abundances below the lowest measured in a comet and OCS is
  detected in both comets but with a low abundance, too.

\item We obtained only shallow constraints on isotopic ratios:
  $^{12}$C/$^{13}$C, $^{14}$N/$^{15}$N, and $^{32}$S/$^{34}$S are compatible
  with values observed in other comets (90, 150 and 23, respectively), although
  the C$^{32}$S/C$^{34}$S seems marginally twice lower than the terrestrial value
  in comet C/2022~E3.
  
\item HNCO and HCOOH are well detected in both comets. Because of recently
  published lifetimes for these molecules that are significantly shorter than
  previous estimates, we have revised
  their inferred abundances in all comets for which detections of these
  species have been reported. The values found in C/2021~A1 (0.07\% and 0.19\%)
  and C/2022~E3 (0.04\% and 0.19\%) are close to their median abundances
  relative to water in other comets, 0.05\% and 0.25\%, respectively.
  Their total abundance is also about half the median abundance of NH$_3$
  in comets. We cannot exclude that these species are produced in the coma by a
  distributed source, and their abundance in the comae
  seems to increase at lower heliocentric distances.
  This favours the hypothesis that HNCO, HCOOH and consequently an important
  fraction of NH$_3$ could come from the decomposition of ammonium salts. 
\end{itemize}

These observations have provided complementary and new insights regarding
the source of several cometary molecules that are unlikely present as such
in cometary ices.
Further investigations at higher spatial resolution (e.g. interferometric
investigations undertaken on those comets under reduction or on future comets),
will be very useful to pin down the production processes and their scale lengths
of these molecules (CS, H$_2$CO, HNCO, HCOOH).


\begin{acknowledgements}
This work is based on observations carried out under projects number
001-21, 100-21 and 097-22 with the IRAM 30-m telescope.
IRAM is supported by the Institut Nationnal des Sciences de l'Univers
(INSU) of the French Centre national de la recherche scientifique (CNRS),
the Max-Planck-Gesellschaft (MPG, Germany) and the Spanish IGN
(Instituto Geográfico Nacional).
We gratefully acknowledge the support from the IRAM staff for its
support during the observations.
The data were reduced and analyzed thanks to the use of the GILDAS,
class software (http://www.iram.fr/IRAMFR/GILDAS).
This research has been supported by the Programme national de 
plan\'etologie de l'Institut des sciences de l'univers (INSU).
The Nan\c{c}ay radio telescope (NRT) is part of the Observatoire
radioastronomique de Nan\c{c}ay, operated by the Paris Observatory,
associated with the French Centre national de la recherche
scientifique (CNRS) and with the University of Orl\'eans.
Part of this research was carried out at the Jet Propulsion Laboratory,
California Institute of Technology, under a contract with the National
Aeronautics and Space Administration (80NM0018D0004).
D.C.L. acknowledges financial support from the National Aeronautics and
Space Administration (NASA) Astrophysics Data Analysis Program (ADAP).
M.~N. Drozdovskaya acknowledges the Holcim Foundation Stipend.
B.~P. Bonev and N. Dello Russo
acknowledge support of NSF grant AST-2009398 and NASA grant 80NSSC17K0705,
respectively.
N.~X. Roth was supported by the NASA Postdoctoral Program at the NASA Goddard
Space Flight Center, administered by Universities Space Research Association
under contract with NASA. M.A. Cordiner was supported in part by the National
Science Foundation (under Grant No. AST-1614471).
S.~N. Milam and M.~A. Cordiner acknowledge the
Planetary Science Division Internal Scientist Funding
Program through the Fundamental Laboratory Research (FLaRe) work package,
as well as the NASA Astrobiology Institute through the Goddard Center for
Astrobiology (proposal 13-13NAI7-0032).
\end{acknowledgements}


\clearpage
\newpage
\begin{appendix}
\FloatBarrier
  \section{Rotational diagrams}
 The logarithm of a quantity proportional to the line intensity (line area
  divided by column density) is plotted
  against the energy of the upper level for each observed rotational
  transition. The upper levels of the transition are given at the top.
  The inverse of the slope of a fitted line provides the
  rotational temperature (given in pink) of the group of lines.
  
\begin{figure}[h]
\centering
\resizebox{\hsize}{!}{
  \includegraphics[angle=270,width=0.9\textwidth]{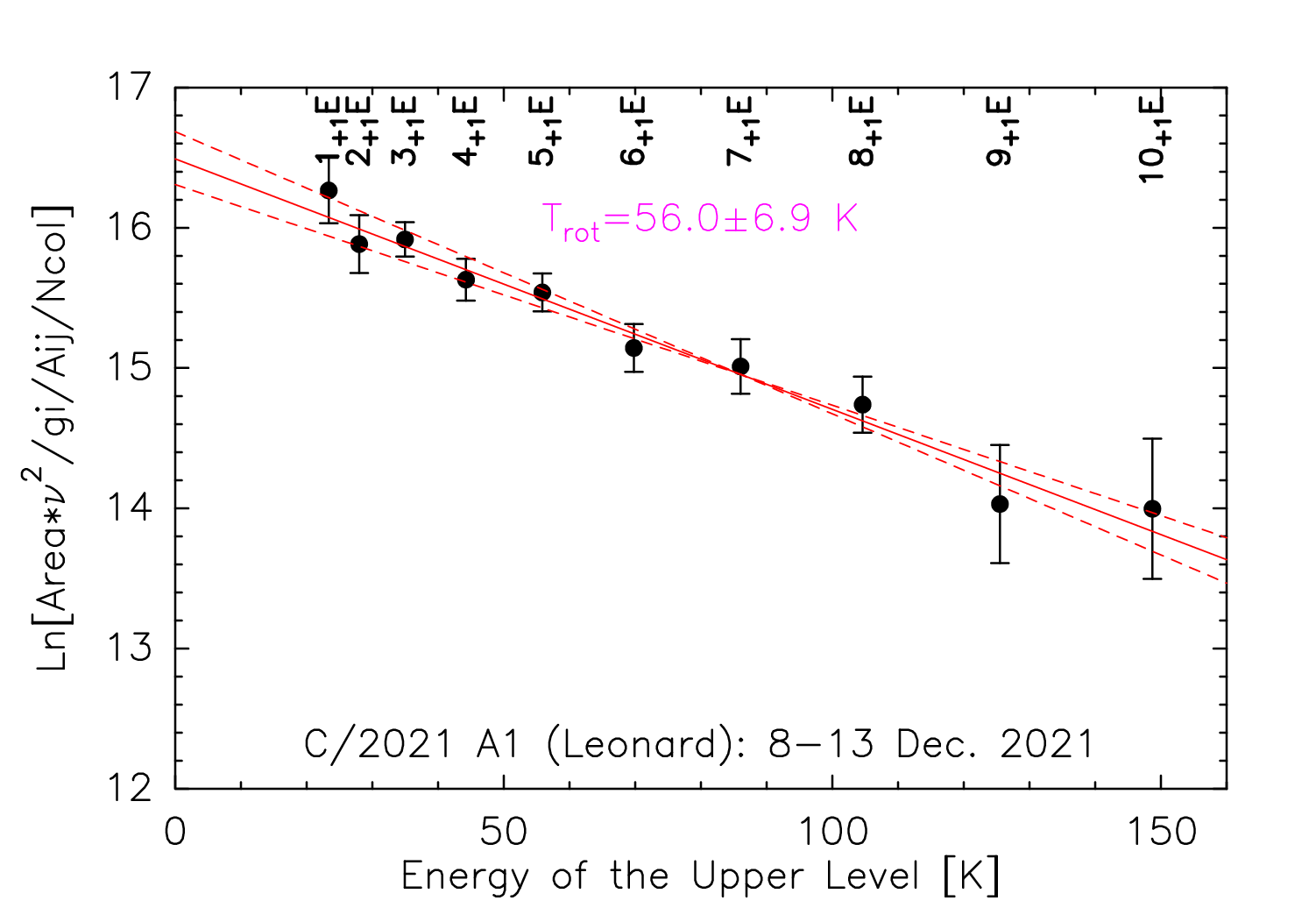}}
\caption{Rotational diagram of the CH$_3$OH lines around 166~GHz observed
  between 8.6 and 13.7 December 2021 in comet C/2021~A1 (Leonard).}
\label{diagrot2021a1-met166}
\end{figure}

\begin{figure}[h]
\centering
\resizebox{\hsize}{!}{
  \includegraphics[angle=270,width=0.9\textwidth]{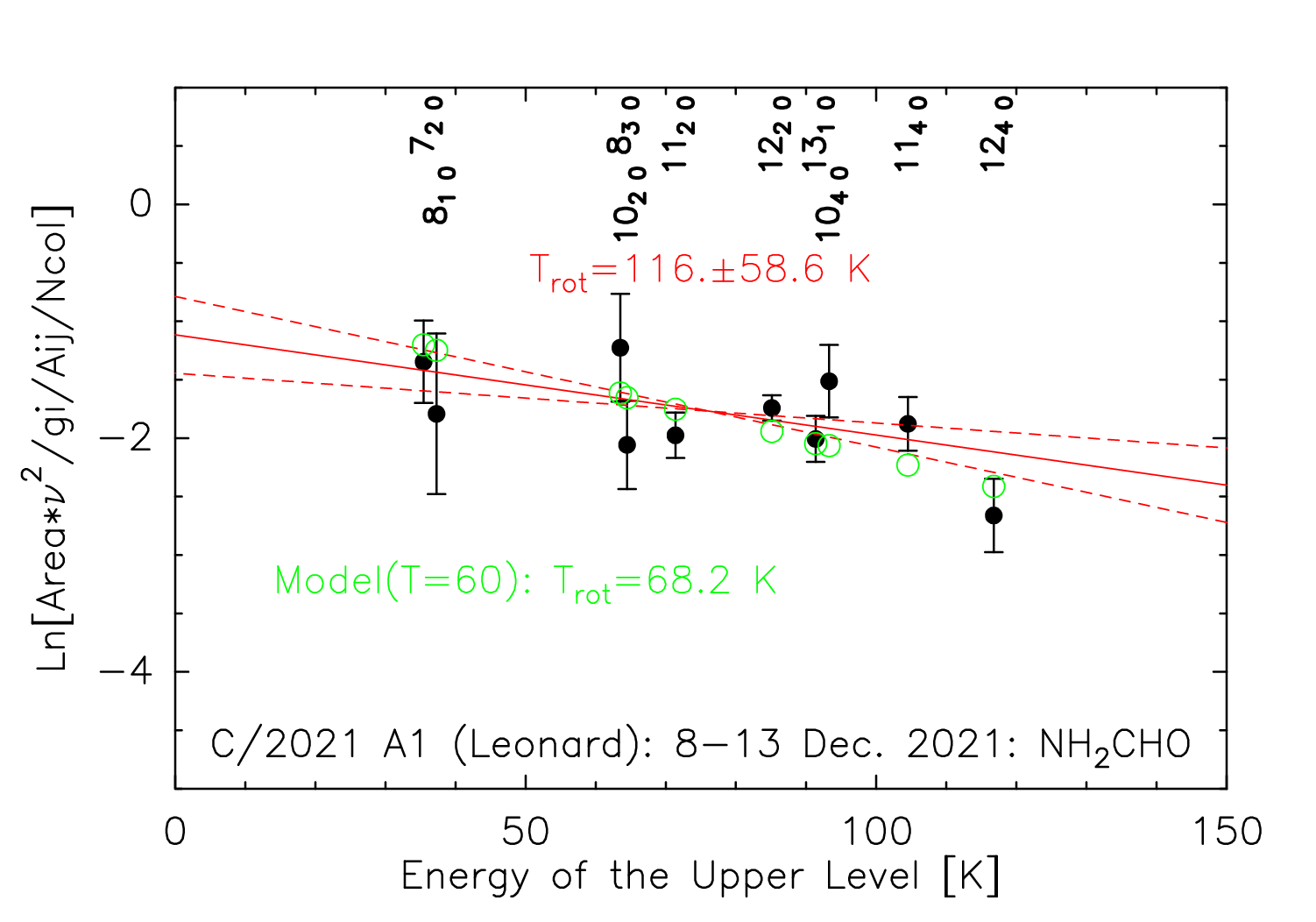}}
\caption{Rotational diagram of all the NH$_2$CHO lines observed
  between 8.6 and 13.7 December 2021 in comet C/2021~A1 (Leonard).
  Lines with same $J$ number and $K_a$= 0 to 2 or 3 to 4 are averaged together.
  Scales as in Fig.~\ref{diagrot2021a1-met166}).
  The values in green are those expected from modelling with $T=60$ K.}
\label{diagrot2021a1-nh2cho}
\end{figure}

\begin{figure}[h]
\centering
\resizebox{\hsize}{!}{
  \includegraphics[angle=270,width=0.9\textwidth]{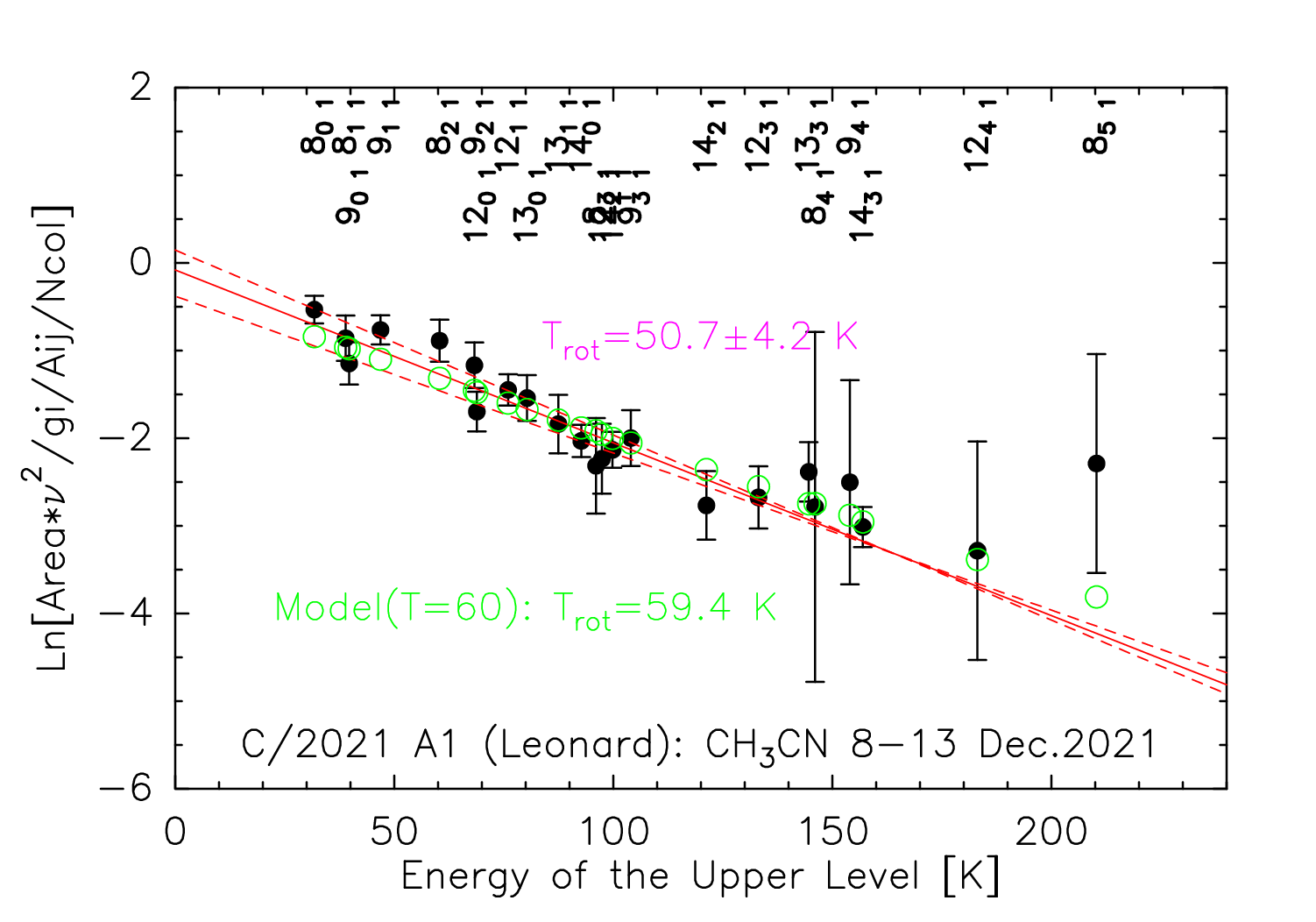}}
\caption{Rotational diagram of the CH$_3$CN lines at 147, 165, 220, 239 and
  257~GHz observed between 8.4 and 13.7 December 2021 in comet C/2021~A1
  (Leonard).
  Scales as in Fig.~\ref{diagrot2021a1-met166}).
  The values in green are those expected from modelling with $T=60$ K.}
\label{diagrot2021a1-ch3cn}
\end{figure}

\begin{figure}[]
\centering
\resizebox{\hsize}{!}{
  \includegraphics[angle=270,width=0.9\textwidth]{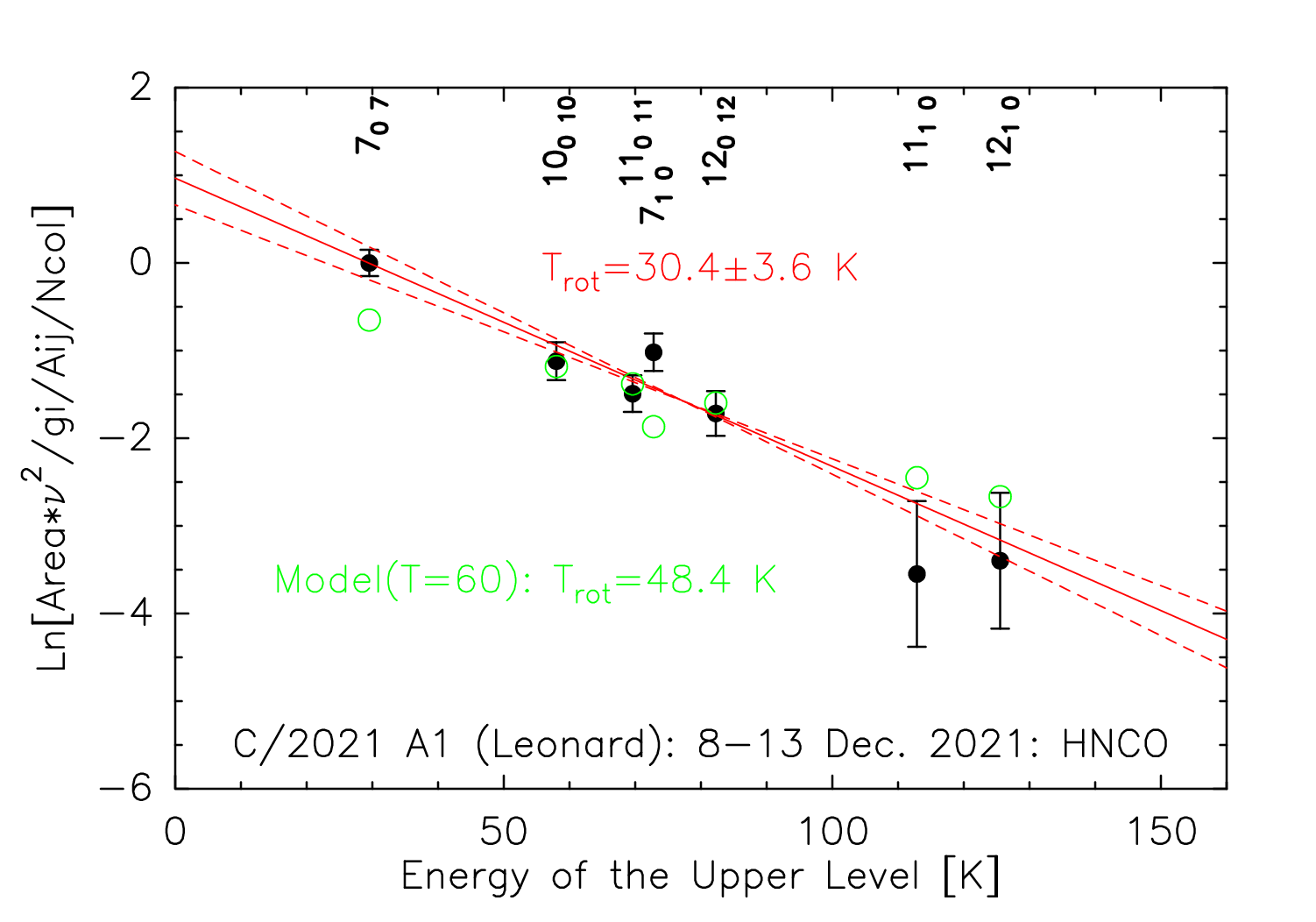}}
\caption{Rotational diagram of the HNCO lines at 1~mm and 2~mm
  observed between 8.4 and 13.7 December 2021 in comet C/2021~A1 (Leonard).
  Scales as in Fig.~\ref{diagrot2021a1-met166}.}
\label{diagrot2021a1-hnco}
\end{figure}

\begin{figure}[]
\centering
\resizebox{\hsize}{!}{
  \includegraphics[angle=270,width=0.9\textwidth]{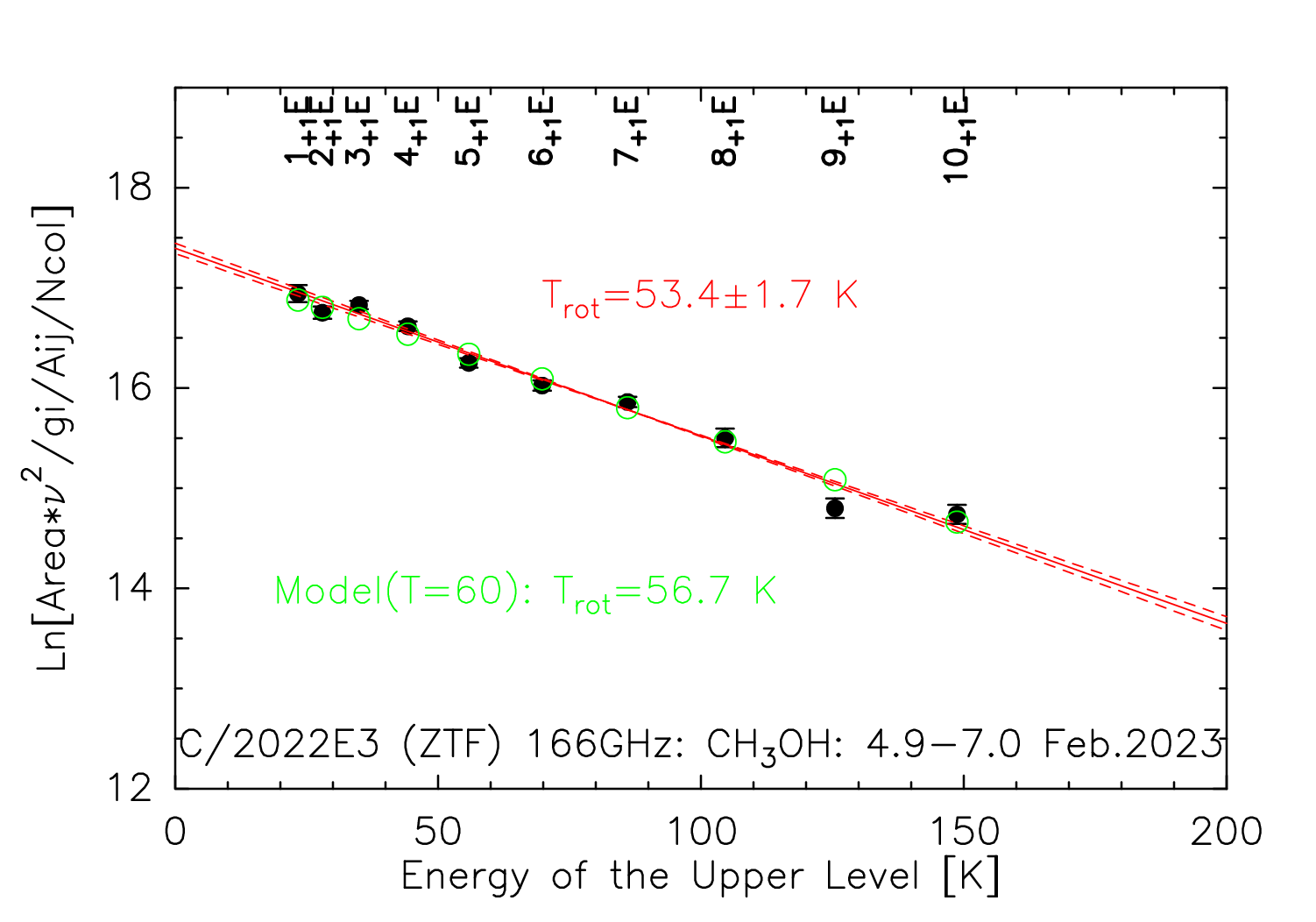}}
\caption{Rotational diagram of the CH$_3$OH lines around 166~GHz observed
  between 4.9 and 7.0 February 2023 in comet C/2022~E3 (ZTF).
  The values in green are those expected from modelling with $T=60$ K.}
\label{diagrotztf-met166}
\end{figure}

\begin{figure}[]
\centering
\resizebox{\hsize}{!}{
  \includegraphics[angle=270,width=0.9\textwidth]{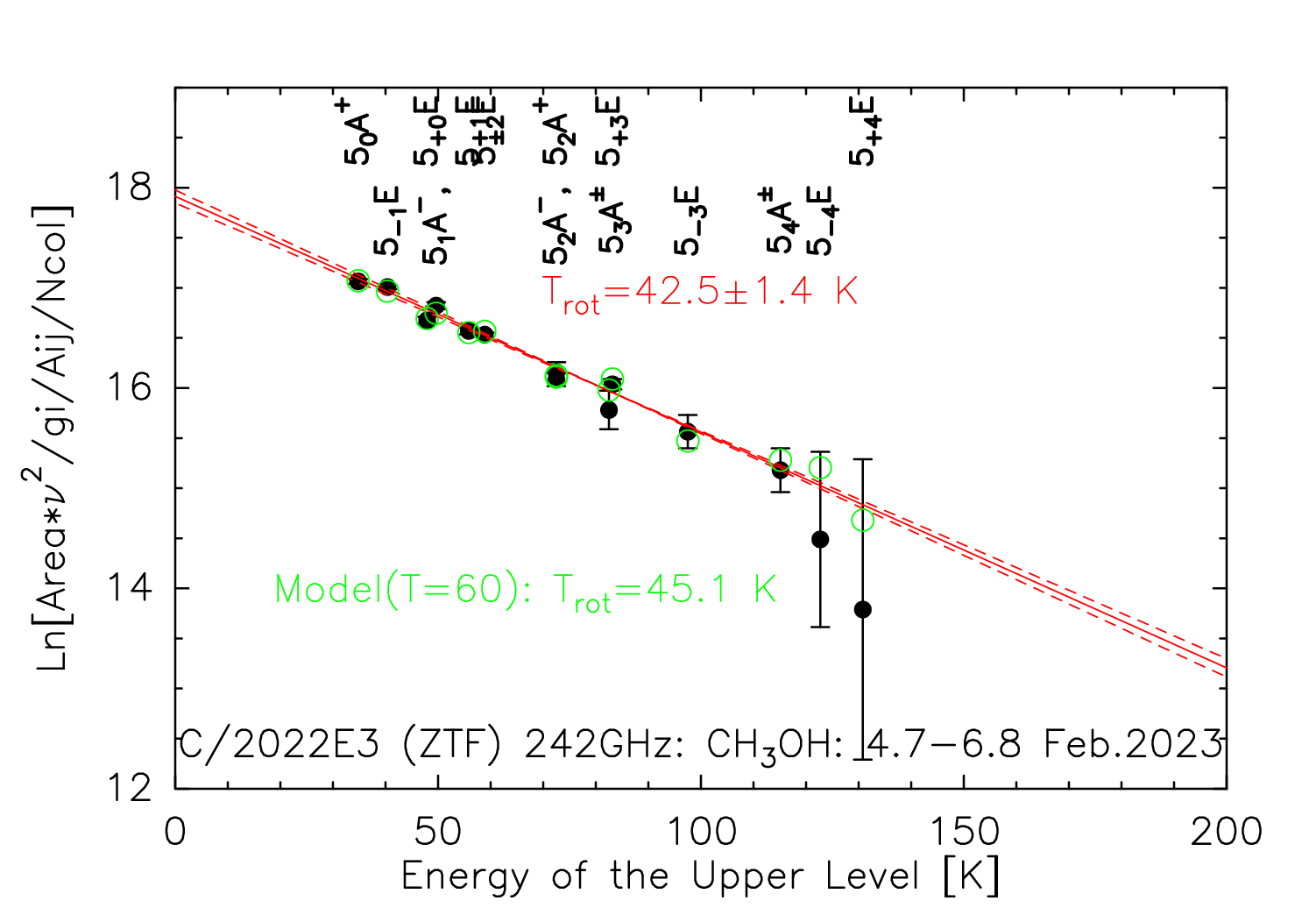}}
\caption{Rotational diagram of the CH$_3$OH lines at 242~GHz observed
  between 4.7 and 6.8 February 2023 in comet C/2022~E3 (ZTF).
  Scales as in Fig.~\ref{diagrotztf-met166}.}
\label{diagrotztf-met242}
\end{figure}

\begin{figure}[]
\centering
\resizebox{\hsize}{!}{
  \includegraphics[angle=270,width=0.9\textwidth]{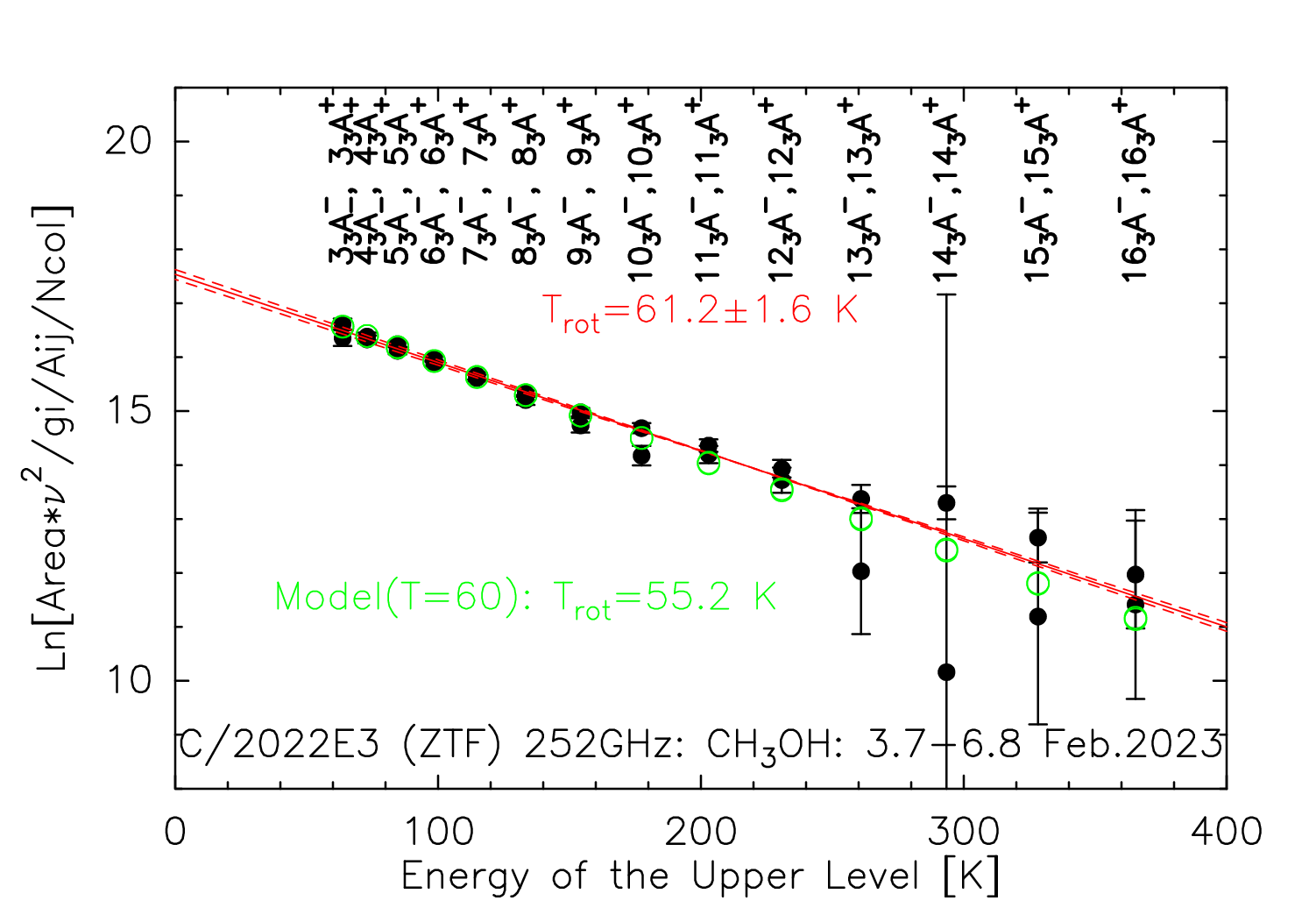}}
\caption{Rotational diagram of the CH$_3$OH lines around 252~GHz observed
  between 3.7 and 6.7 February 2023 in comet C/2022~E3 (ZTF).
  Scales as in Fig.~\ref{diagrotztf-met166}.}
\label{diagrotztf-met252}
\end{figure}

\begin{figure}[]
\centering
\resizebox{\hsize}{!}{
  \includegraphics[angle=270,width=0.9\textwidth]{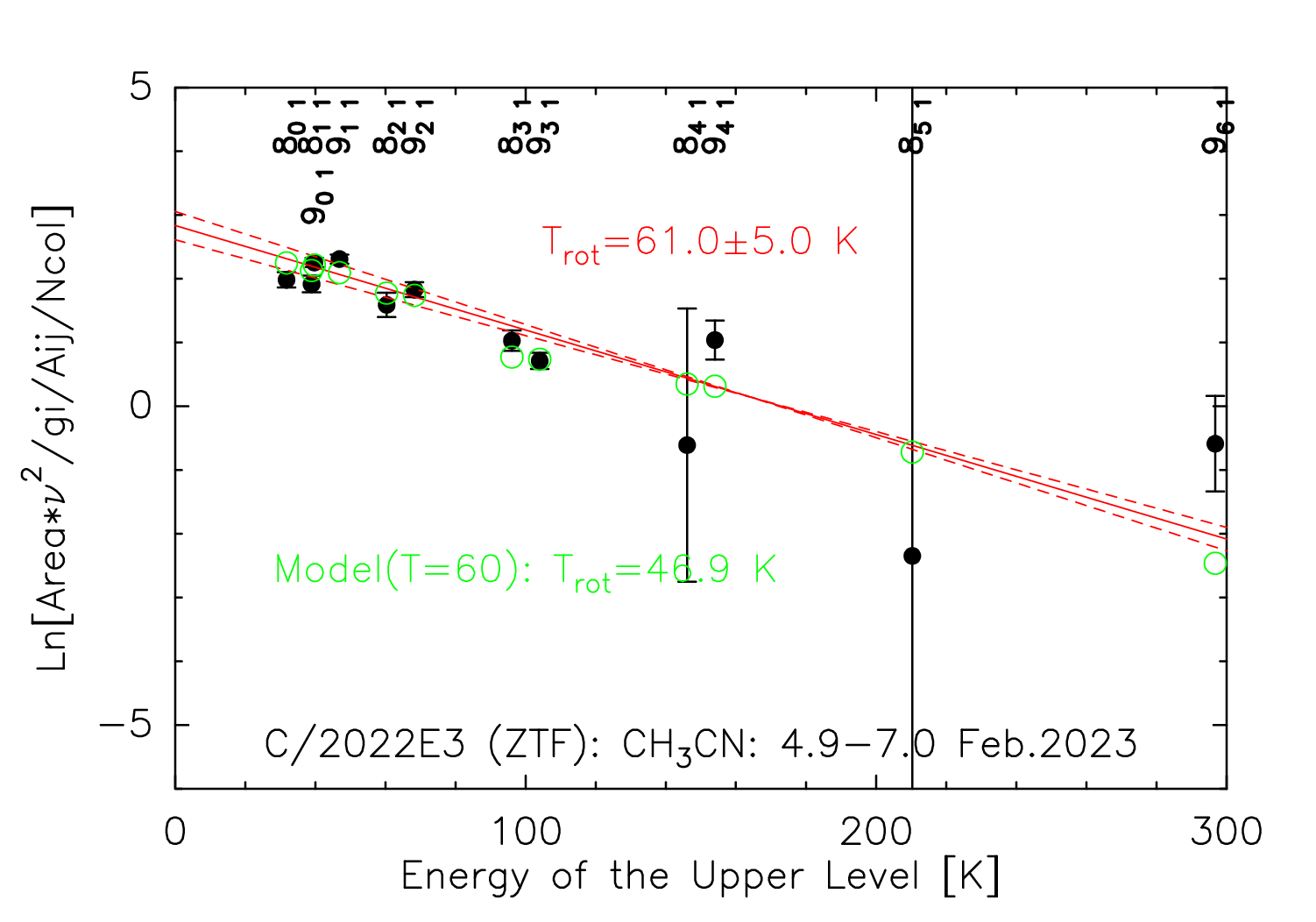}}
\caption{Rotational diagram of the CH$_3$CN lines at 147 and 165~GHz observed
  between 4.9 and 7.0 February 2023 in comet C/2022~E3 (ZTF).
  Scales as in Fig.~\ref{diagrotztf-met166}.}
\label{diagrotztf-ch3cn89}
\end{figure}

\begin{figure}[]
\centering
\resizebox{\hsize}{!}{
  \includegraphics[angle=270,width=0.9\textwidth]{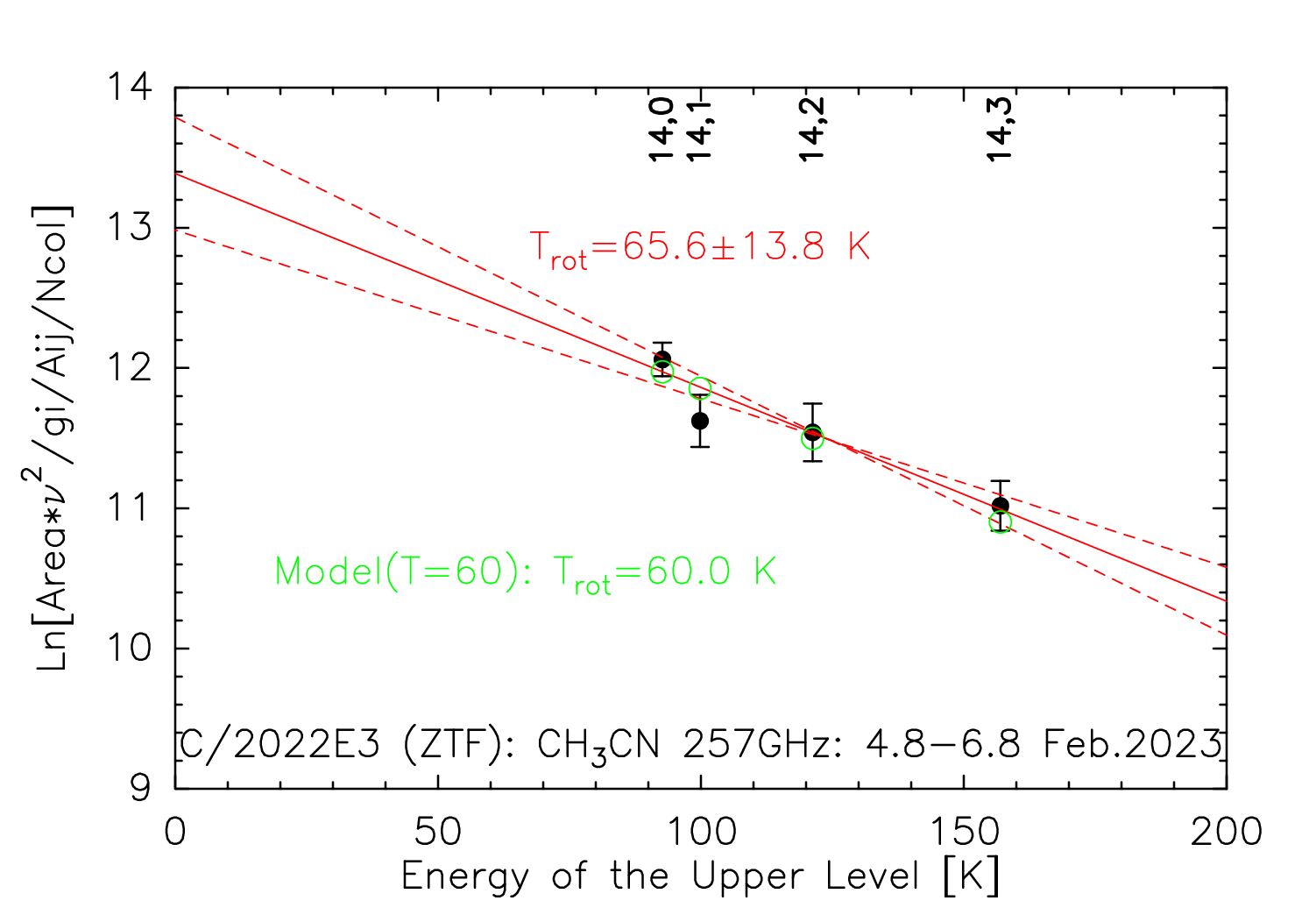}}
\caption{Rotational diagram of the CH$_3$CN lines at 257~GHz observed
  between 3.7 and 6.7 February 2023 in comet C/2022~E3 (ZTF).
  Scales as in Fig.~\ref{diagrotztf-met166}.}
\label{diagrotztf-ch3cn14}
\end{figure}

\begin{figure}[]
\centering
\resizebox{\hsize}{!}{
  \includegraphics[angle=270,width=0.9\textwidth]{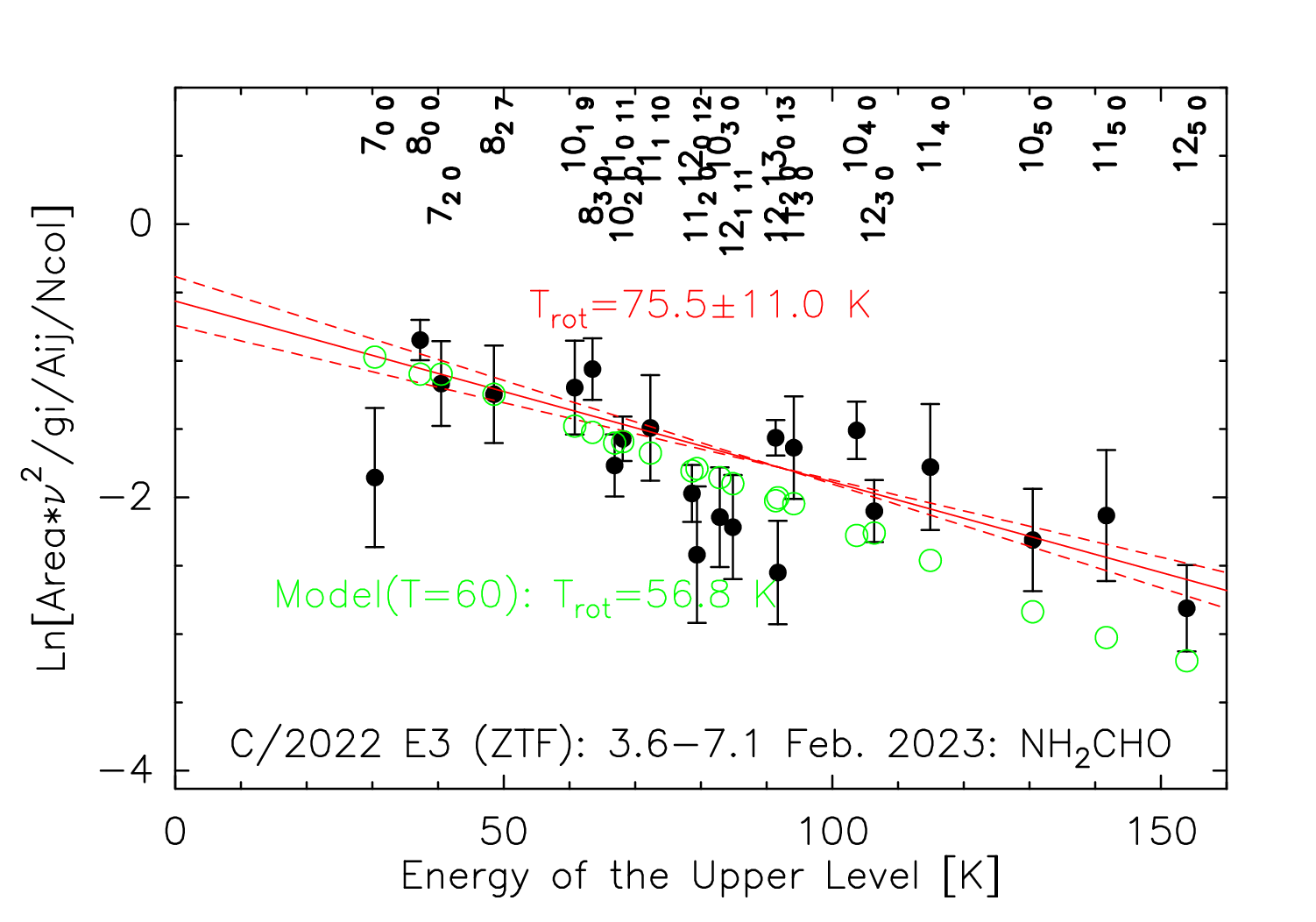}}
\caption{Rotational diagram of all the NH$_2$CHO lines observed
  between 3.6 and 7.1 February 2023 in comet C/2022~E3 (ZTF) at 1~mm and 2~mm.
  Some lines with the same $J$ and $K_a$ levels (and/or similar energy levels
  and Einstein A coefficients) have been grouped together ($K_c$ given as 0
  in the third quantum number of the line).
  Scales as in Fig.~\ref{diagrotztf-met166}.}
\label{diagrotztf-nh2cho}
\end{figure}

\begin{figure}[]
\centering
\resizebox{\hsize}{!}{
  \includegraphics[angle=270,width=0.9\textwidth]{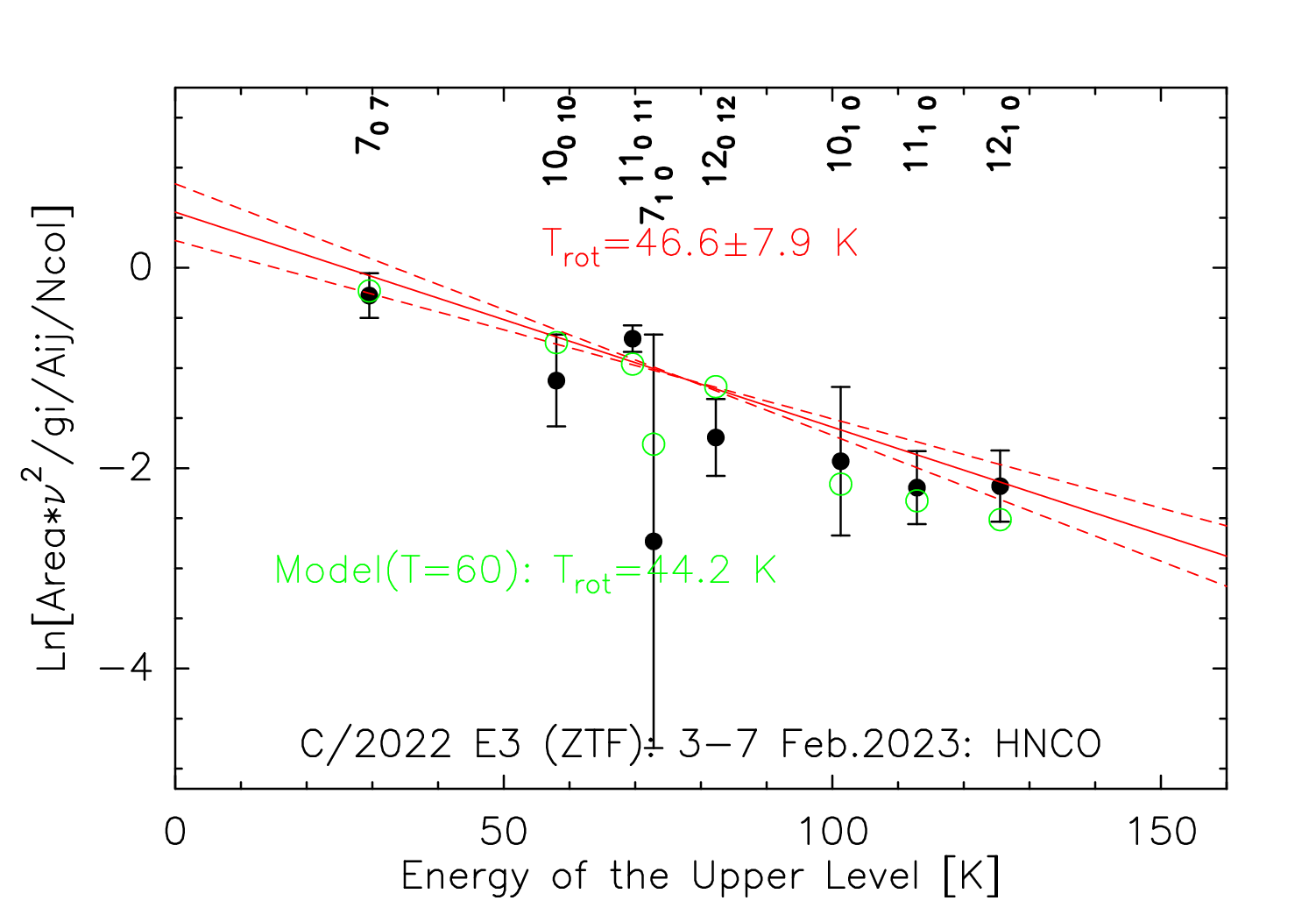}}
\caption{Rotational diagram of all the HNCO lines observed
  between 3.6 and 7.1 February 2023 in comet C/2022~E3 (ZTF).
  The lines with the same $J$ for $K_a=1$ levels have been grouped together.
  Scales as in Fig.~\ref{diagrotztf-met166}.}
\label{diagrotztf-hnco}
\end{figure}

\begin{figure}[]
\centering
\resizebox{\hsize}{!}{
  \includegraphics[angle=270,width=0.9\textwidth]{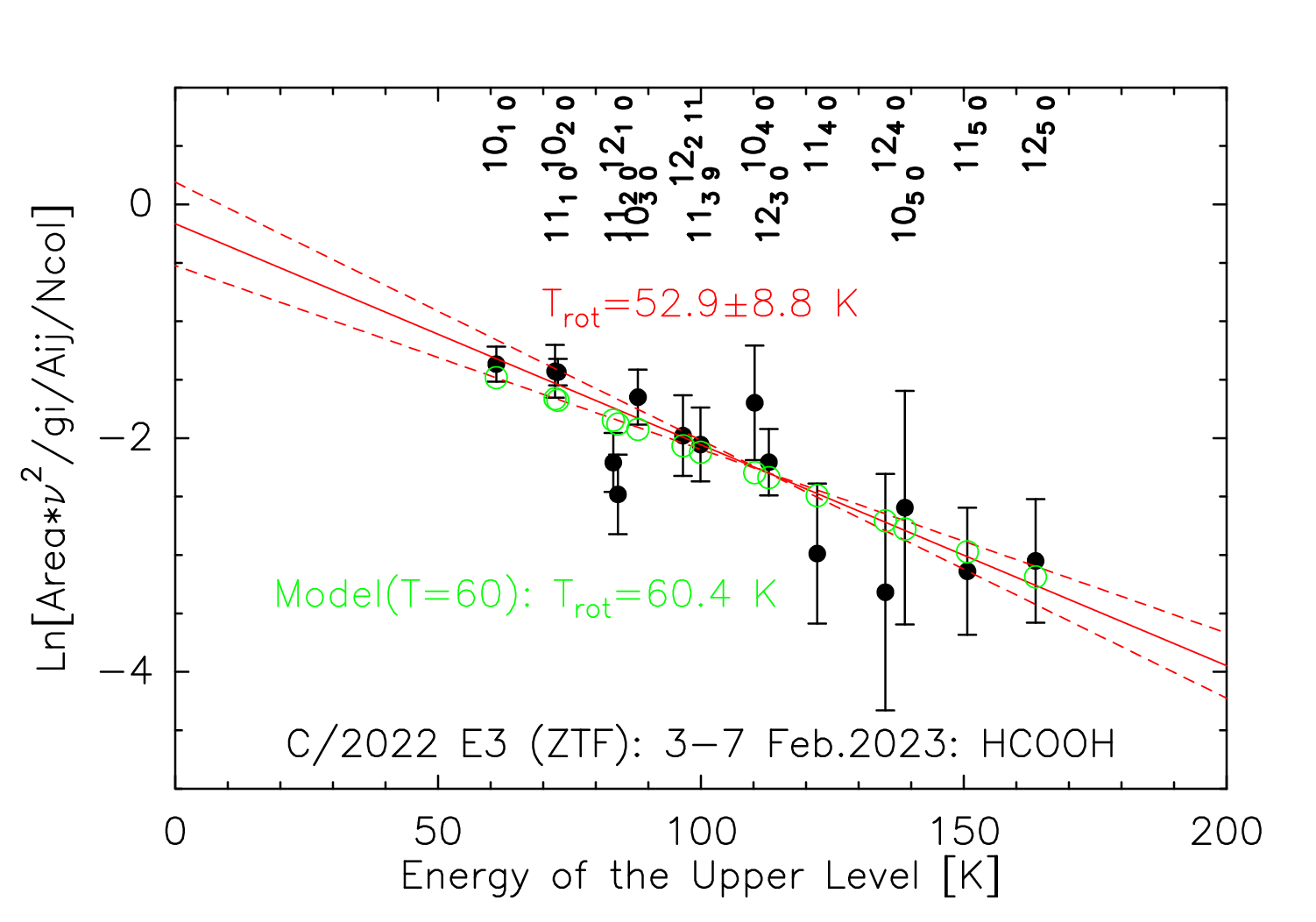}}
\caption{Rotational diagram of all the HCOOH lines observed
  between 3.6 and 7.1 February 2023 in comet C/2022~E3 (ZTF).
  Scales and grouping of lines as in Fig.~\ref{diagrotztf-nh2cho}.}
\label{diagrotztf-hcooh}
\end{figure}

\begin{figure}[]
\centering
\resizebox{\hsize}{!}{
  \includegraphics[angle=270,width=0.9\textwidth]{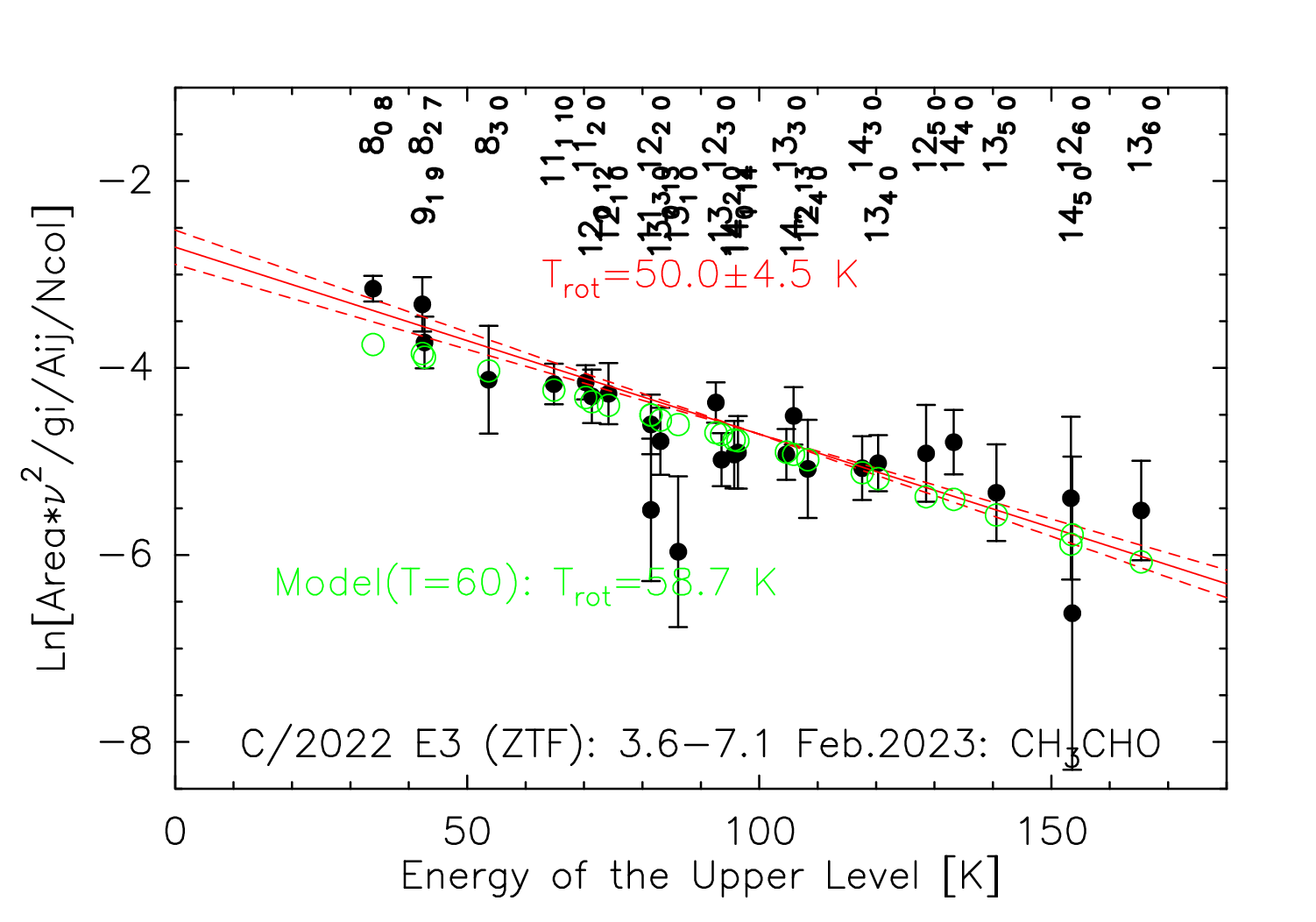}}
\caption{Rotational diagram of all the CH$_3$CHO lines observed
  between 3.6 and 7.1 February 2023 in comet C/2022~E3 (ZTF).
  Scales and grouping of lines as in Fig.~\ref{diagrotztf-nh2cho}.}
\label{diagrotztf-ch3cho}
\end{figure}

\clearpage
\onecolumn
\FloatBarrier
\section{Tables of intensities of averaged lines}

\begin{table*}[h]
\caption[]{Line intensities from IRAM observations: multi-line averages - C/2021~A1 (Leonard).}\label{tabobssumleonard}\vspace{-0.5cm}
  \renewcommand{\arraystretch}{0.95}
\begin{center}
\begin{tabular}{llllrrc}
  \hline\hline
Molecule & \multicolumn{2}{c}{Transitions} & Frequency & Pointing offset & Intensity & Doppler shift \\
         & N\tablefootmark{a} & $J_{Ka,Kc}$\tablefootmark{b} &  range (GHz) & (\arcsec) & (mK~\kms) & (\kms)\\
\hline
NH$_2$CHO   &  4 & $7_{0,7}+7_{1,6}+8_{0,8}+8_{1,8}$  & 147--167 &  1.9 & $9.5\pm2.8$ & $-0.03\pm0.17$ \\
NH$_2$CHO   &  3 & $10_{1,9}+10_{2,8}+10_{2,9}$       & 213--218 &  1.1 & $15.6\pm5.9$ & $+0.32\pm0.27$  \\
NH$_2$CHO   &  5 & $10_{3,Kc}+10_{4,Kc}+10_{5,Kc}$     & 212--213 &  1.0 & $5.8\pm6.1$ &                  \\
NH$_2$CHO   &  5 & $11_{0,11}+11_{1,Kc}+11_{2,Kc}$     & 223--240 &  1.1 & $21.6\pm4.2$ & $+0.02\pm0.13$  \\
NH$_2$CHO   &  5 & $11_{3,Kc}+11_{4,Kc}+11_{5,Kc}$     & 233--234 &  1.1 & $15.9\pm4.1$ & $-0.29\pm0.19$  \\
NH$_2$CHO   &  5 & $12_{0,12}+12_{1,Kc}+12_{2,Kc}$     & 244--261 &  1.7 & $35.1\pm3.8$ & $-0.07\pm0.08$ \\
NH$_2$CHO   &  5 & $12_{3,Kc}+12_{4,Kc}+12_{5,Kc}$     & 255--256 &  1.7 & $12.5\pm3.8$ & $-0.37\pm0.24$  \\
NH$_2$CHO   &  2 & $13_{0,13}+13_{1,13}$              & 264--267 &  1.7 & $31.5\pm6.2$ & $+0.06\pm0.13$  \\
NH$_2$CHO   & 30 & $J_{Ka,Kc}$, J=10--13, Ka=0--5    & 212--267 &  1.5 & $20.1\pm1.9$ & $-0.10\pm0.06$ \\
            & 21 &                                  &          & 10.2 & $11.1\pm5.5$ & $-0.83\pm0.51$ \\
CH$_3$CHO   &  8 & $8_{0,8}+8_{1,8}+8_{2,7}+9_{1,9}$  & 149--168 &  1.9 &  $4.2\pm1.9$ & $+0.06\pm0.24$ \\
CH$_3$CHO   & 10 & $11_{1,10}+11_{2,Kc}+11_{3,Kc}$    & 211--217 &  1.0 & $-1.8\pm4.6$ &           \\
CH$_3$CHO   & 10 & $12_{0,12}+12_{1,Kc}+12_{2,Kc}$    & 224--236 &  1.1 &  $4.3\pm2.8$ & $-0.45\pm0.50$ \\
CH$_3$CHO   & 10 & $13_{0,13}+13_{1,Kc}+13_{2,Kc}$    & 242--255 &  1.7 &  $7.5\pm2.5$ & $+0.09\pm0.21$ \\
CH$_3$CHO   &  8 & $13_{3,Kc}+13_{4,Kc}$              & 250--251 &  1.7 &  $3.3\pm2.9$ & $-0.69\pm0.80$ \\
CH$_3$CHO   &  6 & $14_{0,14}+14_{1,14}+14_{2,13}$     & 261--268 &  1.6 &  $8.6\pm3.6$ & $+0.41\pm0.32$ \\
CH$_3$CHO   &  4 & $14_{3,11}+14_{3,12}$              & 270--271 &  1.7 & $13.2\pm4.3$ & $+0.18\pm0.21$  \\
CH$_3$CHO      & 52 & $J=11-14,Ka=0-3$ + $13_{4,Kc}$ & 211--271 & 1.5 & $5.2\pm1.3$ & $-0.03\pm0.14$ \\
               & 46 &                                  & 211--271 & 10.2 & $1.2\pm3.5$ &             \\

CH$_2$OHCHO  & 11 & $J=7-9$,$Ka$=6,7; $J=21-26$,$Ka$=0,+1 & 213--269 & 1.5 & $10.2\pm3.0$ & $+0.10\pm0.21$  \\
        
C$_2$H$_5$OH & 91 & $J$=4--16          & 214--267 &  1.4 & $3.4\pm1.0$  & $+0.10\pm0.20$  \\

(CH$_2$OH)$_2$ & 15 & $J=14-18,Ka=0-8$ & 147--169 &  1.9 & $4.4\pm1.4$ & $-0.03\pm0.22$ \\
(CH$_2$OH)$_2$ & 28 & $J=20-29,Ka=0-4$ & 210--271 &  1.5 & $4.1\pm1.7$ & $+0.33\pm0.30$ \\

HC$_3$N   &  2 & 17-16 + 18-17         & 155--164 &  1.7 & $6.6\pm4.3$ &  \\
HC$_3$N   &  6 & 24-23 to 29-28        & 218--264 &  1.4 & $13.0\pm3.8$ & $-0.54\pm0.24$  \\

OCS       &  5 & 18-17 to 22-21        & 219--267 &  1.5 & $19.0\pm4.2$ & $+0.17\pm0.14$  \\

HNCO      &  3 & $10_{0,10}+11_{0,11}+12_{0,12}$ & 220--264 &  1.3 & $49.7\pm4.7$ & $+0.01\pm0.07$  \\
          &  2 & $11_{0,11}+12_{0,12}$   & 242--264 &  9.9 & $33\pm14$    & $+0.01\pm0.27$ \\
HNCO      &  6 & $J=10-12, Ka=1$        & 219--263 &  1.7 & $6.2\pm3.2$  & $-0.01\pm0.36$  \\

HCOOH     &  5 & $10_{Ka,Kc}$:Ka=0--2   & 215--232 &  1.1 & $17.1\pm4.2$ & $-0.38\pm0.16$ \\
HCOOH     &  5 & $10_{Ka,Kc}$:Ka=3--5   & 225--226 &  1.2 &  $9.9\pm4.2$ & $-0.31\pm0.28$ \\
HCOOH     &  5 & $11_{Ka,Kc}$:Ka=0--2   & 237--254 &  1.5 & $17.9\pm3.3$ & $+0.19\pm0.11$  \\
HCOOH     &  4 & $11_{Ka,Kc}$:Ka=3--5   & 247--248 &  1.6 & $14.9\pm4.1$ & $+0.58\pm0.23$  \\
HCOOH     &  3 & $12_{Ka,Kc}$:Ka=0--2   & 258--268 &  1.6 & $17.5\pm4.8$ & $-0.17\pm0.17$  \\
HCOOH     &  5 & $12_{Ka,Kc}$:Ka=3--5   & 270--271 &  1.7 & $11.5\pm3.8$ & $-0.37\pm0.24$  \\
HCOOH     & 13 & $J=10-12,Ka<3$        & 215--268 &  1.5 & $16.6\pm2.4$ & $-0.14\pm0.09$  \\
HCOOH     & 10 & $J=10-12:Ka<3$        & 215--268 & 10.1 & $16.8\pm7.4$ & $+0.42\pm0.30$  \\

CH$_2$CO  &  6 & $J_{1,Kc}$:J=11--13    & 220--265 &  1.4 & $7.9\pm3.1$ & $+0.17\pm0.25$ \\

H$_2$CS   &  3 & $7_{17}+7_{16}+8_{18}$  & 237--271 &  1.5 & $9.7\pm4.9$ &         \\

SO        &  5 & $N,J = 5,5+5,6+6,5+6,6+6,7$ & 215--262 &  1.5 & $7.0\pm4.2$  &   \\

SO$_2$    & 18 & $J=4-15, Ka=0-3$      & 222--272 &  1.5 & $<6.8$    & \\

HCOOCH$_3$ & 6 & $J_{Ka,J}$:$J=20-25, Ka=0,1$  & 217--270 &  1.5 & $<14.8$ &  \\

C$_2$H$_5$CN   & 63 & $J=23-31,Ka=0-7$ & 210--266 &  1.5 & $<3.9$ & \\
CH$_3$COCH$_3$ & 101 & $J_{Ka,Kc}$      & 210--272 &  1.6 & $<3.0$ &  \\
CH$_3$COOH     & 52 & $J_{Ka,Kc}$       & 215--271 &  1.6 & $<3.0$  & \\
CH$_3$NH$_2$   & 32 & $J=2-9,Ka=0-2$   & 215--271 &  1.6 & $<4.9$  & \\
\hline
\hline
\end{tabular}
\end{center}
\tablefoot{
\tablefoottext{a}{Number of lines averaged}\\
\tablefoottext{b}{$J_{Ka,Kc}$ quantum numbers of the upper level of the
  transitions, with range of values, or $J_{up}-J_{low}$ for OCS, HC$_3$N}
}
\end{table*}

\begin{table*}
\caption[]{Line intensities from IRAM observations: multi-line averages - C/2022~E3 (ZTF): 3.7--7.0 Feb. 2023.}\label{tabobssumztf}\vspace{-0.5cm}
  \renewcommand{\arraystretch}{0.87}
\begin{center}
\begin{tabular}{llllrrc}
  \hline\hline
Molecule & \multicolumn{2}{c}{Transitions} & Frequency & Pointing offset & Intensity & Doppler shift \\
         & N\tablefootmark{a} & $J_{Ka,Kc}$\tablefootmark{b} &  range (GHz) & (\arcsec) & (mK~\kms) & (\kms)\\
\hline
NH$_2$CHO  &  7 & $7_{Ka,Kc}+8_{Ka,Kc}$ Ka=0--2  & 147--167 &  2.1 & $12.0\pm1.4$ & $-0.09\pm0.08$ \\
NH$_2$CHO  & 15 & $J=10-13, Ka=0-2$    & 211--267 &  1.7 & $23.5\pm1.6$ & $+0.01\pm0.04$ \\
NH$_2$CHO  & 15 & $J=10-12, Ka=3-5$    & 212--256 &  1.6 & $13.1\pm1.8$ & $-0.15\pm0.10$ \\

HCOOH     & 13 & $J=10-12,Ka<3$        & 215--268 &  1.7 & $27.6\pm1.7$ & $-0.04\pm0.04$  \\
HCOOH     & 14 & $J=10-12:Ka=3-5$      & 225--270 &  2.0 & $13.2\pm1.9$ & $+0.06\pm0.09$  \\

CH$_3$CHO   & 11 & $J=8,9, Ka=0-3$     & 149--168 &  2.1 &  $10.9\pm1.1$ & $-0.05\pm0.06$ \\
CH$_3$CHO   &  8 & $J=13,14, Ka=0-1$   & 242--263 &  1.8 &  $11.3\pm2.3$ & $+0.13\pm0.13$ \\
CH$_3$CHO   &  9 & $J=13,14, Ka=4-5$   & 251--270 &  1.6 &  $12.5\pm2.2$ & $-0.02\pm0.11$ \\
CH$_3$CHO   & 16 & $J=13,14, Ka=2-3$   & 249--271 &  1.6 &  $13.3\pm1.7$ & $+0.09\pm0.09$ \\
            &    &                     &          &  5.4 &  $12.5\pm4.4$ & $-0.05\pm0.19$ \\
CH$_3$CHO   & 12 & $J=11,12, Ka=4-5$   & 212--231 &  1.2 &   $9.0\pm1.4$ & $-0.09\pm0.10$ \\
CH$_3$CHO   & 18 & $J=11,12, Ka=0-3$   & 211--236 &  1.2 &  $11.5\pm1.3$ & $-0.09\pm0.06$ \\

C$_2$H$_5$OH & 11 & $J=8-10, Ka=0-2$   & 147--168 &  2.1 &  $6.5\pm1.2$ & $-0.34\pm0.15$ \\
C$_2$H$_5$OH & 97 & $J=4-16, Ka=0-5$   & 209--270 &  2.1 &  $2.5\pm0.6$ & $-0.21\pm0.13$ \\

CH$_2$OHCHO & 8 & $J_{KaJ}:J=15,16 Ka=0,1$   & 154--164 &  2.1 &  $3.1\pm1.6$ &  \\
CH$_2$OHCHO & 26 & $J=7-9,Ka=6,7$         & 220--268 &  1.7 &  $-0.6\pm1.1$ &  \\
           &   & + $J_{KaJ}:J=22-25,Ka=0,1$   &      &       &              & \\
CH$_3$OCHO & 26 & $J_{KaJ} A+E:J=20-25$   & 217--270 &  1.7 &  $4.1\pm4.3$ &  \\

(CH$_2$OH)$_2$ & 22 & $J=14-18 Ka=0-8$   & 147--169 &  2.1 &  $3.6\pm0.8$ & $+0.07\pm0.16$ \\
(CH$_2$OH)$_2$ & 92 & $J=20-29 Ka=0-9$   & 210--271 &  1.7 &  $3.0\pm0.7$ & $-0.19\pm0.16$ \\

HC$_3$N    &  2 & 17-16 + 18-17         & 155--164 &  2.1 & $3.2\pm3.3$ &  \\
HC$_3$N    &  6 & 24-23 to 29-28        & 218--264 &  2.0 & $0.2\pm2.9$ &  \\

OCS        &  5 & 18-17 to 22-21        & 219--267 &  1.8 & $11.9\pm3.2$ & $+0.09\pm0.19$  \\
H$_2$CS    &  3 & $7_{17}+7_{16}+8_{18}$  & 237--271 &  1.9 & $9.0\pm4.4$ &         \\
SO         &  5 & $N,J = 5,5+5,6+6,5+6,6+6,7$ & 215--262 &  1.7 & $7.4\pm3.4$  & $-0.58\pm0.42$  \\

SO$_2$     & 17 & $J=4-15, Ka=0-3$      & 222--272 &  2.0 & $<5.4$    & \\
l-c$_3$H$_2$ & 4 & $J=7-8, Ka=0-1$      & 147--168 &  2.1 & $<5.7$    & \\
l-c$_3$H$_2$ & 9 & $J=10-13, Ka=0-1$    & 210--270 &  1.5 & $<6.3$    & \\

\hline
\hline
\end{tabular}
\end{center}
\tablefoot{
\tablefoottext{a}{Number of lines averaged - for offset positions see Table~\ref{tabqdistriztf}.}\\
\tablefoottext{b}{$J_{Ka,Kc}$ quantum numbers of the upper level of the
  transitions, with range of values, or $J_{up}-J_{low}$ for OCS, HC$_3$N.}
}
\end{table*}

\clearpage
\twocolumn
\FloatBarrier
\section{Tables of daily production rates}

\begin{table}[h]\renewcommand{\tabcolsep}{0.07cm}
\caption[]{Production rates in C/2021~A1 in November-December 2021.}\label{tabqpdayleonard}
\begin{center}
\begin{tabular}{lcccc}
\hline\hline\noalign{\smallskip}
UT date  & Molecule & $r_h$  & Production rate  & Lines\tablefootmark{a} \\
(mm/dd.dd) &      &  (au)  & ($\times10^{25}$~\mols) &  \\
\hline
11/12.19    & HCN   & 1.218  & $1.7\pm0.5$  &  1 \\
11/14.18    & HCN   & 1.187  & $3.1\pm0.4$  &  1 \\
11/16.18    & HCN   & 1.157  & $2.9\pm0.2$  &  1 \\
\hline
12/08.46    & HCN   & 0.826  & $3.4\pm0.2$  &  1 \\
12/09.46    & HCN   & 0.812  & $3.4\pm0.1$  &  1 \\
12/10.32    & HCN   & 0.800  & $3.2\pm0.1$  &  1 \\
12/11.58    & HCN   & 0.785  & $4.6\pm0.2$  &  2 \\
12/12.40    & HCN   & 0.774  & $3.1\pm0.1$  &  1 \\
12/13.40    & HCN   & 0.762  & $3.5\pm0.1$  &  1 \\
12/09.50    & HNC   & 0.812  & $0.28\pm0.05$  &  1 \\
12/10.37    & HNC   & 0.800  & $0.24\pm0.05$  &  1 \\
12/11.62    & HNC   & 0.784  & $0.51\pm0.12$  &  1 \\
12/12.40    & HNC   & 0.774  & $0.22\pm0.08$  &  1 \\
12/13.47    & CH$_3$CN & 0.761  & $0.59\pm0.06$  & 7+(2) \\
12/08.55    & H$_2$S & 0.824  & $3.8\pm1.5$  &  (1) \\
12/10.53    & H$_2$S & 0.798  & $8.1\pm1.1$  &  1 \\
12/12.60    & H$_2$S & 0.772  & $6.0\pm4.6$  &  (1) \\
12/13.63    & H$_2$S & 0.759  & $4.4\pm1.2$  &  1+(1) \\
12/08.55    & CS    & 0.824  & $3.6\pm0.4$  &  1 \\
12/10.53    & CS    & 0.798  & $3.7\pm0.2$  &  1 \\
12/12.48    & CS    & 0.773  & $2.6\pm0.1$  &  1 \\
12/13.59    & CS    & 0.760  & $2.9\pm0.1$  &  2 \\
12/08.55    & H$_2$CO$_d$\tablefootmark{b} & 0.824 & $9.5\pm2.8$  & 1 \\
12/10.53    & H$_2$CO$_d$ & 0.798 & $11.8\pm1.8$  & 1 \\
12/11.52    & H$_2$CO$_d$ & 0.785 & $14.6\pm2.5$  & 1 \\
12/12.60    & H$_2$CO$_d$ & 0.772 & $10.2\pm0.6$  & 2+ \\
12/13.55    & H$_2$CO$_d$ & 0.760 & $8.6\pm1.1$  & 2+(2) \\
12/08.54    & CH$_3$OH    & 0.824 & $35\pm5$      & 4+(8) \\
12/09.50    & CH$_3$OH    & 0.812 & $37\pm7$      &   1 \\
12/10.51    & CH$_3$OH    & 0.798 & $44\pm3$      &  11 \\
12/11.60    & CH$_3$OH    & 0.784 & $43\pm7$      & 3+(16) \\
12/12.49    & CH$_3$OH    & 0.774 & $38\pm1$      & 13+(6) \\ 
12/13.53    & CH$_3$OH    & 0.760 & $28\pm2$      & 16+(14) \\ 
\hline
\end{tabular}
\end{center}
\tablefoot{Subscript ``$_d$'' has been added to the molecules for which
  a daughter Haser density profile is assumed with the parent scale length provided below.\\
  \tablefoottext{a}{Number of lines used for the determination of $Q$,
    in parentheses the number of lines that are not individually detected.}\\
  \tablefoottext{b}{Where we assume that H$_2$CO is produced in the
  coma with a Haser parent scale length of 1500~km in December.}
}
\end{table}

\begin{table}\renewcommand{\tabcolsep}{0.07cm}
\caption[]{Production rates in C/2022~E3 in February 2023.}\label{tabqpdayztf}
\begin{center}
\begin{tabular}{lcccc}
\hline\hline\noalign{\smallskip}
UT date  & Molecule & $r_h$  & Production rate  & Lines\tablefootmark{a} \\
(mm/dd.dd) &      &  (au)  & ($\times10^{25}$~\mols) &  \\
\hline
02/03.73    & HCN   & 1.168  & $4.1\pm0.1$  &  1+ \\
02/04.67    & HCN   & 1.173  & $6.3\pm0.1$  &  1+ \\
02/05.03    & HCN   & 1.175  & $4.9\pm0.3$  &  1+ \\
02/05.67    & HCN   & 1.178  & $5.3\pm0.1$  &  1+ \\
02/06.03    & HCN   & 1.180  & $4.5\pm0.1$  &  1+ \\
02/06.67    & HCN   & 1.183  & $4.7\pm0.1$  &  1+ \\

02/03.73    & HNC$_d$\tablefootmark{b}   & 1.168  & $0.12\pm0.05$  &  1 \\
02/04.67    & HNC$_d$   & 1.173  & $0.34\pm0.11$  &  1 \\
02/05.67    & HNC$_d$   & 1.178  & $0.25\pm0.15$  &  1 \\
02/06.03    & HNC$_d$   & 1.180  & $0.17\pm0.12$  &  1 \\

02/04.91    & CH$_3$CN & 1.174  & $0.87\pm0.05$  & 12 \\
02/05.76    & CH$_3$CN & 1.178  & $0.84\pm0.06$  & 17 \\
02/06.90    & CH$_3$CN & 1.184  & $0.62\pm0.17$  & 8+(4) \\

02/04.91    & H$_2$S & 1.174  &  $9.4\pm0.4$  &  1+ \\
02/05.02    & H$_2$S & 1.175  & $11.6\pm1.1$  &  2+ \\
02/05.90    & H$_2$S & 1.179  &  $7.5\pm0.6$  &  2+ \\
02/07.00    & H$_2$S & 1.185  &  $8.9\pm1.7$  &  2+ \\

02/04.84    & CS$_d$\tablefootmark{c}    & 1.174  & $3.8\pm0.1$  &  2+ \\
02/05.86    & CS$_d$    & 1.179  & $3.7\pm0.3$  &  1+ \\
02/06.90    & CS$_d$    & 1.184  & $3.9\pm0.3$  &  2+ \\

02/04.91    & H$_2$CO$_d$\tablefootmark{d} & 1.174 & $12.0\pm0.9$  & 1+ \\
02/05.00    & H$_2$CO$_d$ & 1.174 & $13.3\pm0.5$  & 2+ \\
02/05.76    & H$_2$CO$_d$ & 1.178 & $13.5\pm1.1$  & 2+ \\
02/05.91    & H$_2$CO$_d$ & 1.179 & $10.2\pm0.4$  & 3+ \\
02/06.97    & H$_2$CO$_d$ & 1.185 &  $8.3\pm0.6$  & 3+ \\

02/03.73    & CH$_3$OH    & 1.168 & $76\pm2$      & 27 \\
02/04.67    & CH$_3$OH    & 1.173 & $115\pm4$     & 27 \\
02/04.91    & CH$_3$OH    & 1.174 & $87\pm1$      & 43 \\
02/05.76    & CH$_3$OH    & 1.178 & $87\pm3$      & 32 \\
02/06.03    & CH$_3$OH    & 1.180 & $75\pm6$      & 15 \\ 
02/06.76    & CH$_3$OH    & 1.184 & $80\pm2$      & 41 \\ 

02/03.73    & NH$_2$CHO   & 1.168 & $1.5\pm0.4$   &  1 \\
02/04.85    & NH$_2$CHO   & 1.174 & $1.0\pm0.1$   & 14 \\
02/05.85    & NH$_2$CHO   & 1.179 & $0.9\pm0.1$   & 12 \\
02/06.91    & NH$_2$CHO   & 1.184 & $1.2\pm0.3$   & 12 \\

02/04.81    & HNCO        & 1.174 & $2.0\pm0.3$   &  3 \\
02/05.81    & HNCO        & 1.178 & $1.4\pm0.6$   &  2 \\
02/06.82    & HNCO        & 1.184 & $1.8\pm0.9$   &  3 \\
\hline
\end{tabular}
\end{center}
\tablefoot{Subscript ``$_d$'' has been added to the molecules for which
  a daughter Haser density profile is assumed with the parent scale length provided below.\\
  \tablefoottext{a}{Number of lines used for the determination of $Q$,
    in parentheses the number of lines that are not individually detected.
    A ``+'' indicates that offset data have been taken into account.}\\
  \tablefoottext{b}{Where we assume that HNC is produced in the
  coma with a Haser parent scale length 2000~km.}
  \tablefoottext{c}{Where we assume that CS is produced in the
  coma with a Haser parent scale length 2000~km.}
  \tablefoottext{d}{Where we assume that H$_2$CO is produced in the
  coma with a Haser parent scale length 1700~km.}
}
\end{table}

\clearpage

\FloatBarrier
\onecolumn
\section{Tables of individual line intensities}
The following tables provide the line integrated intensities in
K~\kms corrected for main beam efficiency and Doppler shifts, when detected.
Data are provided for individual lines, averaged over one day or several days.
Integrated intensities and rms are weighted average when covered
by several backends, and Doppler shift measured with the backend providing the
highest spectral resolution.

\begin{longtable}{llllrrl}
\caption[]{Line intensities from IRAM observations: C/2021~A1 (Leonard)\label{tabobsleonard}} \\
\hline\hline
Date\tablefootmark{a} & Molecule & Transition & Frequency\tablefootmark{b} & offset\tablefootmark{c} & Intensity & Doppler shift \\
(yyyy/mm/dd.dd) &  &  & (MHz) & (\arcsec) &  (K~\kms) & (\kms)  \\
\hline
\endfirsthead
\caption{Continued.}\vspace{-0.3cm}\\
\hline
\hline
Date & Molecule & Transition & Frequency & Pointing offset & Intensity & Doppler shift \\
(yyyy/mm/dd.dd) &  &       & (GHz)     & (\arcsec) &       (K~\kms) & (\kms) \\
\hline
\endhead
\hline
\endfoot
\hline
\endlastfoot
2021/11/12.19 & HCN        &    3-2    & 265886.434 &  1.3 & $0.308\pm0.085$ & $-0.16\pm0.15$ \\
              &            &           &            &  2.7 & $0.026\pm0.181$ & \\
2021/11/14.18 & HCN        &    3-2    & 265886.434 &  1.4 & $0.524\pm0.070$ & $-0.03\pm0.08$ \\
              &            &           &            &  4.8 & $0.382\pm0.138$ & $-0.01\pm0.17$ \\
              &            &           &            &  6.1 & $0.342\pm0.201$ & \\
2021/11/16.18 & HCN        &    3-2    & 265886.434 &  1.4 & $0.521\pm0.030$ & $-0.03\pm0.03$ \\
              &            &           &            &  5.1 & $0.423\pm0.059$ & $-0.04\pm0.08$ \\
              &            &           &            &  7.3 & $0.273\pm0.067$ & $+0.00\pm0.12$ \\
\hline\hline
\end{longtable}
\tablefoot{
  \tablefoottext{a}{Dates are rounded to the nearest tenth of a day for averages over several days.}
  \tablefoottext{b}{From \citet{CDMS} and \citet{JPLmol}. For HCN the strongest hyperfine
    component are in the integration window and integrated intensity takes into account the weaker ones
  according to their statistical weight.}
  \tablefoottext{c}{Average radial pointing offset.}
}

\begin{longtable}{llllrrl}
\caption[]{\label{tabobsztf} Line intensities from IRAM observations: C/2022~E3 (ZTF)}\\
\hline\hline
Date\tablefootmark{a} & Molecule & Transition & Frequency\tablefootmark{b} & offset\tablefootmark{c} & Intensity & Doppler shift \\
(yyyy/mm/dd.dd) &  &  & (MHz) & (\arcsec) &  (K~\kms) & (\kms)  \\
\hline
\endfirsthead
\caption{Continued.}\vspace{-0.3cm}\\
\hline\hline
Date & Molecule & Transition & Frequency & Pointing offset & Intensity & Doppler shift \\
(yyyy/mm/dd.dd) &  &       & (GHz)     & (\arcsec) &       (K~\kms) & (\kms) \\
\hline
\endhead
\hline
\endfoot
\hline
\endlastfoot
2023/02/03.73 & HCN        &    3-2    & 265886.434 &  1.1 & $2.110\pm0.018$ & $-0.06\pm0.01$ \\
              &            &           &            &  5.1 & $1.822\pm0.051$ & $-0.08\pm0.02$ \\
              &            &           &            &  9.0 & $1.228\pm0.048$ & $-0.01\pm0.02$ \\
              &            &           &            & 10.5 & $0.932\pm0.036$ & $-0.10\pm0.02$ \\
              &            &           &            & 14.0 & $0.852\pm0.055$ & $-0.04\pm0.04$ \\
2023/02/04.67 & HCN        &    3-2    & 265886.434 &  0.9 & $3.061\pm0.034$ & $-0.06\pm0.01$ \\
              &            &           &            &  9.1 & $1.829\pm0.114$ & $-0.01\pm0.04$ \\
              &            &           &            & 11.0 & $1.419\pm0.121$ & $-0.13\pm0.05$ \\
\hline\hline
\end{longtable}
\tablefoot{
  \tablefoottext{a}{Dates are rounded to the nearest tenth of a day for averages over several days.}
  \tablefoottext{b}{From \citet{CDMS} and \citet{JPLmol}. For HCN the strongest hyperfine
    component are in the integration window and integrated intensity takes into account the weaker ones
  according to their statistical weight.}
  \tablefoottext{c}{Average radial pointing offset.}
}
\newpage
\FloatBarrier
\section{Spectra of the full frequency range}

\begin{figure}[ht]\vspace{-0.0cm}
  \centering
\includegraphics[angle=0, width=16cm]{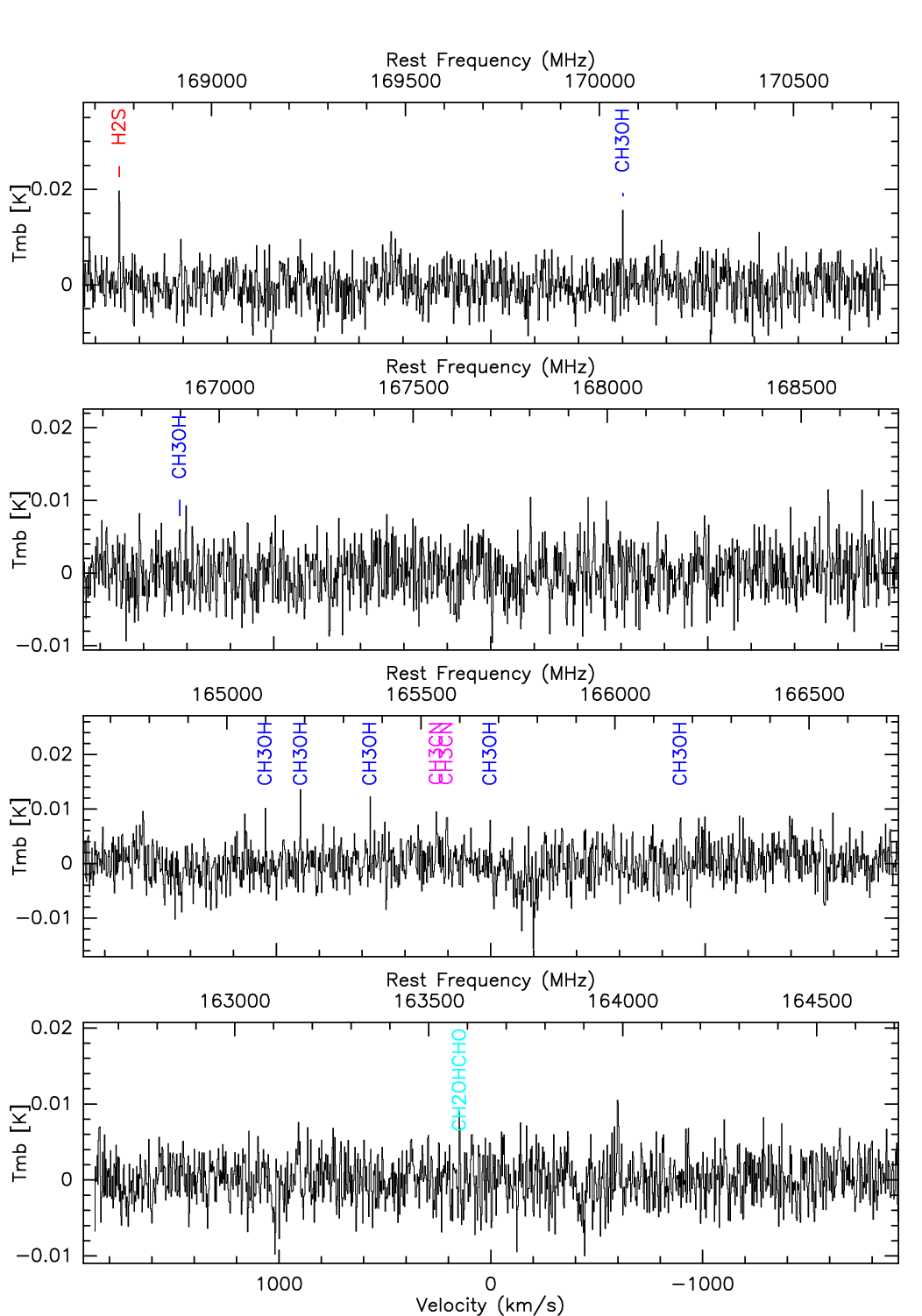}
\caption{C/2021~A1 (Leonard) 2mm spectrum.
  Vertical scale is main beam brightness temperature adjusted to the
  lines or noise level.
  The frequency scale in the rest frame of the comet is indicated on the upper
  axis. A velocity scale with a reference at the centre of each band is
  indicated on the lower axis.}
\label{figsurveyleonard2mm2}
\end{figure}

\begin{figure}[ht]\vspace{-0.0cm}
  \centering
\includegraphics[angle=0, width=16cm]{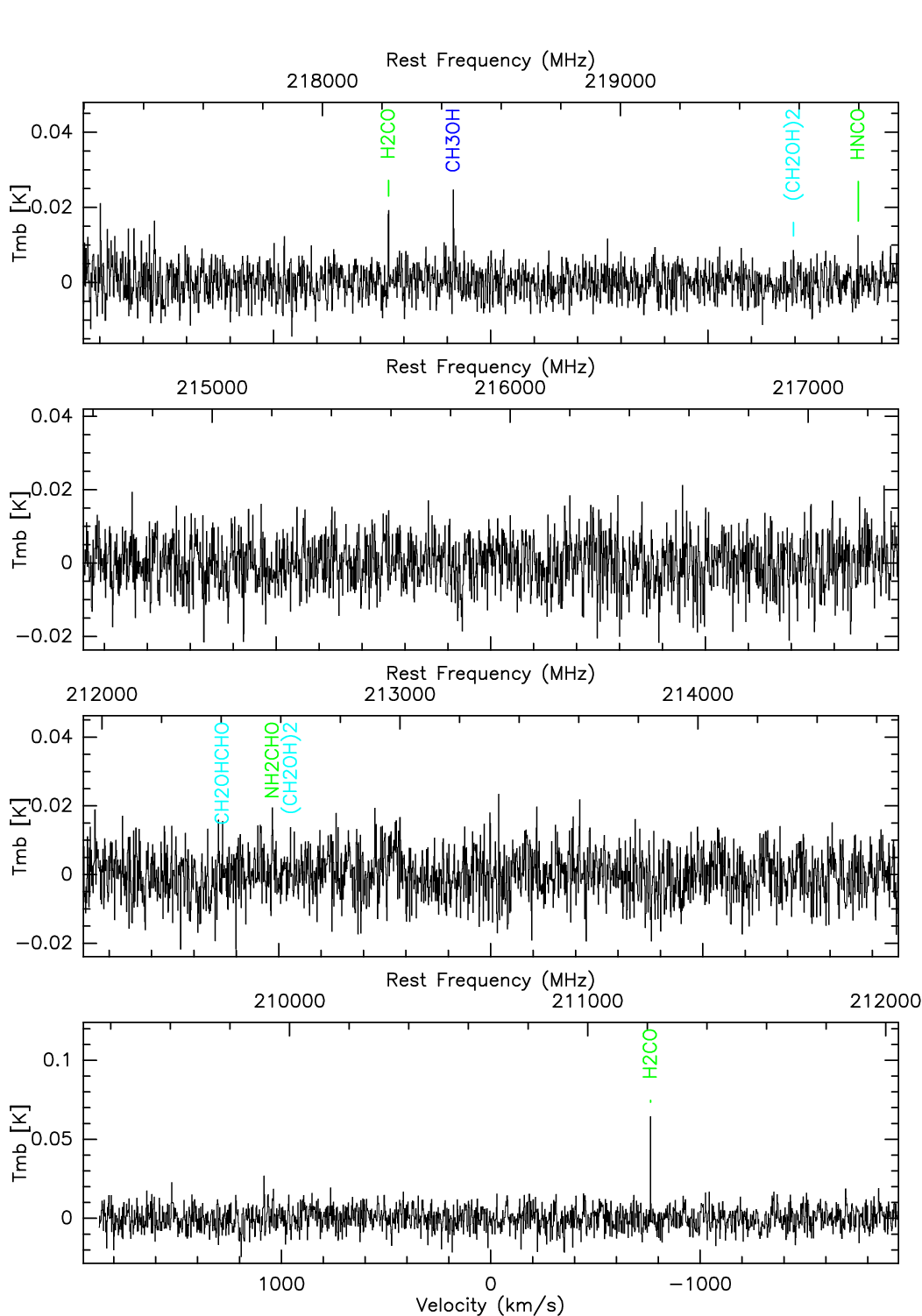}
\label{figsurveyleonard1mm1}
\end{figure}

\begin{figure}[ht]\vspace{-0.0cm}
  \centering
\includegraphics[angle=0, width=16cm]{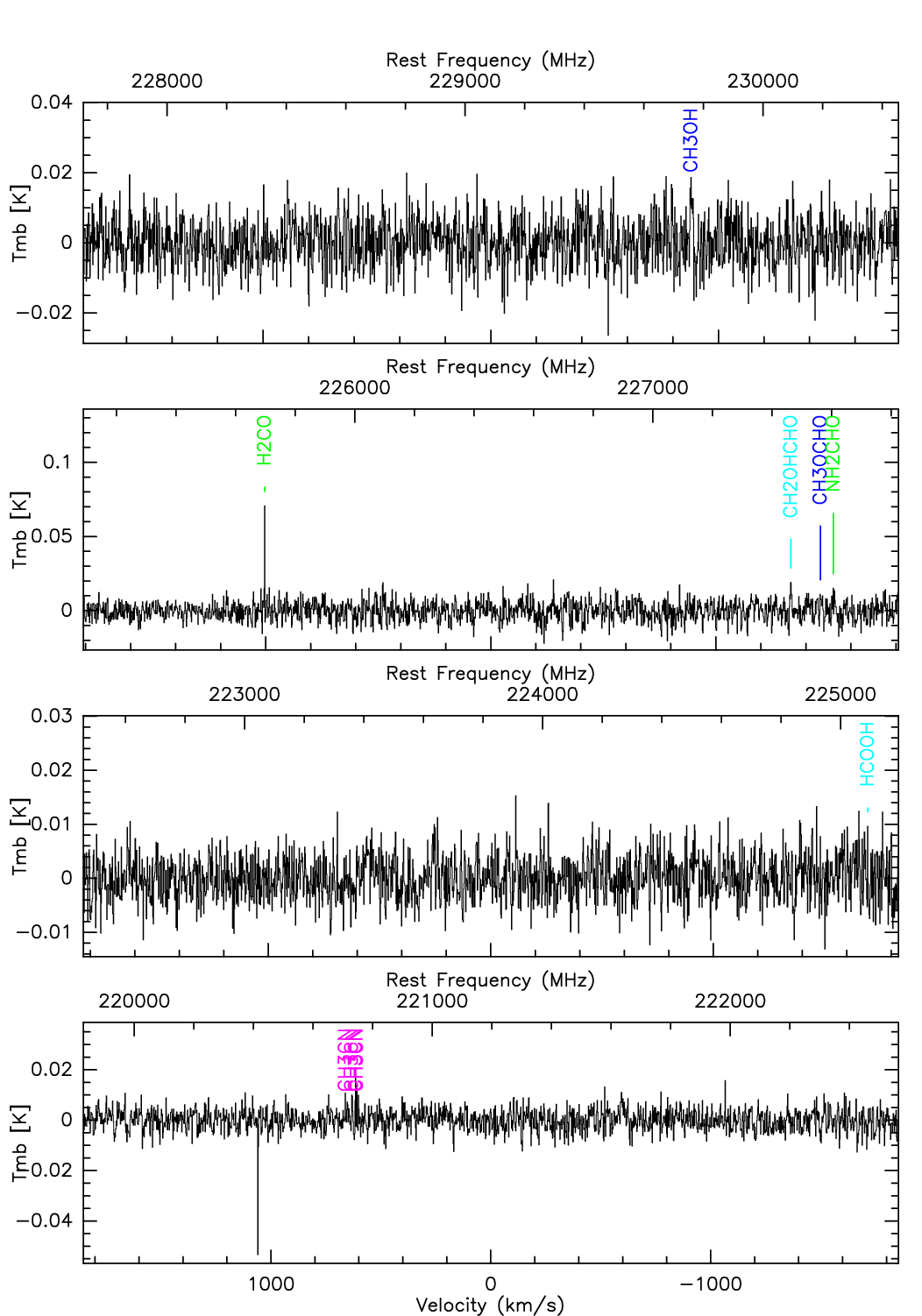}
\label{figsurveyleonard1mm2}
\end{figure}

\begin{figure}[ht]\vspace{-0.0cm}
  \centering
\includegraphics[angle=0, width=16cm]{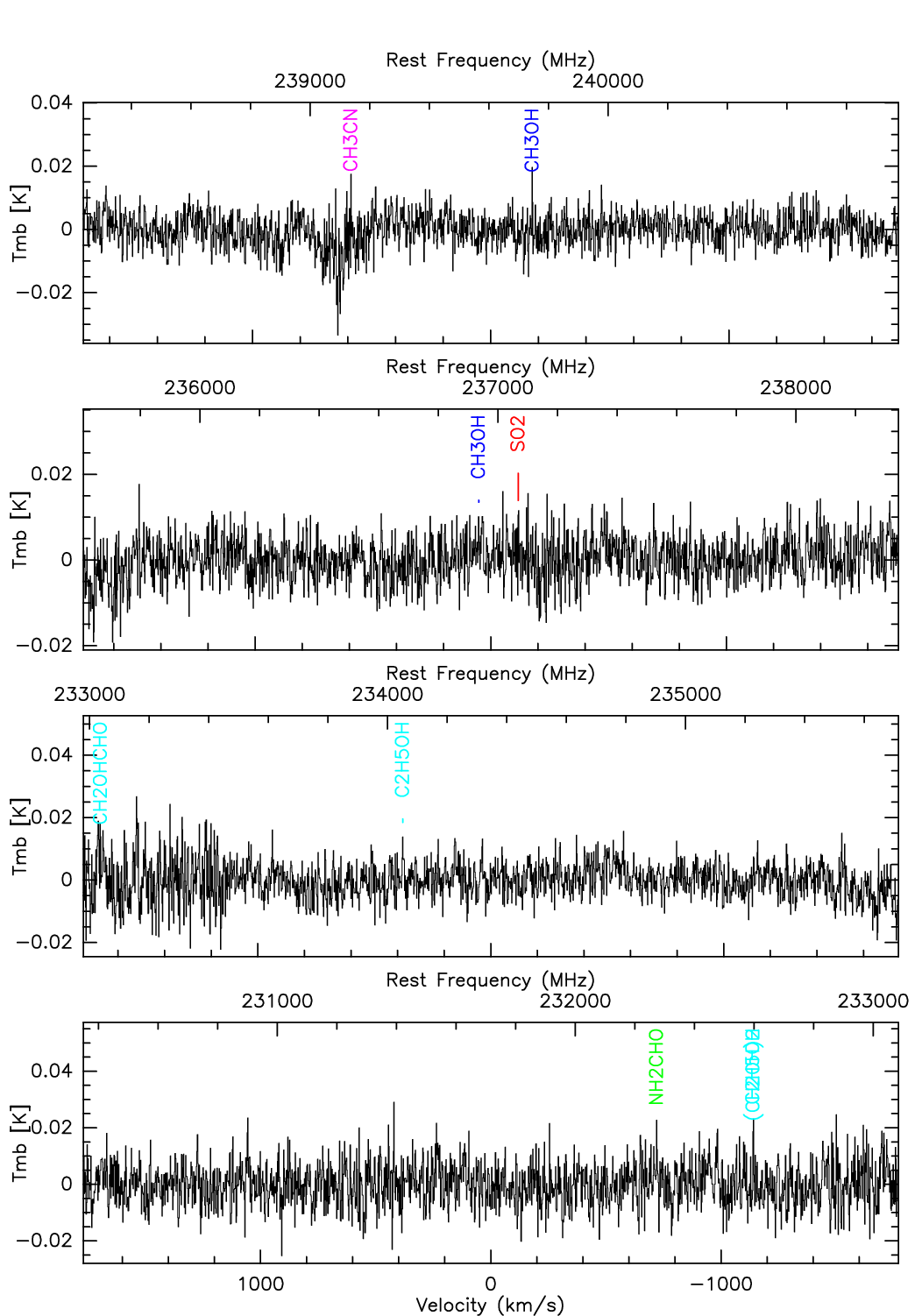}
\label{figsurveyleonard1mm3}
\end{figure}

\begin{figure}[ht]\vspace{-0.0cm}
  \centering
\includegraphics[angle=0, width=16cm]{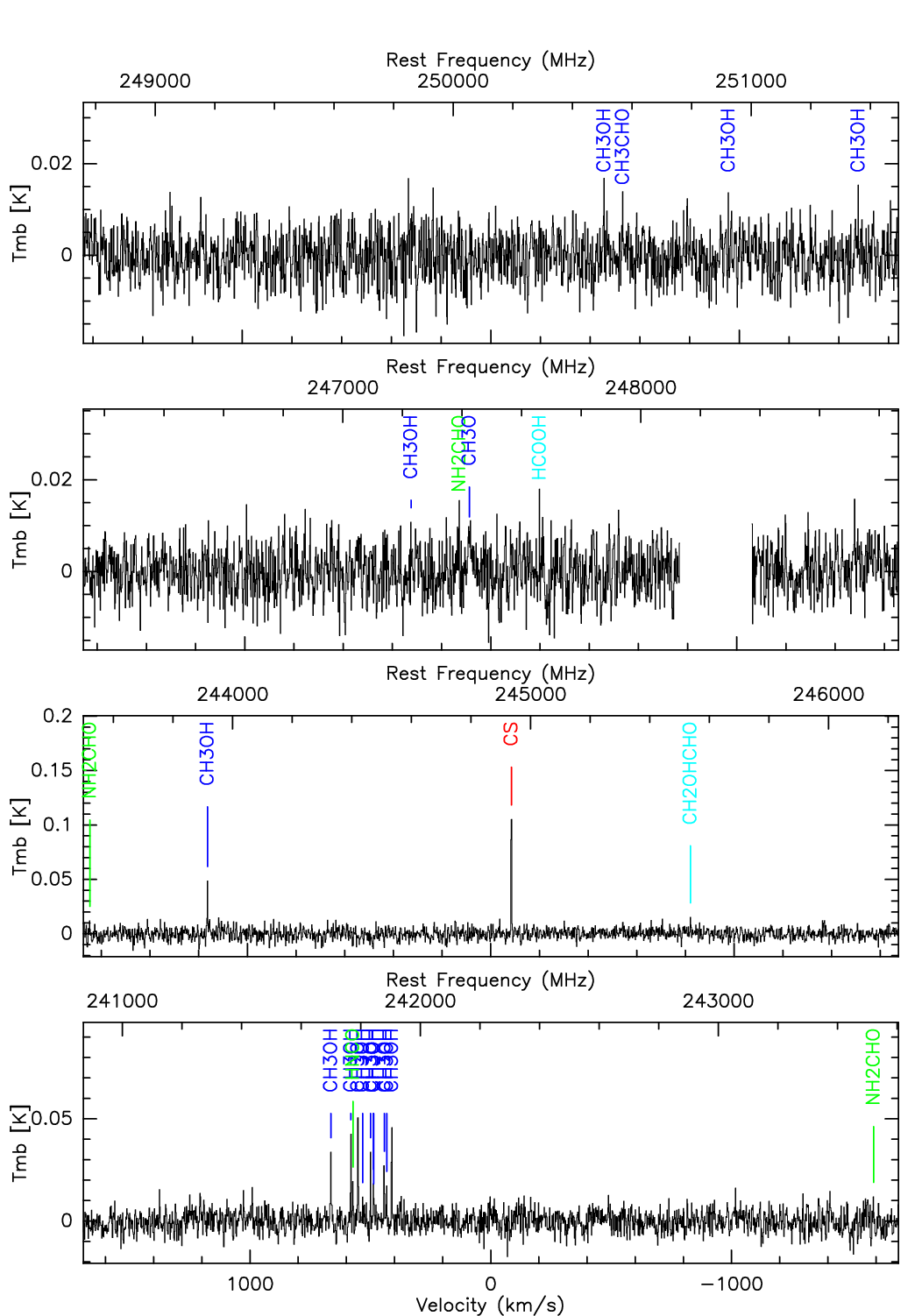}
\label{figsurveyleonard1mm4}
\end{figure}

\begin{figure}[ht]\vspace{-0.0cm}
  \centering
\includegraphics[angle=0, width=16cm]{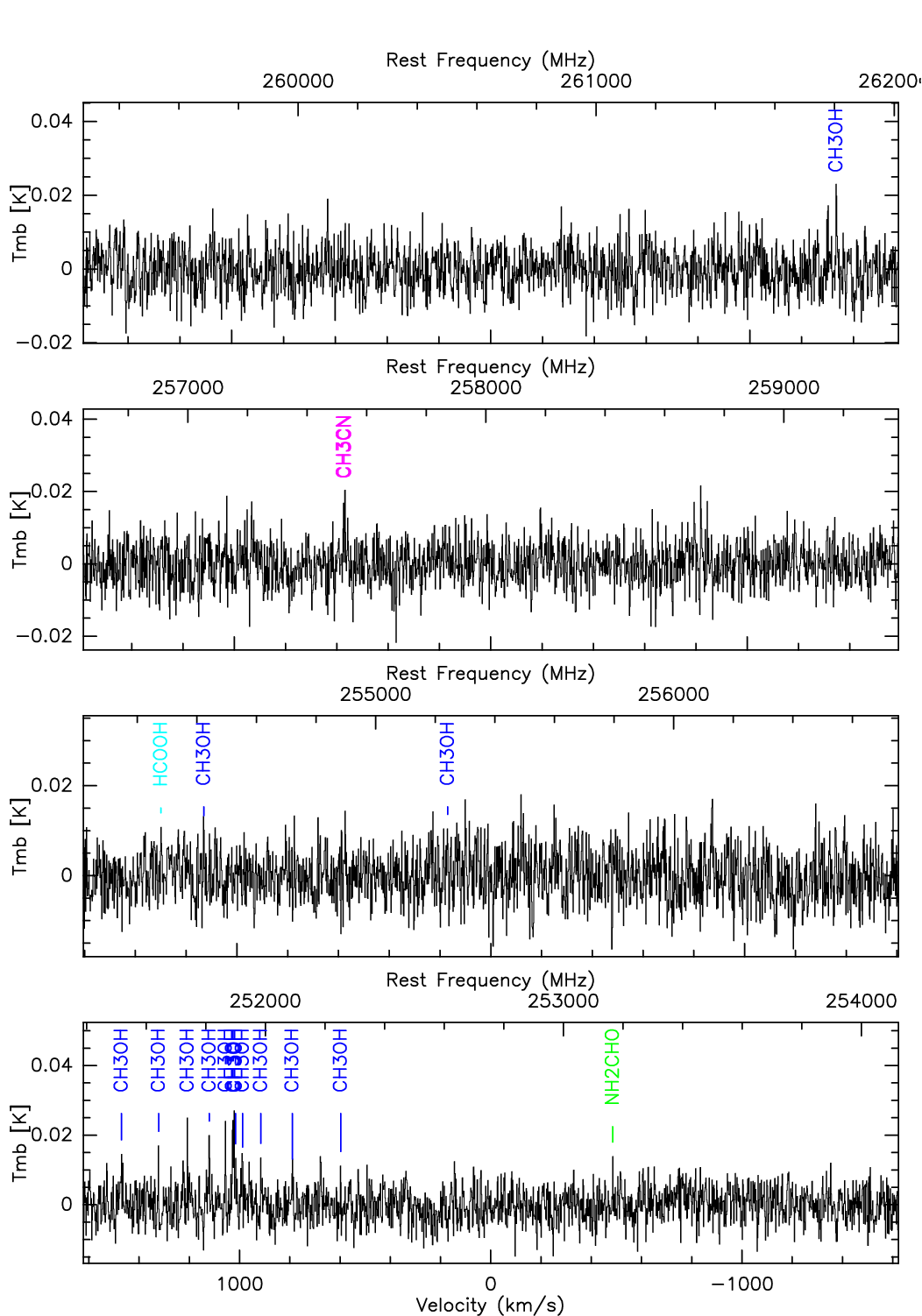}
\label{figsurveyleonard1mm5}
\end{figure}

\begin{figure}[ht]\vspace{-0.0cm}
  \centering
  \includegraphics[angle=0, width=16cm]{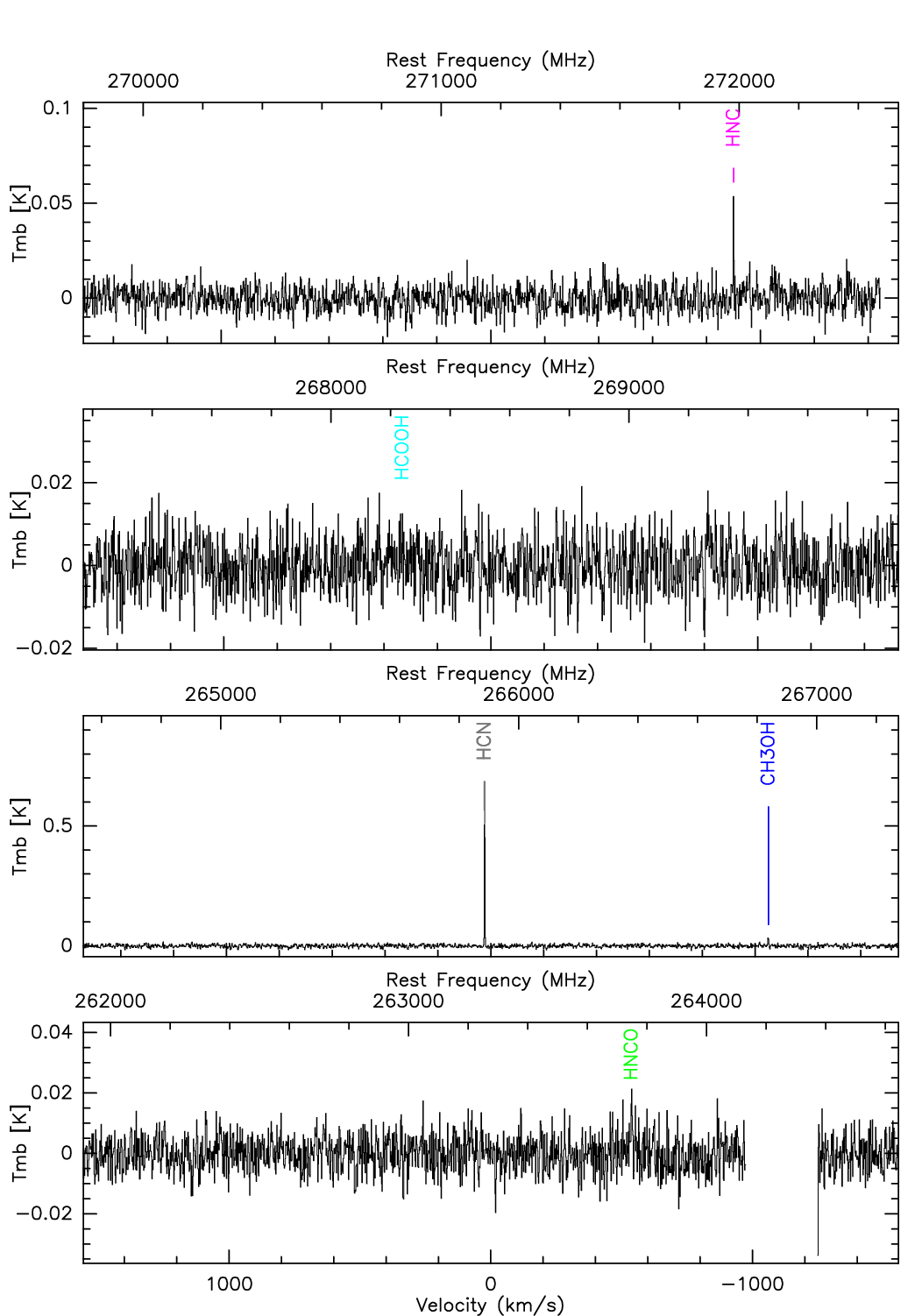}
 \caption{C/2021~A1 (Leonard) 1mm spectrum.
  Vertical scale is main beam brightness temperature adjusted to the
  lines or noise level.
  The frequency scale in the rest frame of the comet is indicated on the upper
  axis. A velocity scale with a reference at the centre of each band is
  indicated on the lower axis.}
 \label{figsurveyleonard1mm6}
\end{figure}

\begin{figure}[ht]\vspace{-0.0cm}
  \centering
\includegraphics[angle=0, width=16cm]{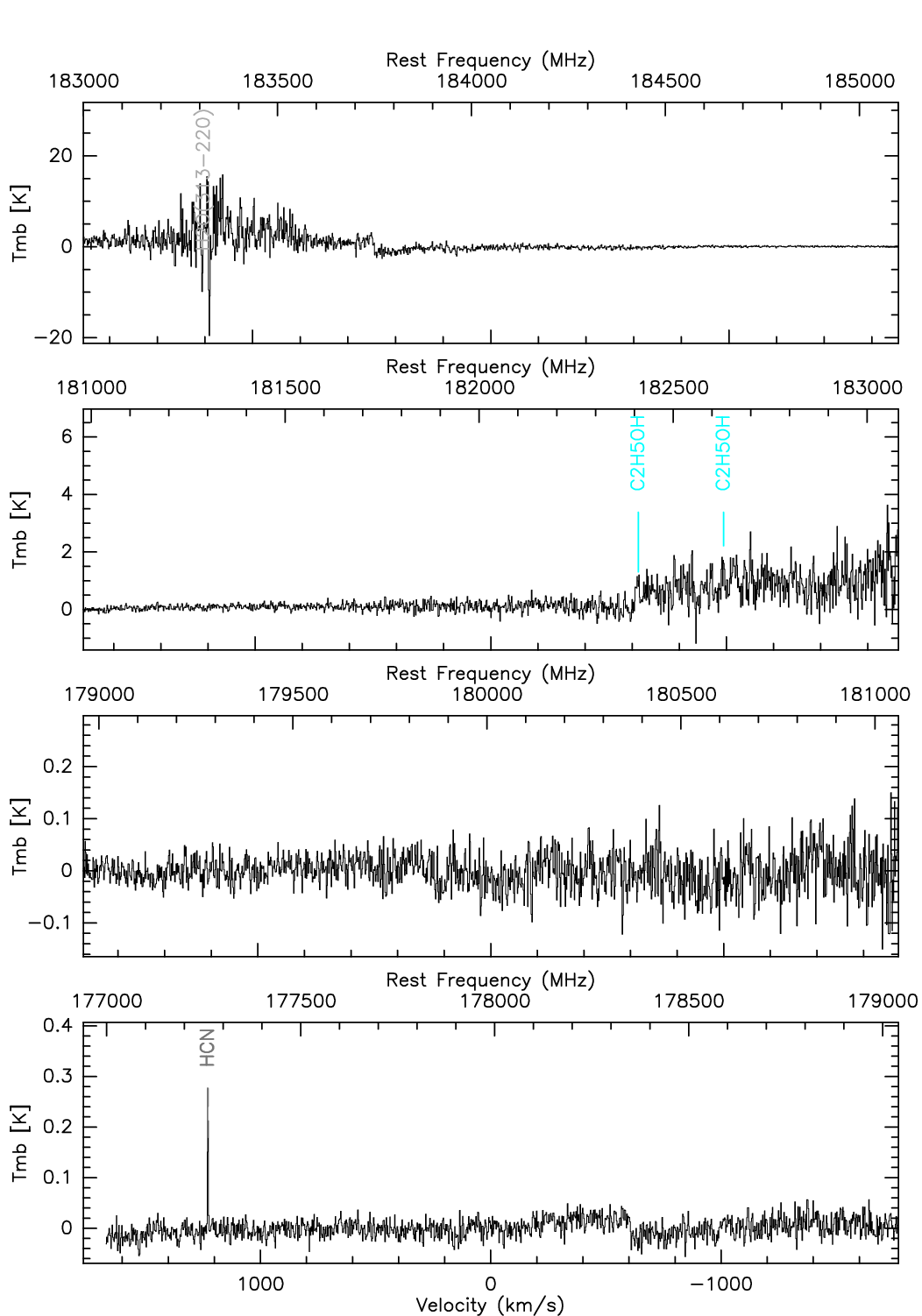}
\caption{C/2022~E3 (ZTF) 2mm spectrum.
  Vertical scale is main beam brightness temperature adjusted to the
  lines or noise level.
  The frequency scale in the rest frame of the comet is indicated on the upper
  axis. A velocity scale with a reference at the centre of each band is
  indicated on the lower axis.}
\label{figsurveyztf2mm3}
\end{figure}

\begin{figure}[ht]\vspace{-0.0cm}
  \centering
\includegraphics[angle=0, width=16cm]{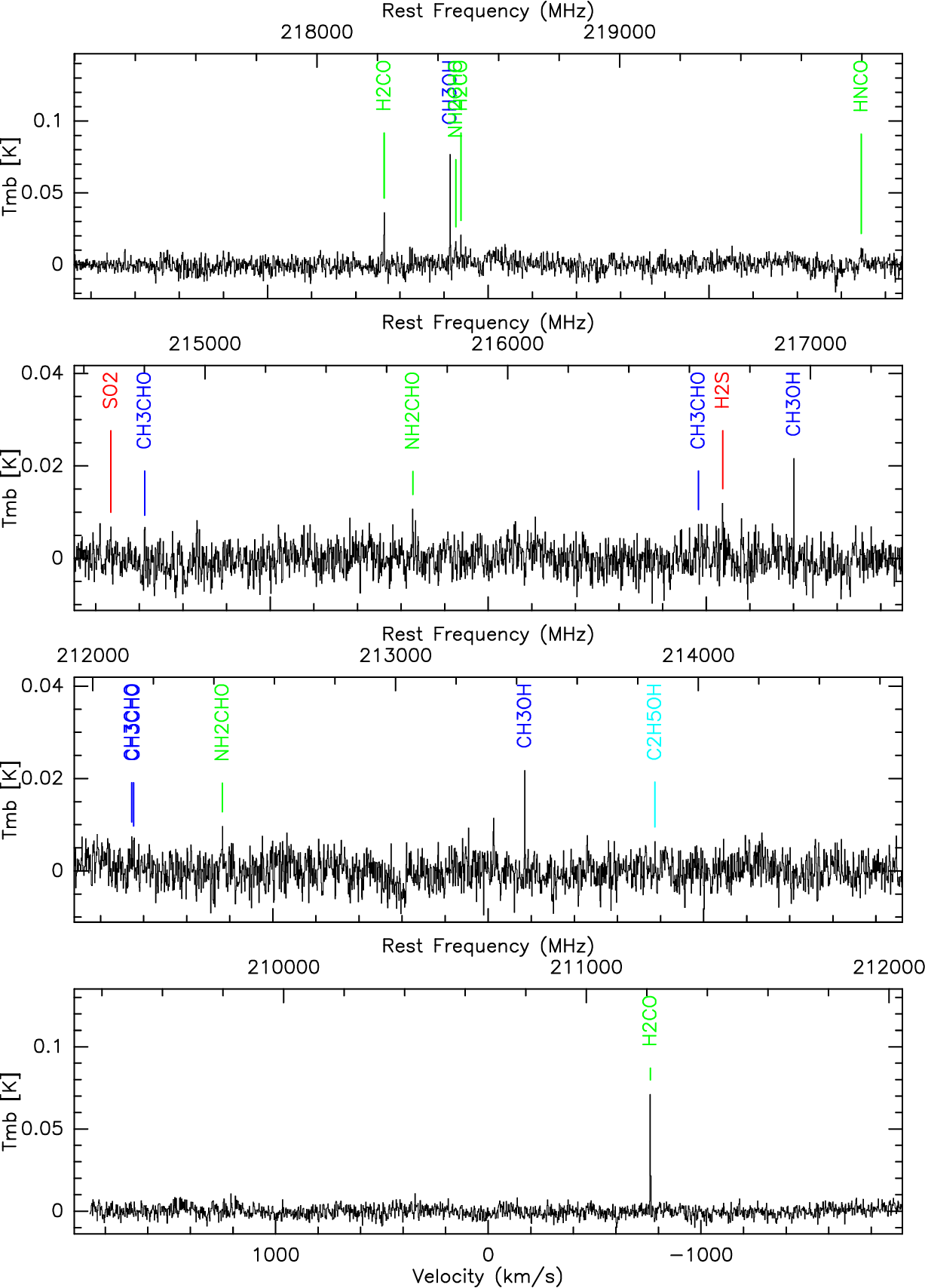}
\label{figsurveyztf1mm1}
\end{figure}

\begin{figure}[ht]\vspace{-0.0cm}
  \centering
\includegraphics[angle=0, width=16cm]{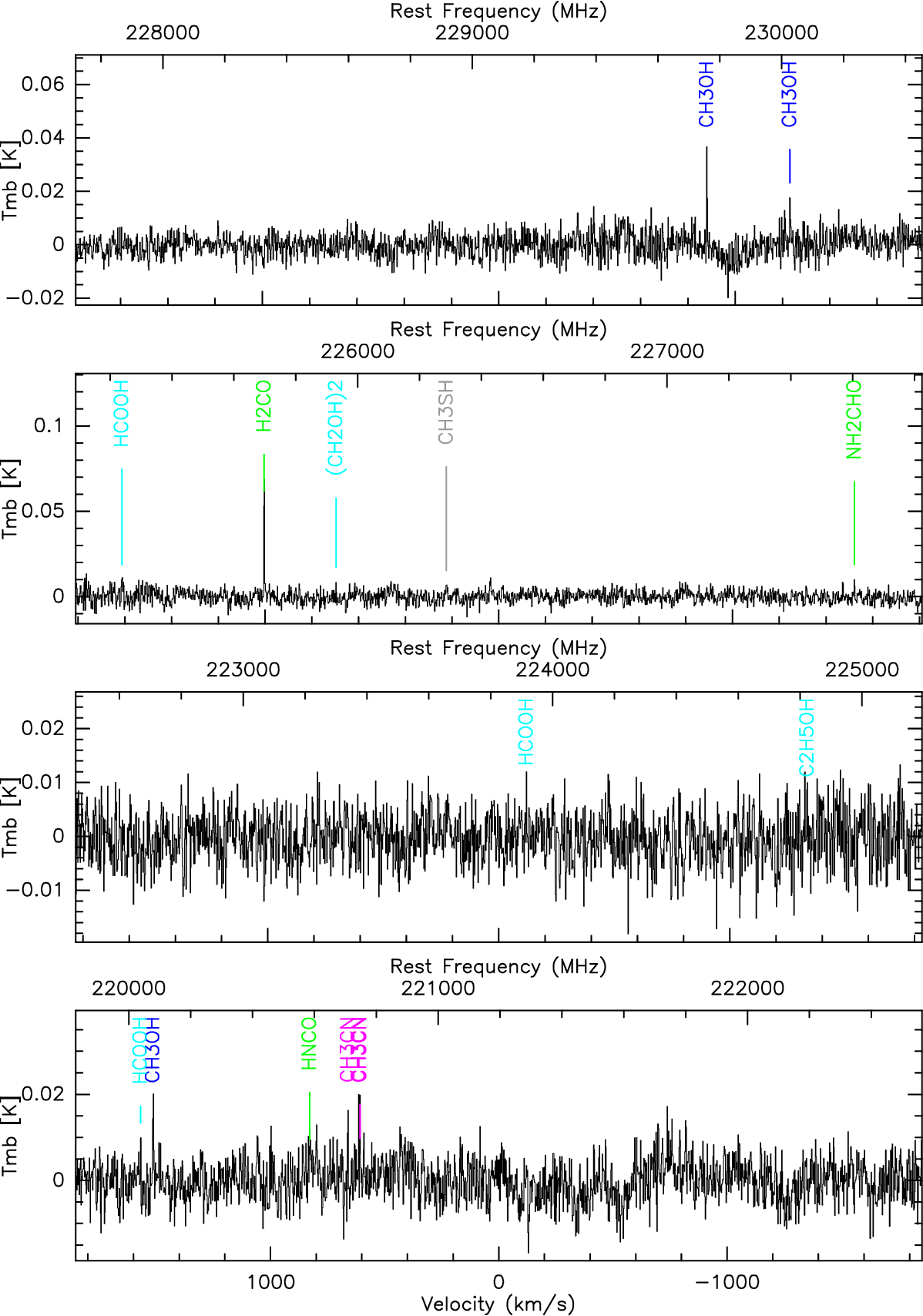}
\label{figsurveyztf1mm2}
\end{figure}

\begin{figure}[ht]\vspace{-0.0cm}
  \centering
\includegraphics[angle=0, width=16cm]{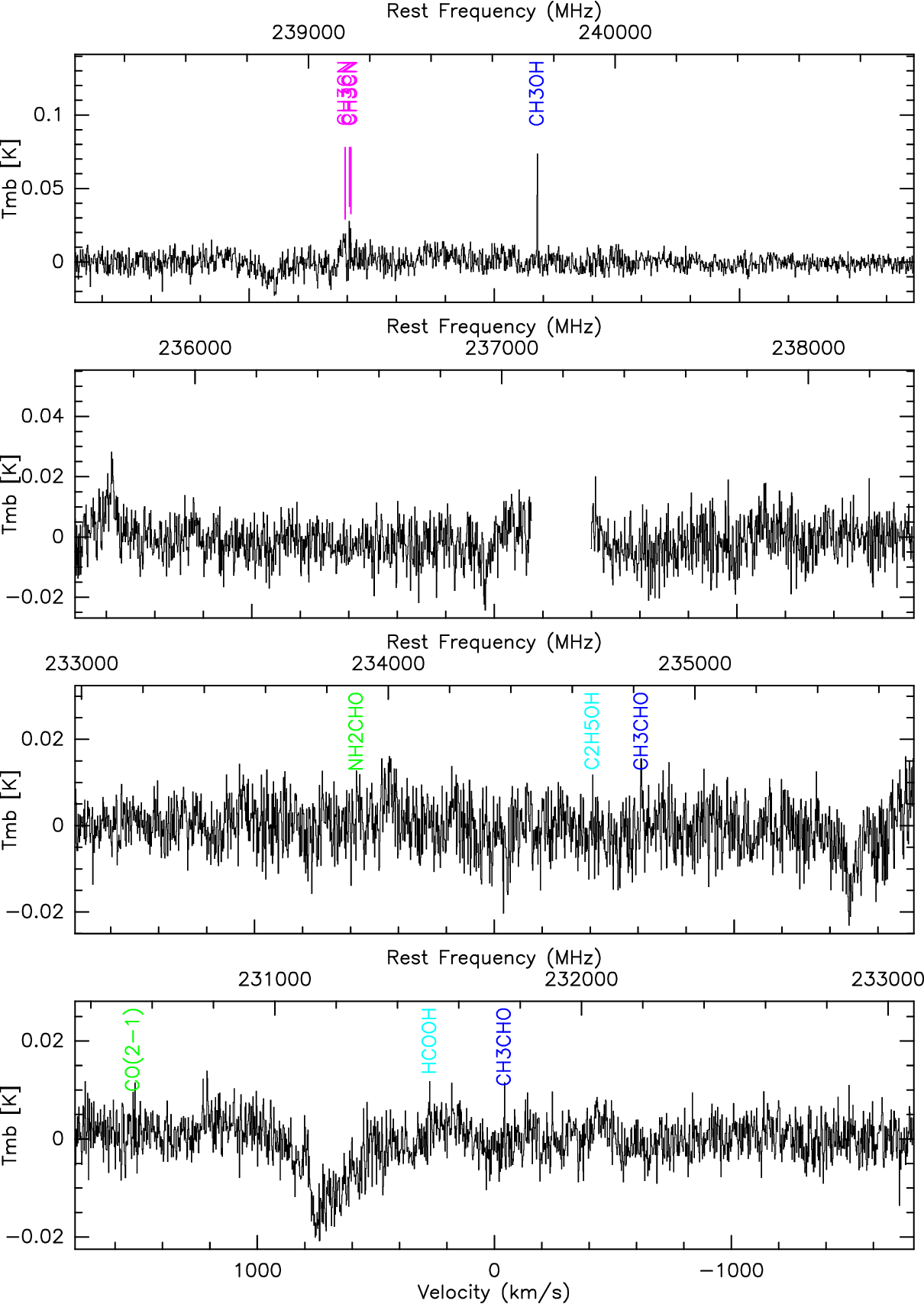}
\label{figsurveyztf1mm3}
\end{figure}

\begin{figure}[ht]\vspace{-0.0cm}
  \centering
\includegraphics[angle=0, width=16cm]{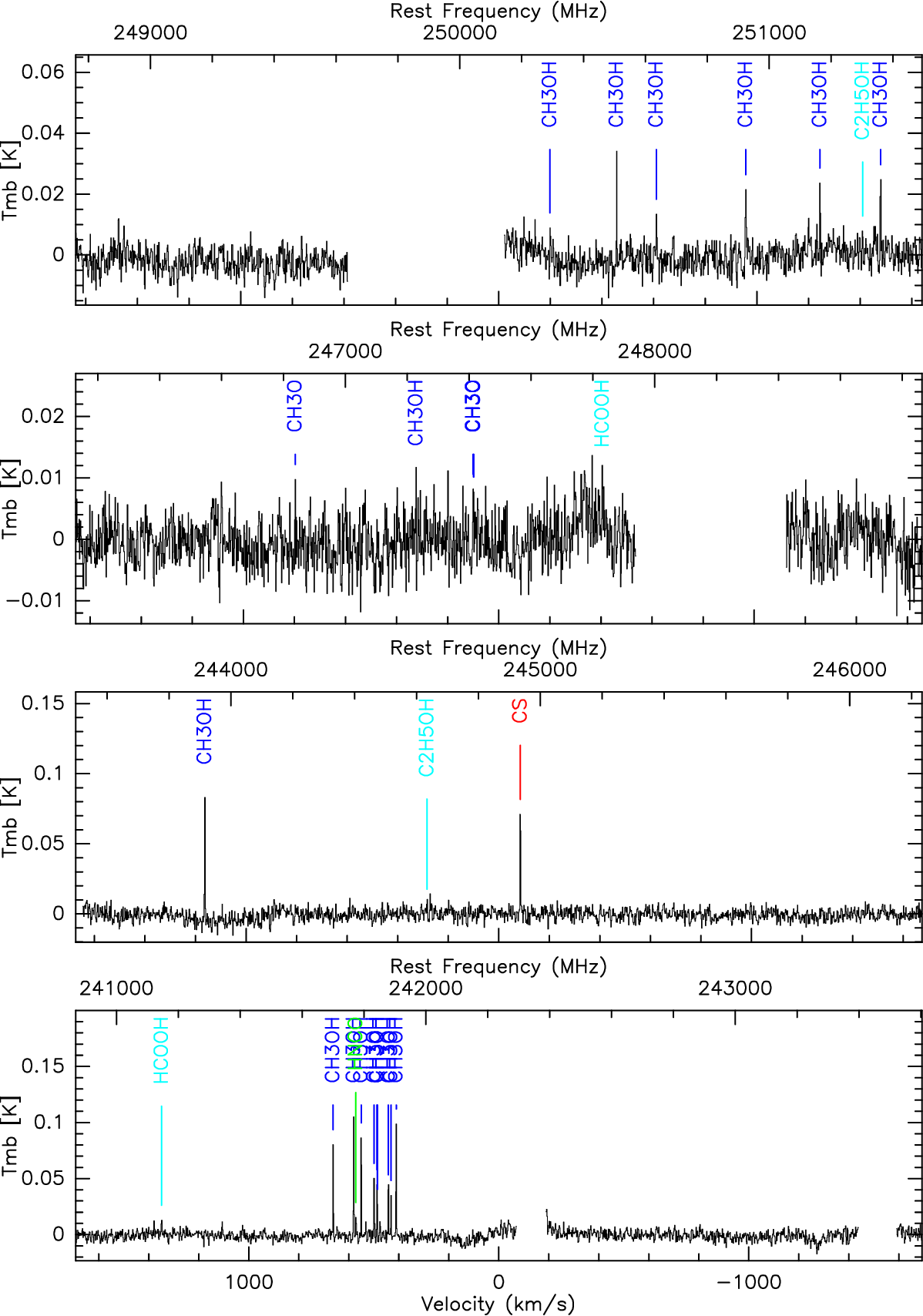}
\label{figsurveyztf1mm4}
\end{figure}

\begin{figure}[ht]\vspace{-0.0cm}
  \centering
\includegraphics[angle=0, width=16cm]{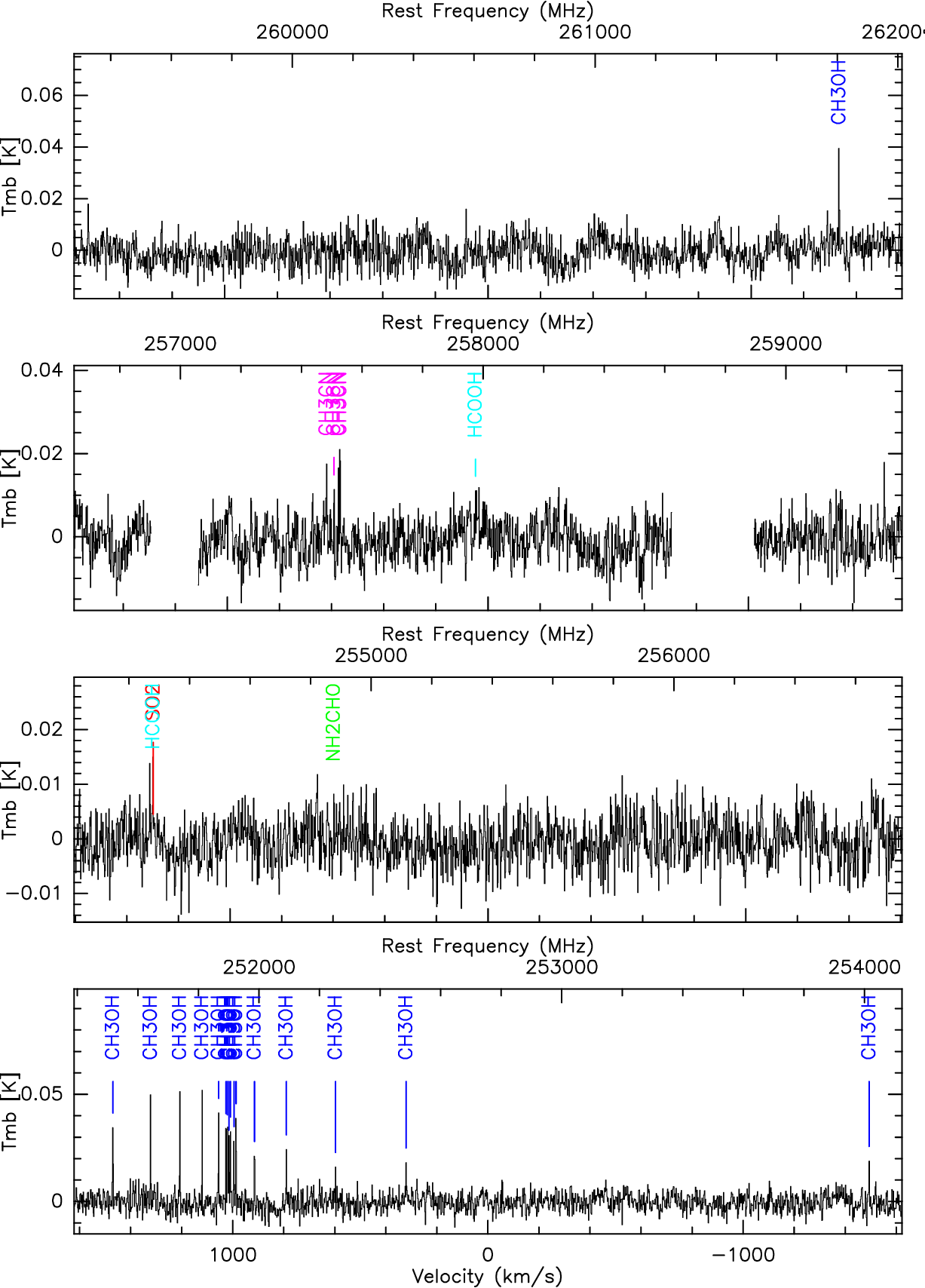}
\label{figsurveyztf1mm5}
\end{figure}

\begin{figure}[ht]\vspace{-0.0cm}
  \centering
  \includegraphics[angle=0, width=16cm]{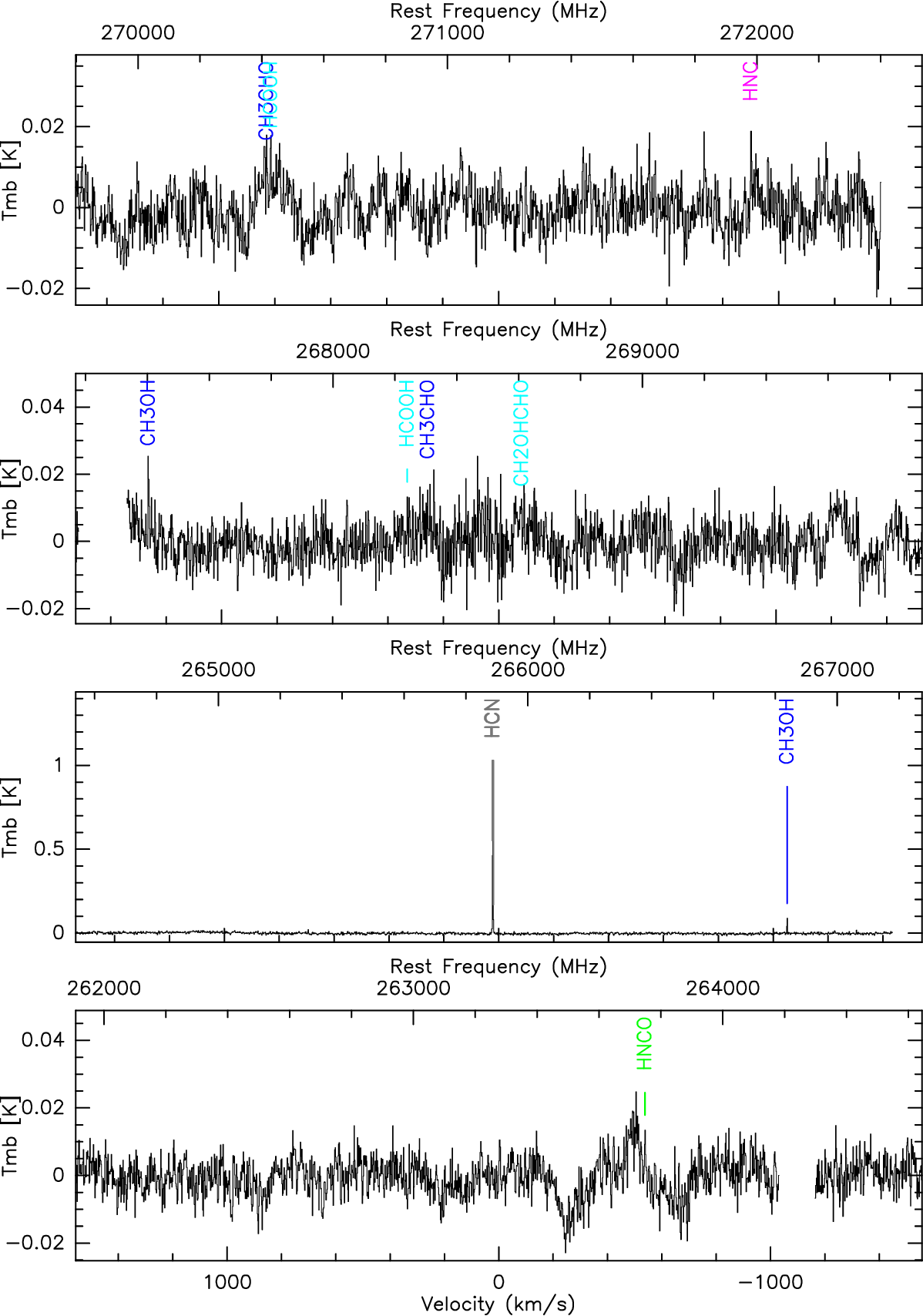}
  \caption{C/2022~E3 (ZTF) 1mm spectrum.
  Vertical scale is main beam brightness temperature adjusted to the
  lines or noise level.
  The frequency scale in the rest frame of the comet is indicated on the upper
  axis. A velocity scale with a reference at the centre of each band is
  indicated on the lower axis.}
\label{figsurveyztf1mm6}
\end{figure}

The following pages present the 147--185~GHz ($\lambda=2$mm) and 210--272~GHz
($\lambda=1$mm) FTS spectra of comets C/2021~A1 (Leonard)
obtained between 8.4 and 13.7 Dec. 2021
(\ref{figsurveyleonard2mm2}--\ref{figsurveyleonard1mm6})
and C/2022~E3 (ZTF) obtained between 3.7 and 7.0 February 2023
(\ref{figsurveyztf2mm3}--\ref{figsurveyztf1mm6})
with the IRAM 30-m telescope.
  The wavelength domains covered are plotted by series of $\approx2-2.5$~GHz
  windows with a spectral resolution smoothed to 0.78~MHz.
  Only spectra obtained within 2\arcsec~  of the nucleus were taken into
  account, and the strongest lines are identified.
  We note the much higher noise level around the frequency
  of the atmospheric H$_2$O line at 183310~MHz targeted in comet C/2022~E3.
  In the case of comet C/2022~E3, spectra having been mostly obtained in PSW
  mode, residuals of atmospheric lines are present in the spectrum and the
  strongest ones have been blanked.
  Identification of lines at the position of stronger peaks ($\approx3\sigma$)
  have been added to the spectra. Not all correspond to real detections and
  alternatively some detections are missed in the automatic process due to
  variable noise across the bandwidth plotted.
  
\begin{figure}[ht]\vspace{-1.0cm}
  \centering
\includegraphics[angle=0, width=16cm]{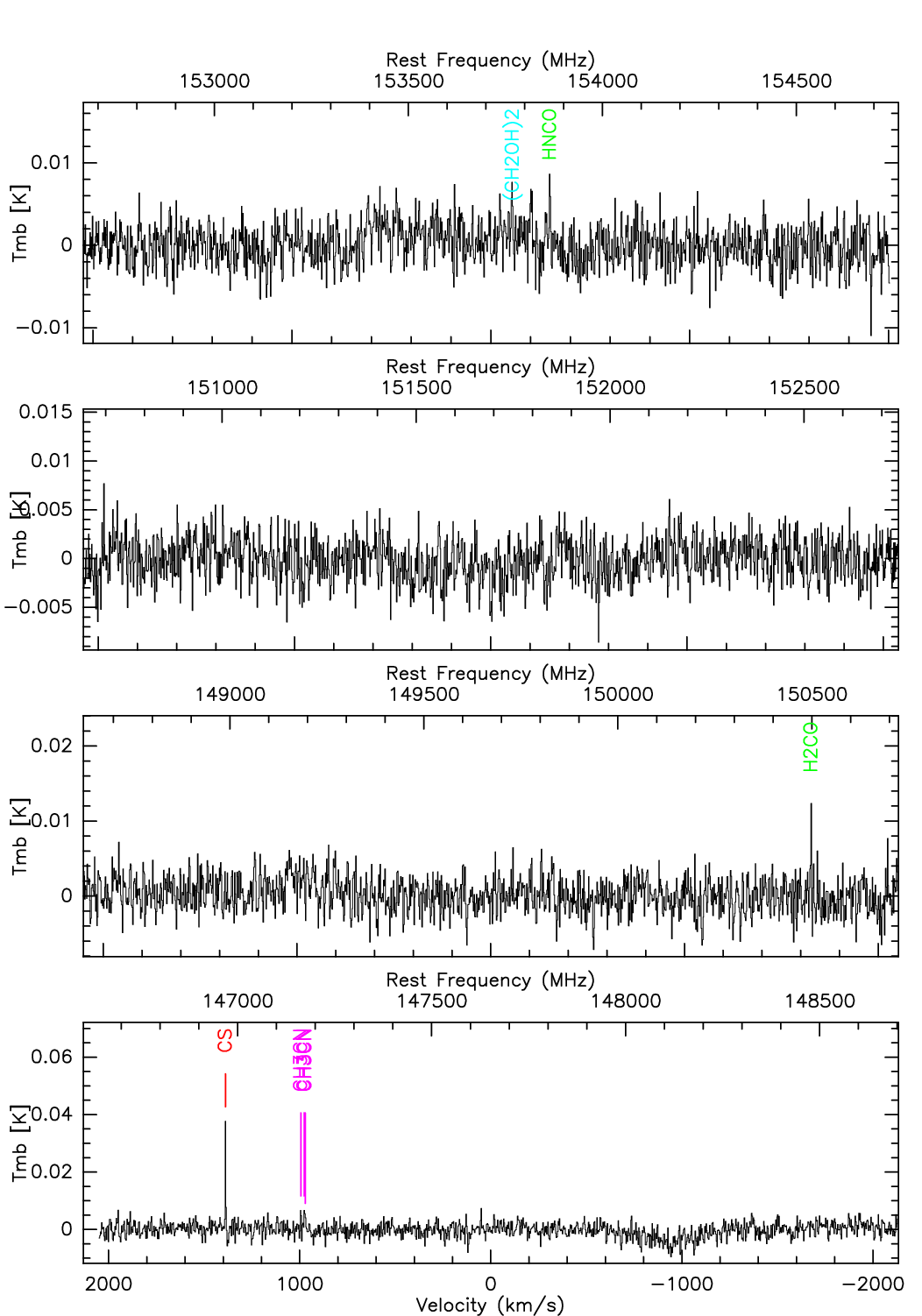}
\label{figsurveyleonard2mm1}
\end{figure}


\begin{figure}[ht]\vspace{-1.0cm}
  \centering
\includegraphics[angle=0, width=16cm]{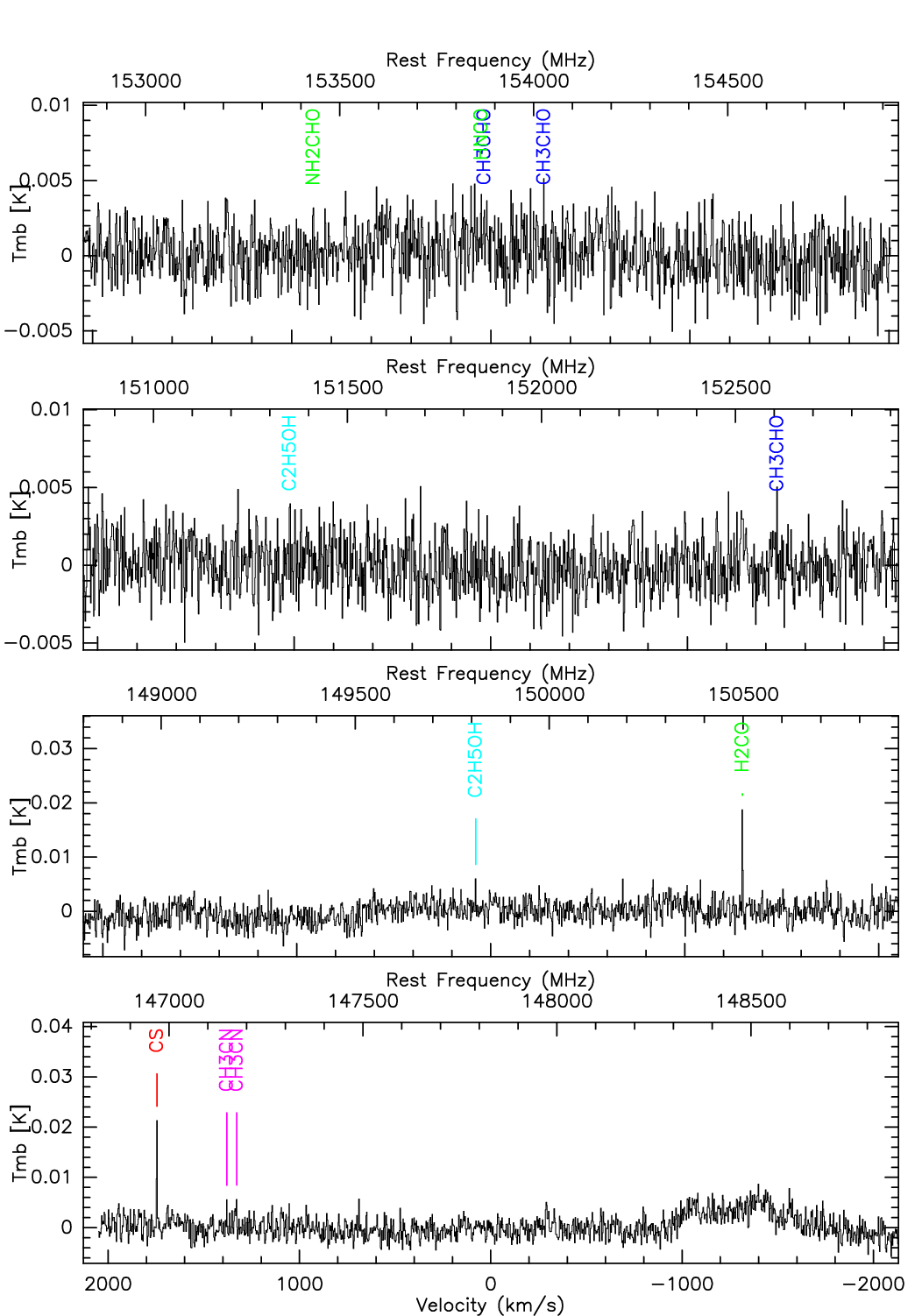}
\label{figsurveyztf2mm1}
\end{figure}
\begin{figure}[ht]\vspace{-0.0cm}
  \centering
\includegraphics[angle=0, width=16cm]{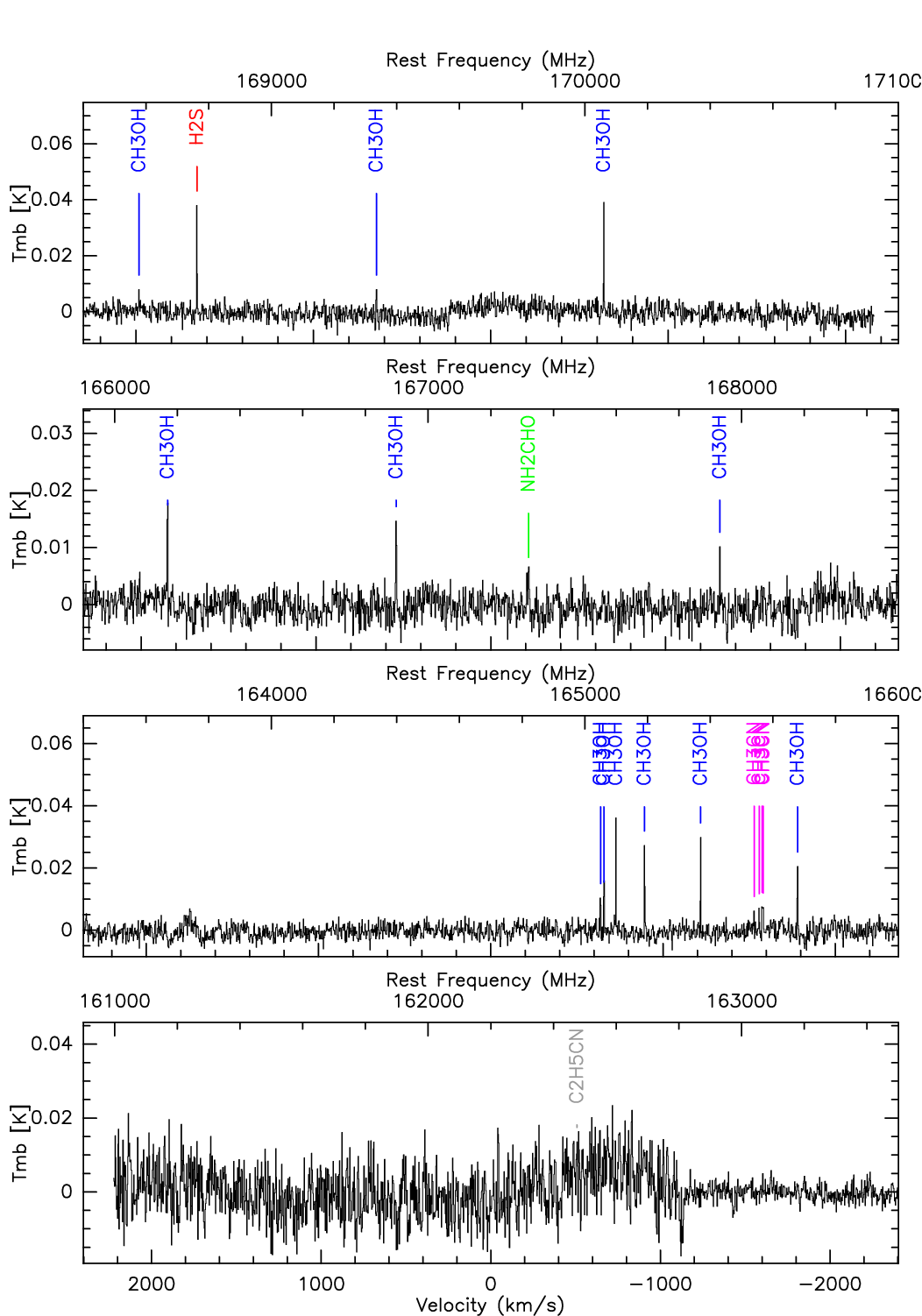}
\label{figsurveyztf2mm2}
\end{figure}


\end{appendix}

\end{document}